\documentclass[journal,10pt]{IEEEtran}
\usepackage[noadjust]{cite}
\usepackage{graphicx} 
\usepackage{multirow}
\usepackage{float}
\usepackage{multicol}
\usepackage{lipsum}
\usepackage[numbers,square, comma, sort & compress]{natbib}
\usepackage{amssymb,amsmath,enumerate}
\usepackage{float}
\usepackage{dblfloatfix}
\usepackage{array}
\newcolumntype{C}[1]{>{\centering\let\newline\\\arraybackslash\hspace{0pt}}m{#1}}
\ifCLASSOPTIONcompsoc
\usepackage[caption=false,font=footnotesize,labelfont=sf,textfont=sf]{subfig}
\else
\usepackage[caption=false,font=footnotesize]{subfig}
\fi

\usepackage{hhline}

\usepackage{color}

\begin{document}

 \title{Spectral Coexistence of LDACS and DME: Analysis via Hardware Software Co-Design in Presence of Real Channels and RF Impairments}
	


\author{Niharika Agrawal, S. J. Darak and Faouzi Bader
	\IEEEcompsocitemizethanks{\IEEEcompsocthanksitem Niharika Agrawal, and S. J. Darak are with the Department of ECE, IIIT-Delhi, India. E-mail: \{niharikaa, sumit\}@iiitd.ac.in. 	
		
		Faouzi Bader is with the SCEE/IETR - CentraleSup´elec, campus of Rennes, France and ISEP - Institut Supérieur d'Electronique de Paris, France. E-mail: faouzi.bader@isep.fr. 
	}
}
\maketitle

\begin{abstract}
To meet the exponentially increasing air traffic, L-band (960-1164 MHz) digital aeronautical communication system (LDACS) has been introduced. The LDACS aims to exploit
the vacant spectrum between incumbent Distance Measuring Equipment (DME) signals and envisioned to follow an orthogonal frequency division multiplexing (OFDM) approach to support high-speed delay-sensitive multimedia services. This paper deals with the design and implementation of end-to-end LDACS the transceiver on the Zynq System on Chip platform, consisting of FPGA as programmable logic (PL) and ARM as processing
system (PS). We consider OFDM based LDACS and improve it further using windowing and/or filtering. We propose a hardware-software co-design approach and analyze various transceiver configurations by dividing it into PL and PS. We demonstrate the flexibility offered by such a co-design approach to choose the configuration as well as word-length for a given area, delay and power constraints. The transceiver is also integrated with the programmable analog front-end to validate its functionality in the presence of various RF impairments and wireless channels and interference specific to the LDACS environment. Via in-depth performance analysis concerning parameters such as out-of-band attenuation, DME interference, bit-error-rate, word-length, and complexity, we demonstrate wide bandwidth filtered OFDM as an attractive solution for the next generation LDACS.


\end{abstract}


\section{Introduction}
\begin{figure}[h!]
		\vspace{-0.2cm}
	\centering
	\captionsetup{justification=centering}
	\includegraphics[scale=0.275]{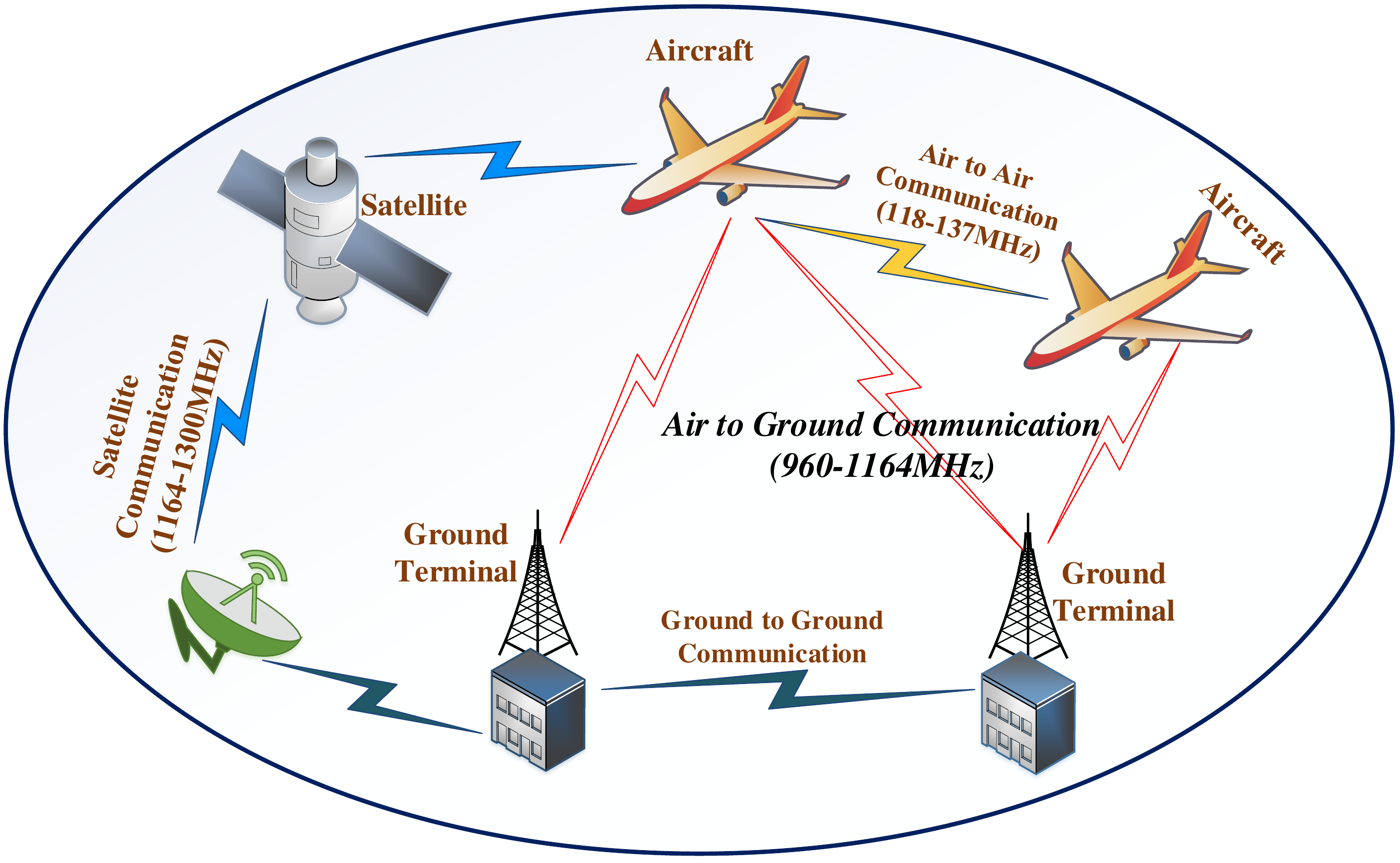}
	\vspace{-0.2cm}
	\caption{Various communication links in the future communication infrastructure (FCI).}
	\label{a2g}
		\vspace{-0.2cm}
\end{figure}

International Civil Aviation Organization (ICAO) envisioned the need for Future Communication Infrastructure (FCI) for aeronautical systems to support exponentially increasing air traffic and enable a wide range of services from voice data to multimedia \cite{STAT, Schnell, SESAR}. The FCI is expected to be deployed in communications, navigation, and surveillance (CNS) applications as well. Research projects such as Next Generation Air Transportation System (NextGen) and Single European Sky air traffic management (ATM) Research (SESAR) \cite{SESAR} have been given the mandate to propose and demonstrate the FCI prototype. As shown in Fig.~\ref{a2g}, FCI comprises several data links such as air-to-ground communication (A2GC), air-to-air communication, ground-to-ground communication, satellite-ground communication, and vice-versa. The A2GC link enables two-way communication between aircrafts and ground terminal, and it is the most critical data link in the FCI. 
The ICAO standardization committee has proposed to switch the A2GC link from the narrow band (118-137 MHz) to wider $L$-band (960-1164 MHz), and the corresponding system is referred to as $L$-band digital aeronautical communications system (LDACS).

 \begin{figure}[!b]
 		\vspace{-0.75cm}
	\centering
	\captionsetup{justification=centering}
	\includegraphics[width=\linewidth]{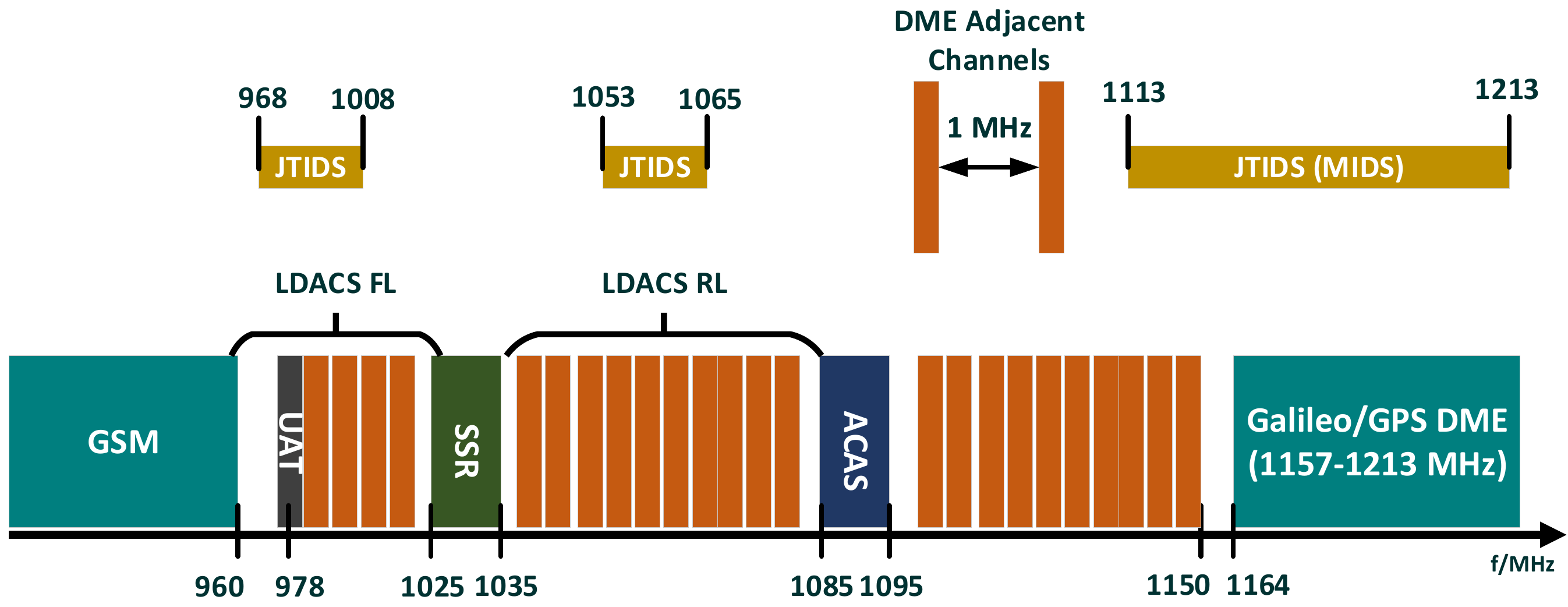}
	\vspace{-0.75cm}
	\caption{Various incumbent users and spectrum occupancy in $L$-band.}
	\label{lband}
\end{figure}

The $L$-band spectrum allocation, shown in Fig.~\ref{lband}, indicates that it has been occupied by various incumbent users such as distance measuring equipment (DME), Multi-functional Information Distribution System (MIDS), Joint Tactical Information Distribution System (JTIDS), Universal Access Transceiver (UAT), Secondary Surveillance Radar (SSR)/Airborne Collision Avoidance System (ACAS), etc. Based on various spectrum measurement studies, ICAO has identified multiple 1~MHz vacant bands between adjacent DME signals for LDACS. To exploit these bands for the A2GC link, ICAO proposed preliminary LDACS transceiver specifications based on orthogonal frequency division multiplexing (OFDM) transceivers. The OFDM has advantages such as low complexity, simple channel equalization, and multi-antenna support. However, the drawbacks such as large out-of-band emission (OOBE), limited flexibility, and stringent synchronization requirements limit the LDACS transmission bandwidth to at most 498 kHz (less than 50\% spectrum utilization) due to significant interference to incumbent DME signals. Thus, ICAO expects further research on various windowing and filtering techniques to improve spectrum utilization and feasibility of OFDM based LDACS in complex channel environments encountered in the A2GC link during multiple stages of flight \cite{Zavani, Abdoli}. From the architecture perspective, most of the LDACS transceivers are analyzed via simulations, and their performance analysis on fixed-point hardware in the presence of various RF impairments and wireless channels/interference has not been done yet.
 
The main objective of the proposed work is to design and implement end-to-end LDACS transceiver on heterogeneous Zynq System on Chip (ZSoC) platform, consisting of FPGA as PL and Advanced RISC Machines (ARM) as PS. We also provide detailed performance analysis for the parameters such as windowing, filtering, OOBE, DME interference, bit-error-rate (BER), word-length, area, delay, and power. The contributions of the paper can be summarized as:

\begin{enumerate}    
\item We design and implement fixed-point OFDM based LDACS and analyze the effect of windowing and/or filtering approaches. Based on the analysis, we suggest enhancements to existing LDACS specifications to improve spectrum efficiency.

\item Since each transceiver block can be realized on PS as well as PL; we provide the architecture for efficient sequential execution on PS and efficient parallel execution on the PL.  

\item We propose a novel hardware-software co-design approach and implement various transceiver configurations by dividing it into PL and PS. We demonstrate the flexibility offered by such a co-design approach to choose the configuration, pipelining, and word-length for a given OOBE, BER, area, delay, and power constraints. 

\item In the end, various configurations are integrated with programmable analog front-end (AFE) to validate the transceiver functionality in the presence of various RF impairments and wireless channels/interference specific to the LDACS environment. 
\end{enumerate}

The first three contributions are a significant extension of our work in \cite{sasha}. In this paper, we design and implement four more transceiver configurations than \cite{sasha}. The performance analysis presented here is detailed as we consider the effect of word-length,  pipelining, LDACS specific channels, DME interference as compared to only power spectral density in \cite{sasha}. Furthermore, we integrate the proposed transceiver with programmable AFE and analyze the effects of RF impairments.

The remaining paper is organized as follows. Section II describes the work done previously in this area. Hardware-Software requirements for the transceiver models are discussed in section III. In section IV and V, the transceiver architecture followed by its variants implementation using hardware-software co-design on ZSoC along with pipelining and AD9361 integration are presented. Experimental results are analyzed in section VI. Section VII concludes the paper.

\begin{figure*}[!b]
\centering
\vspace{-0.2cm}
	\captionsetup{justification=centering}
	\includegraphics[width=0.95\linewidth]{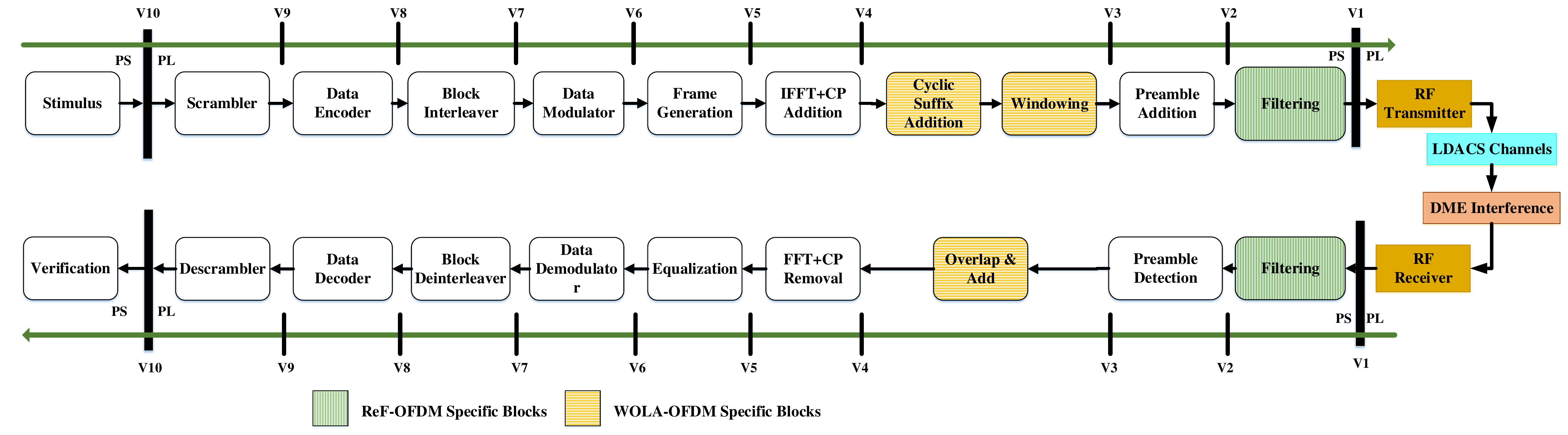}
	\vspace{-0.25cm}
	\caption{Block diagram showing different configurations of the LDACS transceiver along with windowing and filtering blocks.}\vspace{-0.2cm}
	\label{bl_diag}
\end{figure*}

\section{Literature Review}
Various works dealing with the performance analysis and feasibility of OFDM based LDACS transceivers for a wide range of CNS applications are discussed in \cite{Ta1, Ta2, Ta3}. In this section, we focus on the design, implementation, and validation of the LDACS transceiver as well as potential alternatives to improve its performance.

The implementation of various blocks in the conventional OFDM based LDACS on homogeneous platforms such as field-programmable gate arrays (FPGA) or application-specific integrated circuits (ASIC) have been discussed in \cite{Ta5, Ta6, Ta7, Ta8, Ta9, Ta10}. The primary focus of these works was strictly on the synchronization and channel estimation techniques for the LDACS environment. In \cite{Ta5}, a novel correlation-based synchronization approach for large carrier frequency offsets is proposed, and its implementation on the FPGA has shown to consume lower area and power without compromising on the BER performance \cite{Ta6}. In \cite{Ta7}, partial reconfiguration capability of the FPGA is used to design a flexible LDACS transceiver. It offers a significant improvement in the area and power consumption, but provided gains cannot be extended for implementation on the ASIC. In \cite{Ta8}, a novel sensing method for sensing the active LDACS transmissions via a multiplier-less correlation-based approach is proposed. It offers improved performance, especially at a low signal-to-noise ratio (SNR), and lowers power consumption than other architectures. On the receiver side, reconfigurable low complexity filter and filter bank architectures for channelization and spectrum sensing have been proposed in \cite{Ta9, Ta10}. Such architectures are based on a frequency response masking approach, and they enable LDACS ground stations to receive and/or sense single as well as multiple frequency bands simultaneously. The major drawback of these works is that they do not consider end-to-end transceiver design.

The homogeneous platforms have limitations of flexibility and scalability and may not be suitable for various real-time decision-making tasks. Hence, recently heterogeneous platforms consisting of processors and hardware such as FPGA or ASIC on a single chip are being explored. One such platform is ZSoC consisting of ARM and FPGA on a single chip, and it is being envisioned for various wireless communication applications \cite{Jonathan, ryan, Baris} as well as autonomous driving, medical applications. For example, a Cognitive Radio Accelerated with Software and Hardware (CRASH) is introduced in \cite{john} and authors analyzed three possible configurations of spectrum sensing and decision-making blocks: 1) Both blocks on the FPGA, 2) Both blocks on the processor and 3) Spectrum sensing on the FPGA and decision making on processor. Their experiments show that the third approach offers superior performance over the others. Similarly, cognitive radio exploiting the partial reconfiguration capability of the FPGA and decision-making capability of ARM is demonstrated in \cite{Shree}. Precisely, the processor controls the functionality of the FPGA based on real-time network and spectrum status and allows dynamic switching between channelization and spectrum sensing blocks. Similarly, the hardware-software co-design approach for the IEEE 802.11a transceiver system is discussed in \cite{Be1, Be2}. However, such study and analysis have not been done yet for LDACS transceivers.

Various alternatives have been discussed to improve the OOBE of the OFDM based LDACS. In \cite{Ta4}, filter bank multi-carrier (FBMC) based LDACS transceiver is presented, which offers better OOBE and hence, higher vacant spectrum utilization than OFDM. However, the need of sub-carrier filtering at the transmitter and receiver significantly increases the complexity of the FBMC. Since the architecture of FBMC is significantly different from that of OFDM, the single transceiver cannot support both waveforms on a single chip unless they are stacked in parallel. Furthermore, the extension of FBMC for a multi-antenna transceiver system, a default configuration offering high data rates and superior performance in challenging environmental conditions, is difficult. Generalized Frequency Division Multiplexing (GFDM) \cite{gfdm} is another alternative to OFDM, but it has not been analyzed for LDACS yet. Furthermore, due to concern regarding the area and power consumption of the transceiver, ICAO prefers windowing and filtering approaches to improve OOBE of the OFDM based LDACS \cite{Ta3}. In \cite{NA}, we proposed a reconfigurable filtered OFDM (Ref-OFDM) using a reconfigurable linear phase digital filter. Proposed architecture offers better OOBE than OFDM and GFDM as well as enables dynamical switch between various transmission bandwidths using a single prototype filter. Also, it has a lower complexity than FBMC and GFDM, making it an attractive solution for next-generation LDACS. 

This paper deals in detail with the efficient hardware realization of end-to-end LDACS transceiver on the heterogeneous platform and such work has not been discussed yet in the literature. Furthermore, existing works lack in-depth analysis of the effect of windowing and filtering on the performance of LDACS in the presence of various RF impairments, realistic LDACS channels, and DME interference. The proposed work aims to overcome these drawbacks, thereby contributing to ICAO LDACS standardization activities.


\section{Transceiver Architecture} \label{txa}
In this section, we present the detailed architecture of the proposed transceiver and extensions via windowing and filtering. We also discuss the design of AFE along with various LDACS specific channels as well as interference. The detailed block diagram of the transceiver is shown in Fig.~\ref{bl_diag}.

\subsection{Stimulus and Verification Blocks}
The stimulus block at the transmitter reads the input data bits to be transmitted. They are either stored on on-board ZSoC memory or they can be transmitted from the laptop over Ethernet (ENET). For illustration, we consider the total 864 data bits divided into 36 distinct frames of 24 bits each. Frame formation is done using simple counters and multiplexers. The verification block receives the frame and reads the corresponding data bits for subsequent performance analysis. Both blocks are implemented on the PS.

\subsection{Digital Baseband Processing Blocks of Transceiver} \label{DB}
Various baseband signal processing blocks of the transceiver are shown in Fig.~\ref{bl_diag}. The blocks such as scrambler, inter-leaver, data encoder, data modulator, frame generation, IFFT followed by CP addition, and preamble addition are desired signal processing blocks for the OFDM transmitter. The receiver consists of similar blocks that perform the operations in the reverse direction. The OOBE performance of the transceiver can be improved further using windowing or filtering or both. For windowing operation, two new blocks, 1) Cyclic suffix addition, and 2) Windowing, are added before preamble addition. Similarly, at the receiver, we need overlap and add block. For filtering operation, new filtering blocks are added at the transmitter as well as the receiver. The detailed explanation of various blocks in Fig.~\ref{bl_diag} is given later in \ref{OFDM}.

Each transceiver block can be realized on the PS or PL. In Fig.~\ref{bl_diag}, we consider 10 possible configurations, $V1, V2,.., V10$. Each configuration offers a unique boundary between PS and PL. 
We discuss these configurations in detail later in Section IV. Here, we focus on the functionality and architecture of each block for the serial implementation on the PS as well as parallel implementation on the PL.

\subsubsection{\textbf{Orthogonal Frequency Division Multiplexing (OFDM)}} \label{OFDM}
\begin{figure}[H]
	\centering\vspace{-0.2cm}
	\captionsetup{justification=centering}
	\includegraphics[scale=0.31]{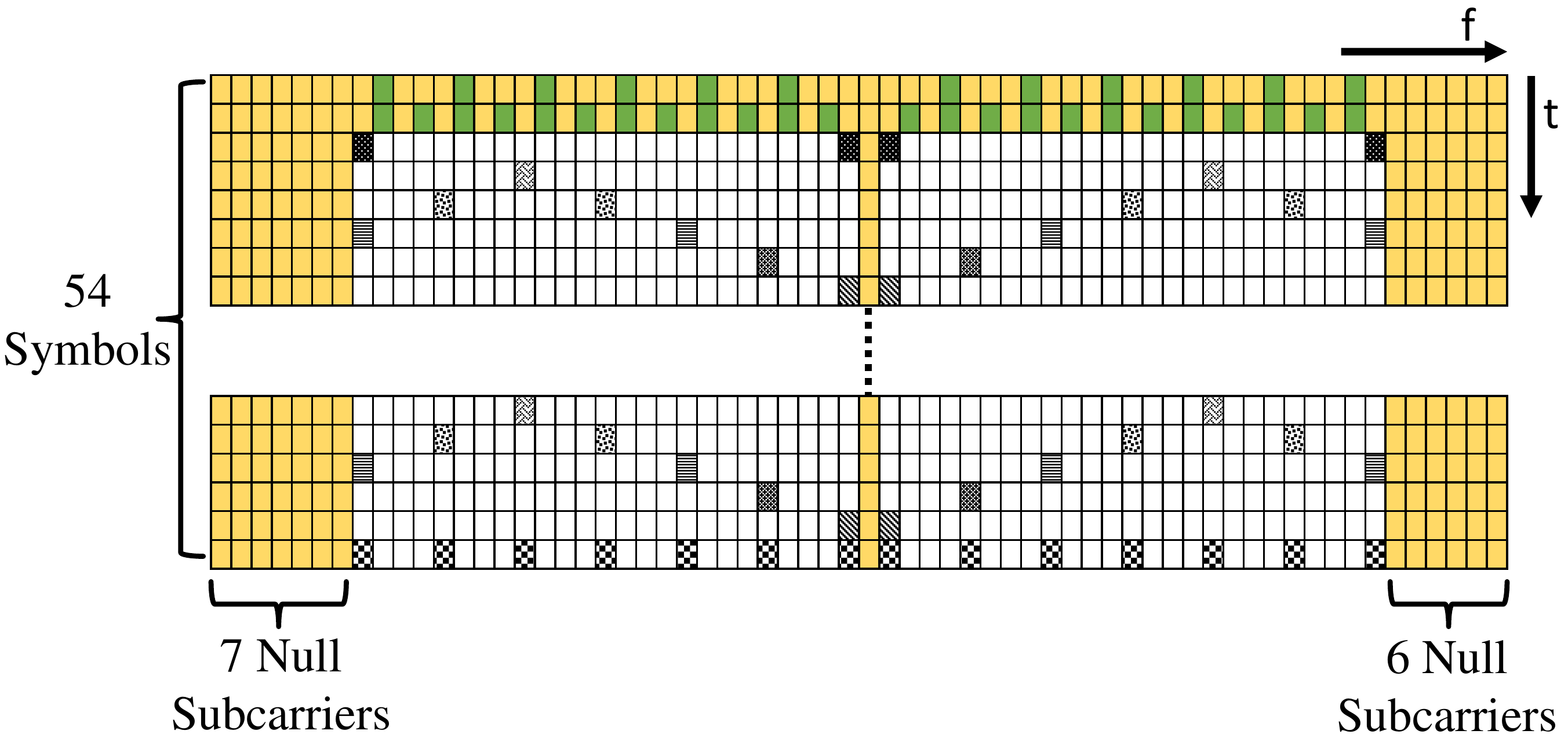}
	\vspace{-0.3cm}
	\caption{LDACS frame with 54 symbols and corresponding symbol-to-subcarrier mapping.}
	\label{SC_map}
	\vspace{-0.3cm}
\end{figure}

The OFDM based transmitter consists of blocks such as scrambler, convolutional encoder, interleaver, binary phase-shift keying (BPSK) modulator, Inverse Fast Fourier Transform (IFFT) and cyclic prefix adder. The scrambler does the bitwise XOR operation on the incoming input data and a random scrambling sequence generated by the linear feedback shift register (LFSR). The same sequence is used to descramble the data at the receiver. This is followed by a convolutional encoder which uses the generator polynomial of $g_{0} = 133$ and $g_{1} = 171$. These correspond to a rate 1/2 code with a maximum free distance of 7. Thus, the output of the convolution encoder is twice the length of the input. The interleaver performs two-step permutation on coded data and used to handle burst errors. The interleaved data is then converted to complex samples using the BPSK modulator to obtain 48 samples. Note that any other modulation scheme such as QPSK, 16 QAM, or 64 QAM can also be used. These samples are then mapped to 64 points IFFT, as shown in Fig.~\ref{SC_map}. As per the LDACS specifications, 64 subcarriers are used, out of which 50 are active subcarriers carrying data and pilot symbols. The number of subcarriers carrying pilots and data in each symbol is not fixed and depends on the symbol index. For example, the number of data subcarriers in the symbol with index 0 and 1 are 36 and 48, respectively, while the number of pilot subcarriers are 14 and 2, respectively. One LDACS frame comprises 54 symbols, as shown in Fig.~\ref{SC_map} and the pilots at each symbol follow specific patterns except symbols with indices 0, 51, 52, and 53. Please refer to \cite{appendix} for more details on the architecture which performs such a symbol to subcarrier mapping in SoC. {\color{red} For the simplicity of representation, we discuss the architecture of various blocks of the transceiver assuming the transmission of OFDM symbol with index 1 (or 6, 11, \ldots , 46) of LDACS frame consisting of 48 data, two pilots, 1 DC, and 13 Null subcarriers. Note that depending on the symbol index, appropriate control signals are generated to meet the symbol mapping requirements of the LDACS frame, as shown in Fig.~\ref{SC_map}. The corresponding details are omitted to maintain the brevity of paper and avoid repetitive description.} Also, we consider the forward link of LDACS since forward and reverse links differ only in the symbol-to-sub-carrier mapping which is not a challenging task.

To avoid inter-symbol interference, a cyclic prefix (CP) of length 11 is added to the OFDM symbol. In the end, preambles are added, which aim the receiver for synchronization. The preamble consists of both short training sequence (STS) and long training sequence (LTS). The STS is used for timing acquisition, coarse frequency acquisition, and diversity selection, while LTS is used for channel estimation and fine frequency acquisition [7, 8].  For the length of 160 samples, LTS is repeated twice, while STS is repeated ten times. In the end, the signal is transmitted over the wireless channel via AFE and antenna. 

\begin{figure}[!b]
	\centering
		\captionsetup{justification=centering}
	\subfloat[]{\includegraphics[scale=0.45]{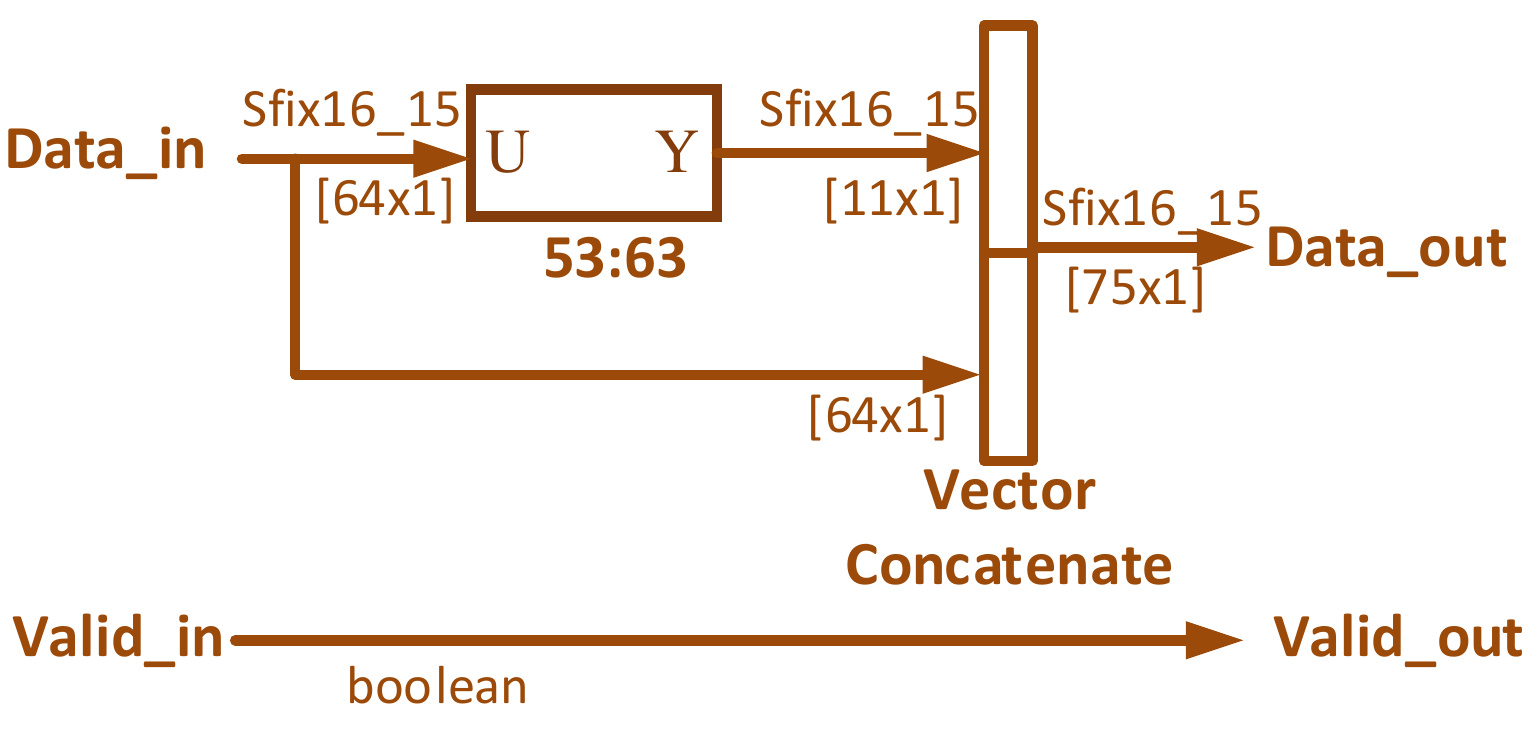}}%
	\hspace{0.2cm}
	\subfloat[]{\includegraphics[width=\linewidth]{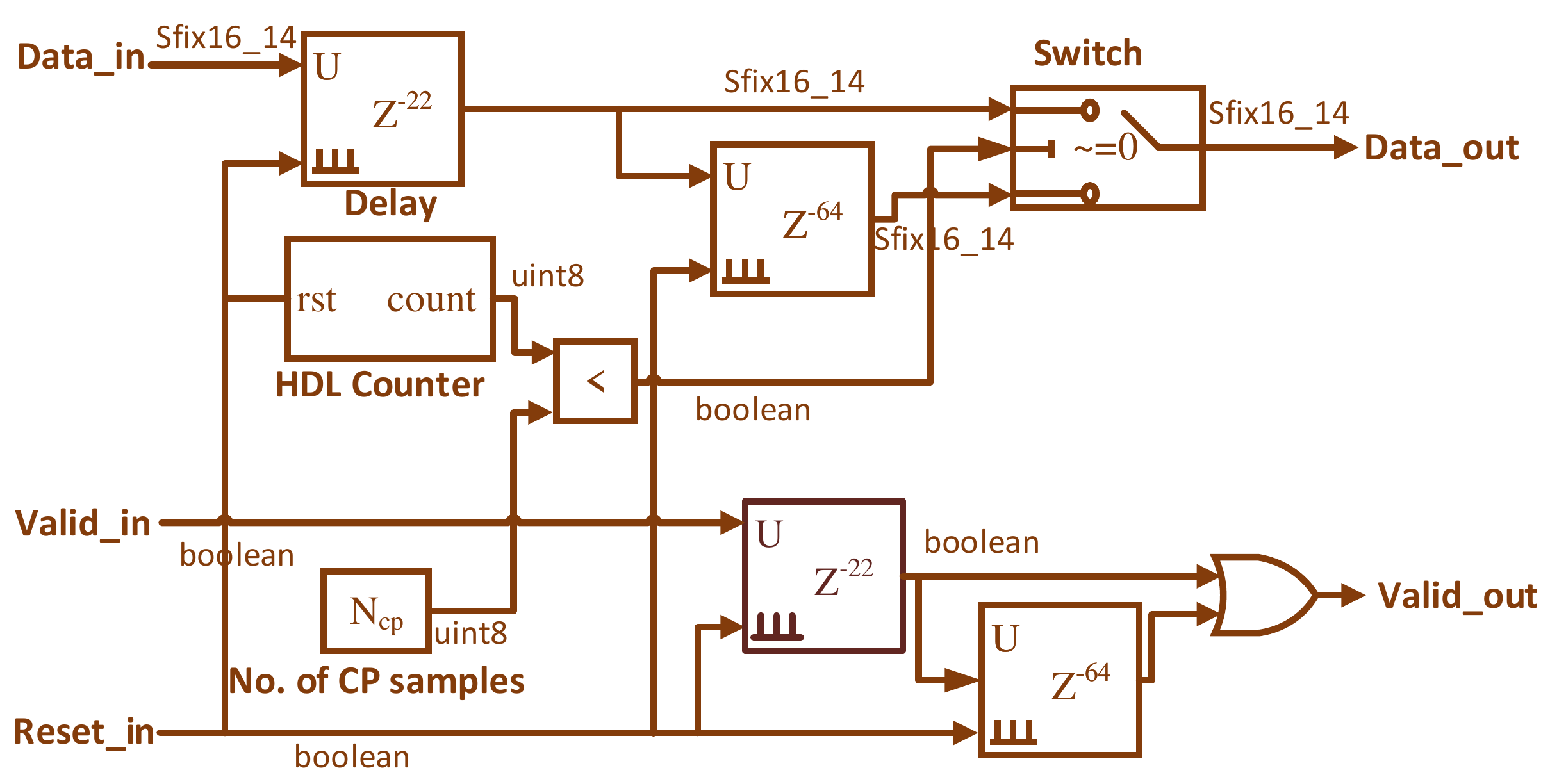}}%
	\vspace{-0.2cm}
		\caption{(a) PS and (b) PL implementations of OFDM cyclic prefix addition.}
	\label{impl}
\end{figure}

The difference in processing modes of PL (Sample mode) and PS (frame mode) leads to a difference in the implementation of each block of the transceiver in the two modes. Due to limited space constraints, we discuss the architecture of a few blocks here while remaining blocks are discussed in detail in Supplementary \cite{appendix}. The PS implementation of the CP addition involves only vector concatenation due to frame-based processing. As shown in Fig.~\ref{impl}(a), the last 11 samples of the IFFT output are appended in the beginning as CP. On the other hand, PL implementation of the same involves additional counter and registers to store the samples to be added as CP. As shown in Fig.~\ref{impl}(b), we need two registers of length 2CP (22) and $N$ (64) along with Mod-$N$ counter. For easier understanding, we consider the illustrative example of a frame consisting of 4 samples with 1 CP sample. In this case, we need the first register of size 2 and the second register of size 4. In the first clock cycle, input sample, $a_0$, is loaded into the first register, and hence the content of two registers are \{$a_0,0$\} and \{$0,0,0,0$\}. At the fifth clock cycle, content of two registers will be \{$a_4,a_3$\} and \{$a_2,a_1,a_0,0$\}. In the next clock cycle, frame reset (reset\_in) happens since we have received all samples of a frame, and hence the content of two registers will be \{$0,a_3$\} and \{$a_2,a_1,a_0,0$\}. From the next cycle onward, output valid is always 1 and we get the first output which is $a_3$ from the first register and content of register becomes \{$b_0,0$\} and \{$a_3,a_2,a_1,a_0$\}. Here, $b_0$ is the first sample of a new frame. Subsequently, the next four outputs are taken from the second registers. In this way, we get the output as $a_3,a_0,a_1,a_2,a_3$. Similarly, in next four clock cycles, the output will be  $b_3,b_0,b_1,b_2,b_3$. As discussed before, valid and reset signals are used to synchronize the transfer of data between any two adjacent blocks and needs to be handled carefully in each block. For instance, as shown in Fig.~\ref{impl}(b), a valid signal involves 22 and 64 tapped delays, similar to the ones used in the data signal.

\subsubsection{\textbf{WOLA-OFDM}}
\begin{figure}[!h]
	\vspace{-0.2cm}
\centering
	\captionsetup{justification=centering}
	\includegraphics[width=\linewidth]{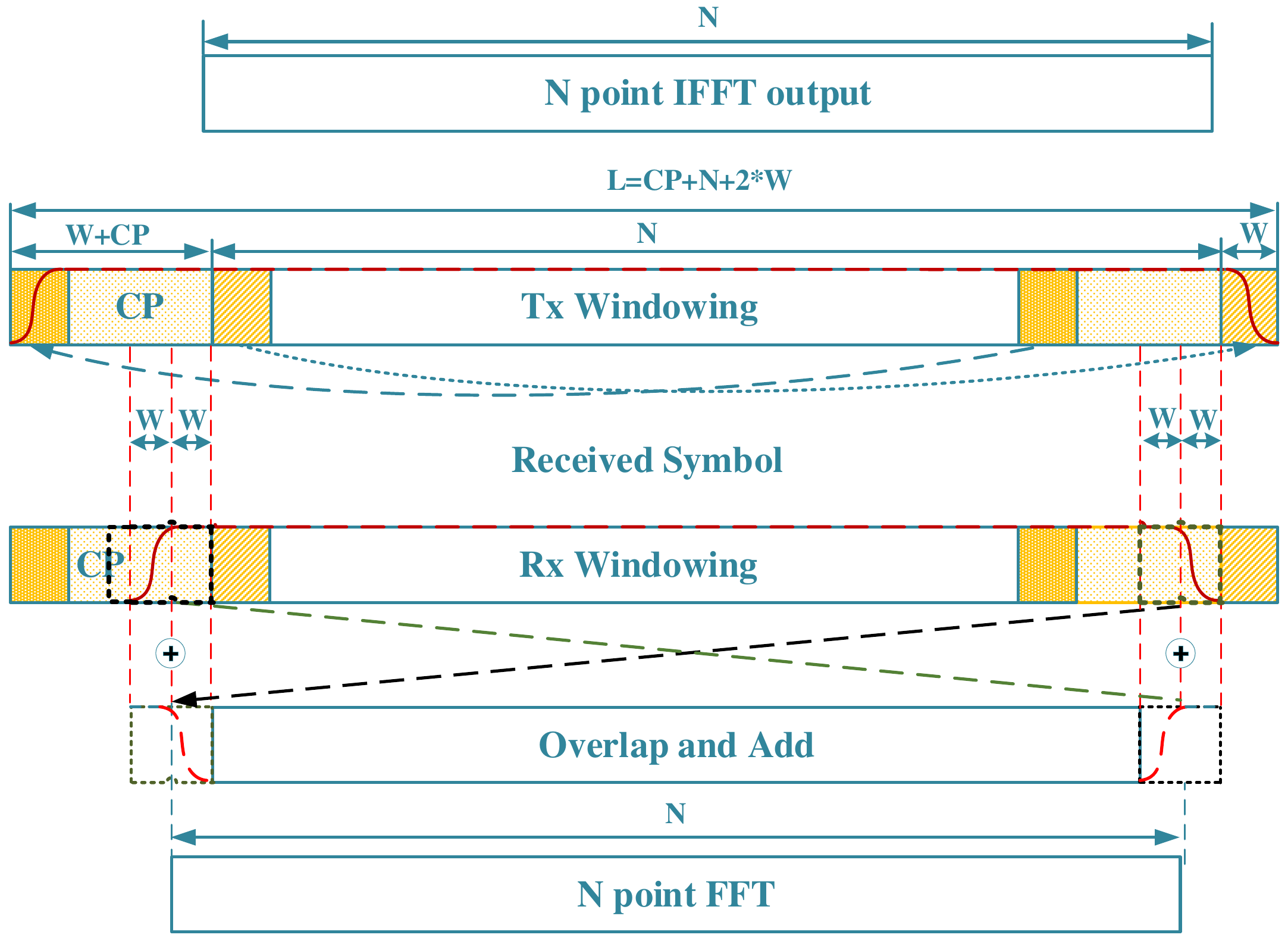}
		\vspace{-0.2cm}
	\caption{Cyclic prefix and cyclic suffix processing along with windowing for WOLA-OFDM.}
	\label{wola}
		\vspace{-0.2cm}
\end{figure}

In WOLA-OFDM, the conventional rectangular window is replaced by a windowing pulse with soft edges to improve the out-of-band emission of CP-OFDM \cite{wol}. This soft edge windowing is applied in the time domain via point-to-point multiplication between the output of the CP block and window function. The additional sequence of operations at the transmitter are as follows:

\begin{enumerate}
    \item Cyclic Extention: The CP addition is slightly different in WOLA-OFDM than CP-OFDM. As shown in  Fig.~\ref{wola}, the CP is formed by appending the last $CP+W$ samples of a given symbol (output of IFFT) to its beginning, and the cyclic suffix (CS) is formed by appending the first $W$ samples of a given symbol in its end. Therefore, the length of the WOLA-OFDM time domain symbol is extended from $N$ to $N+CP+2W$, as shown in Fig. \ref{wola}.
    \item Windowing: After the cyclic extension, a Root Raised Cosine (RRC) window of length $L=N+CP+2W$ is applied in the time domain. For LDACS, we have $N=64$, $ CP=11$ and $W=8$, and corresponding window length is $L=91$ with the taper region of length $W$. 
\end{enumerate}

Such windowing at the transmitter demands additional signal processing at the receiver to suppress the asynchronous inter-user interference. As shown in Fig. \ref{wola}, the additional steps at the receiver are as follows:

\begin{enumerate}
    \item {\color{red}The starting and ending samples of length $W + \left \lfloor \frac{CP}{2} \right \rfloor = 13$ and $W = 8$ respectively are discarded, and the RRC windowing is applied at the retrieved data. The window length at the receiver is not same as the transmitter and the receiver window length is taken as $N + \left \lceil \frac{CP}{2} \right \rceil = 70$.}
    \item Two adjacent received WOLA-OFDM symbols are overlapped with each other and then added to the next symbol to retrieve the 64 main samples. The overlap and add process is applied to minimize the effects of windowing on the useful data, as shown in Fig. \ref{wola}. 
\end{enumerate}

\begin{figure}[!h]
	\vspace{-0.2cm}
	\centering
	\subfloat[]{\includegraphics[width=\linewidth]{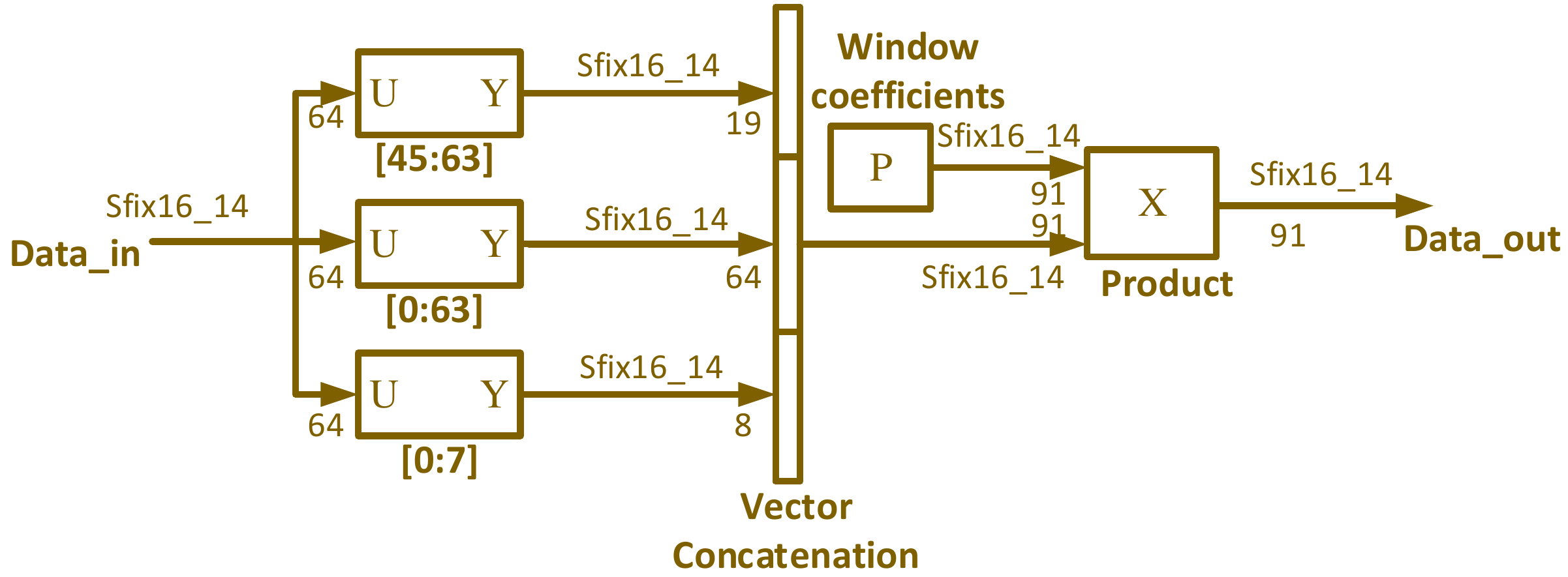}}%
	\hspace{0.3cm}
	\subfloat[]{\includegraphics[width=\linewidth]{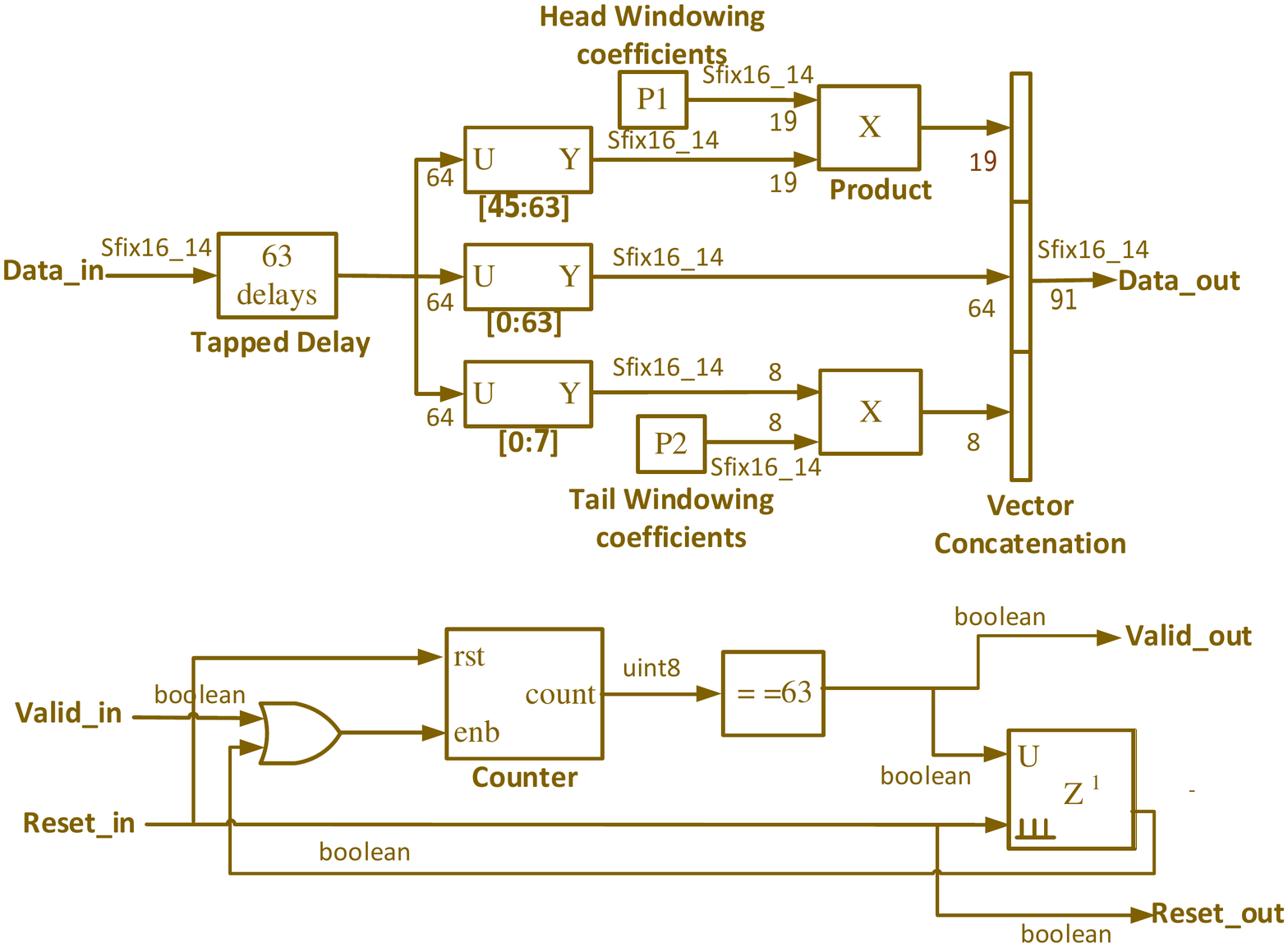}}%
		\vspace{-0.2cm}
		\caption{(a) PS and (b) PL implementation of time-domain windowing.}
	\label{impl1}
		\vspace{-0.2cm}
\end{figure}

The PS and PL implementation of windowing is shown in Fig. \ref{impl1} (a) and Fig. \ref{impl1} (b), respectively. The PS implementation at the transmitter is straightforward due to a frame-based approach in which a time domain multiplication of the input data with the windowing coefficients is performed, as shown in Fig. \ref{impl1} (a). 

In PL implementation, the data is coming in the form of samples, therefore to add the cyclic prefix, suffix, and windowing samples, all 64 samples (1 frame) are collected with the help of 63 tapped delays. For PL implementation of windowing, we exploit the parallel operation by dividing the windowing into head and tail sections. Consider $P1$, and $P2$ denote the windowing coefficients for head and tail sections, respectively. The $P1$ is of length $W+CP$ in which the first $W$ samples corresponds to the first $W$ RRC windowing coefficients $(P)$ while remaining samples are fixed to 1. The $P2$ is of length $W$, and it corresponds to the last $W$ RRC windowing coefficients, $(P)$. In the end, the cyclic prefix, cyclic suffix, and the data are concatenated, and total 91 samples are selected for transmission over the air. 

The input valid signal increments the counter value, and the counter counts till 63 i.e., a total of 64 samples. Once we have received the whole frame of 64 samples (without adding cyclic prefix and suffix), the output valid signal will become one. The output valid signal is generated for one clock cycle for the output frame of size 91 (similar to the size of the transmitted data (91 samples)).

\begin{figure}[!h]
	\centering
		\vspace{-0.2cm}
	\subfloat[]{\includegraphics[width=0.95\linewidth]{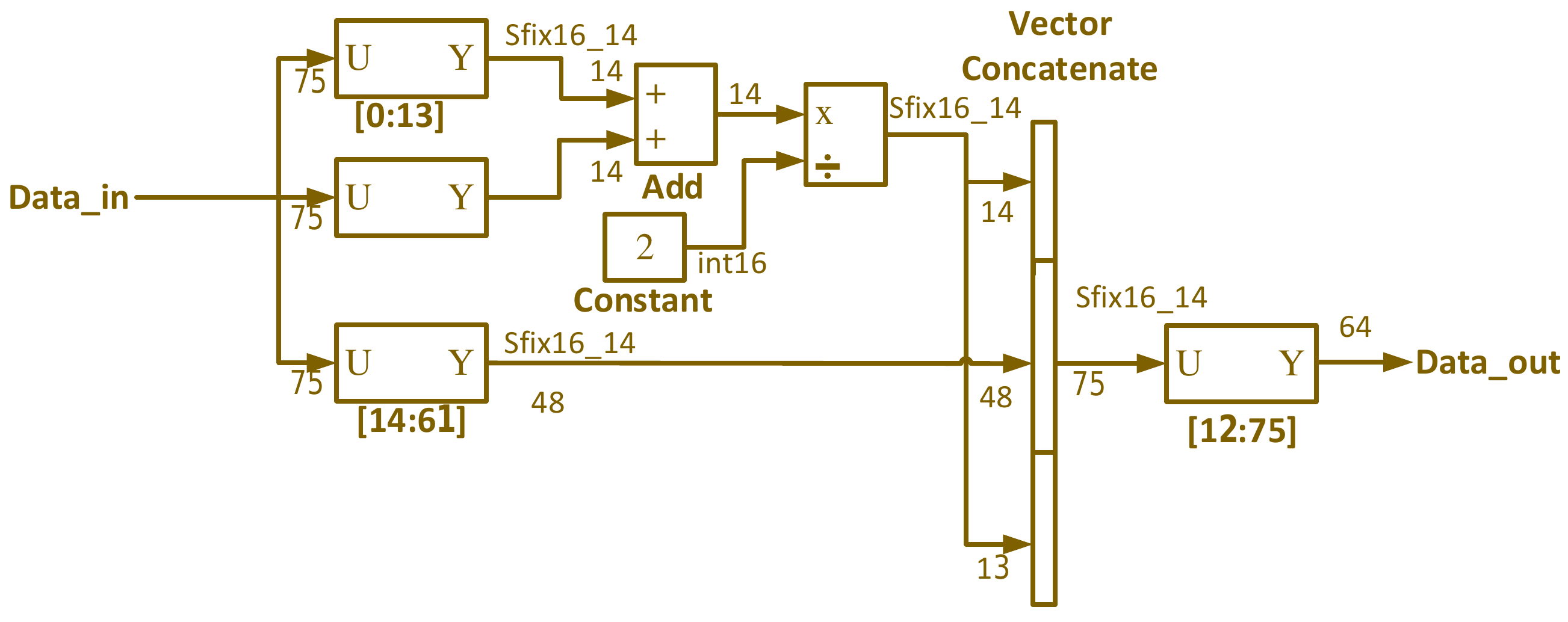}}%
	\hspace{0.2cm}
	\subfloat[]{\includegraphics[width=0.95\linewidth]{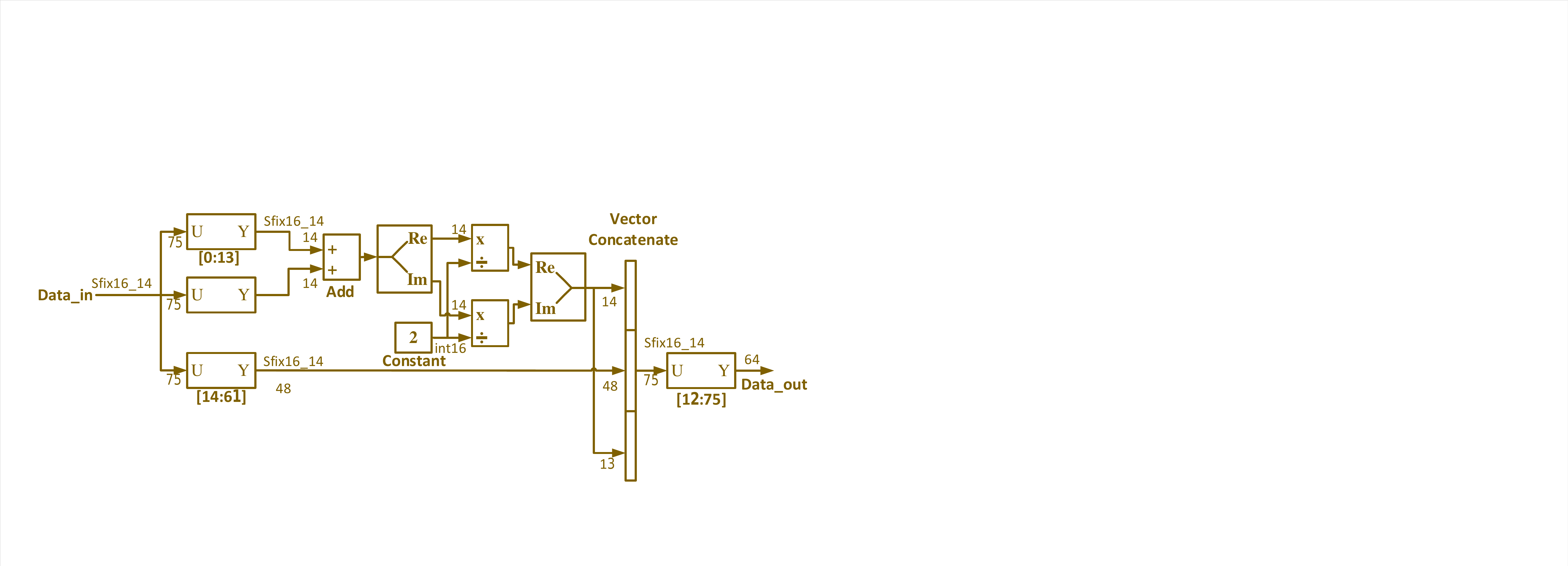}}%
		\vspace{-0.2cm}
		\caption{(a) PS and (b) PL implementation of overlap and add processing.}
	\label{oal}
		\vspace{-0.2cm}
\end{figure}

At the receiver, windowing is implemented in the same manner as the transmitter. Additionally, the {\color{red} overlap and add processing is performed on the $N + \left \lceil \frac{CP}{2} \right \rceil = 70$ windowed samples} by directly extracting the desired samples from the received frame and then concatenate it to the beginning and ending of the symbol. The PS and PL implementation is the same for overlap and add processing, as presented in Fig.~\ref{oal} (a) and (b). 

\subsubsection{\textbf{Filtered OFDM}}
\begin{figure}[H]
	\vspace{-0.2cm}
	\centering
		\captionsetup{justification=centering}
	\subfloat[]{\includegraphics[width=0.95\linewidth]{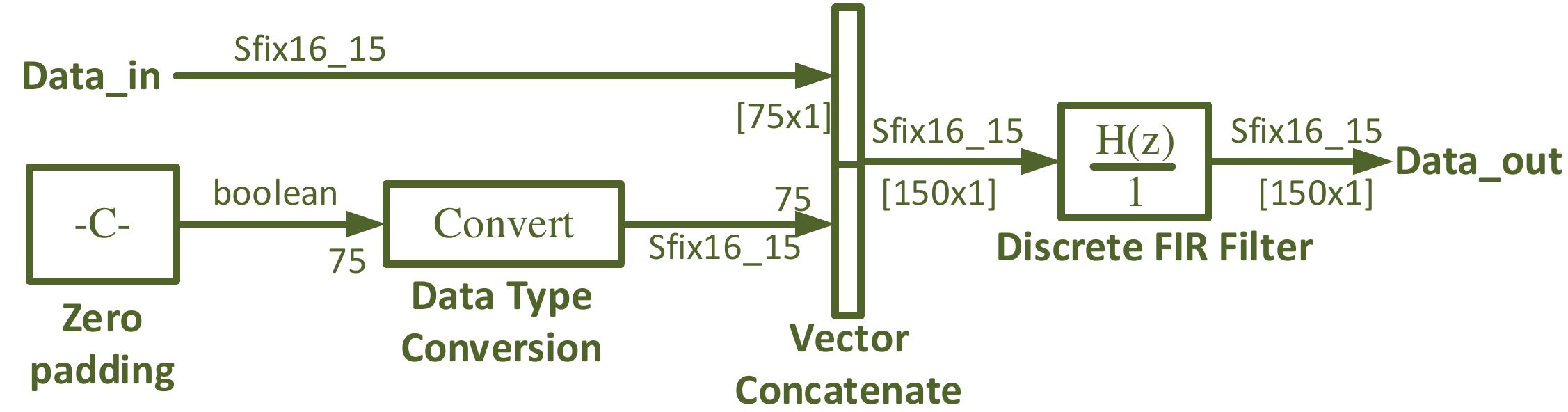}}%
	\hspace{0.2cm}
	\subfloat[]{\includegraphics[scale=0.4]{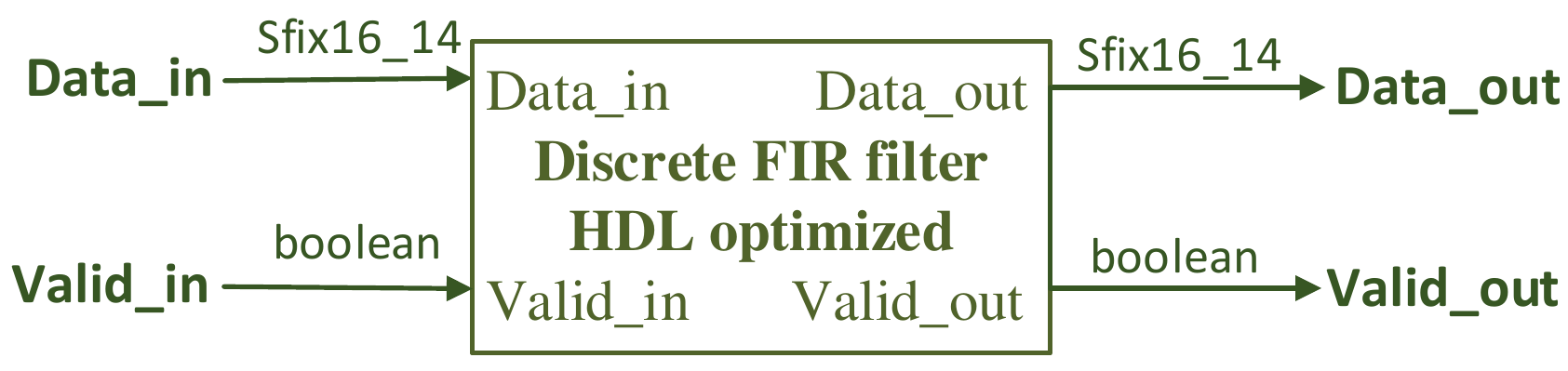}}%
		\vspace{-0.2cm}
		\caption{(a) PS and (b) PL implementation of Filter.}
	\label{impl2}
		\vspace{-0.2cm}
\end{figure}

The FOFDM uses a linear phase finite impulse response filter instead of time-domain windowing for further improvement in out-of-band emission. In \cite{NA}, we have shown that FOFDM enables higher transmission bandwidth compared to bandwidth limitation to 498 kHz in the OFDM based LDACS system. It also enables the transmission in non-contiguous bands and the sharing of adjacent frequency bands among asynchronous users. However, the filter needs to be carefully designed and implemented as it may lead to higher inter-symbol and inter-carrier interference. In the proposed FOFDM transceiver, we consider LDACS with 480 kHz of bandwidth with a sampling frequency of 1.1 MHz and hence, we designed a linear phase low-pass filter of order 150 with a normalized cut-off frequency of 0.86 and the transition bandwidth of 0.02 generated using park McClellan approach \cite{SJ, SJ2}. The PS and PL implementation of the FIR filter is shown in fig.~\ref{impl2} (a) and \ref{impl2} (b) respectively.

The filter specifications and implementation are identical at the transmitter and receiver. For the implementation of the filter, we have directly used the HDL optimized model provided by Xilinx. In case of PS implementation, we need additional zero-padding to handle delay balancing and selector to choose the desired filtered data. For PL implementation of the filter, we have studied the effect of word-length on the performance of the transceiver. Please refer to Section \ref{res} for more details. 

\subsection{Analog Front End: RF Transmitter and Receiver}

\begin{figure}[!b]
\centering
	\vspace{-0.2cm}
    \captionsetup{justification=centering}
    \includegraphics[scale=0.3]{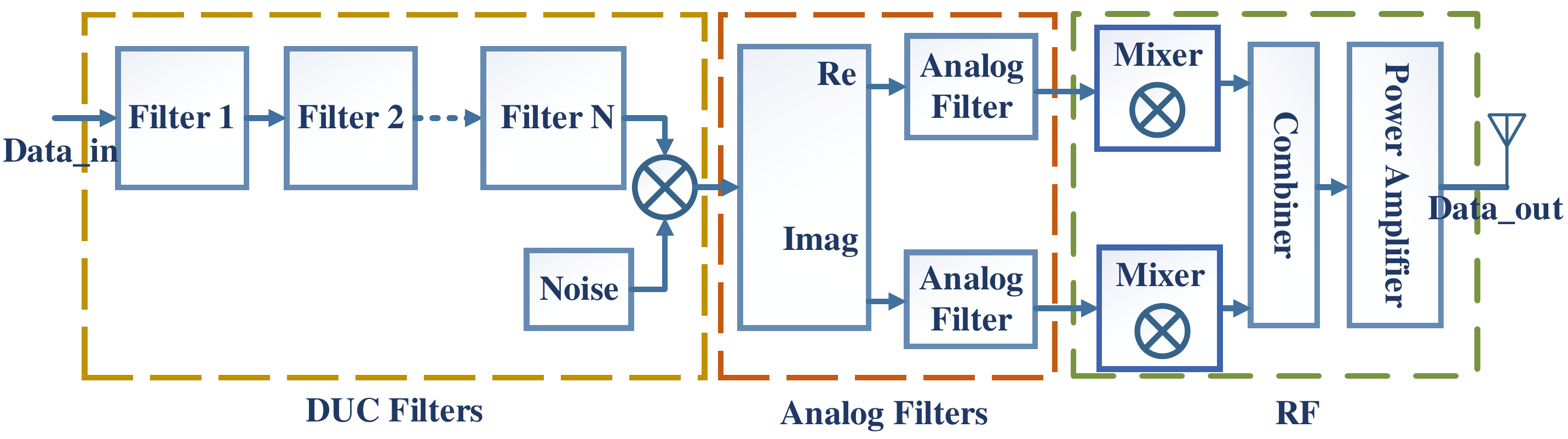}
    \centering
    	\vspace{-0.2cm}
    \caption{Analog Front End: RF Transmitter.}
    \label{tx}
    	\vspace{-0.2cm}
\end{figure}

The output of the transmitter is passed to the AFE for over-the-air transmission in $L$-band. The AFE is designed using the RF models provided by Analog Devices for use in MATLAB/Simulink. The transmitter consists of Digital up-conversion (DUC) filters, analog filters and RF front-end as shown in Fig.~ \ref{tx}. The digital up-conversion filter is a series of digital FIR filters that convert the baseband signal to an intermediate frequency (IF) signal. The sample rate of the DUC filter should be the same as the input signal. The digital filter also introduces the noise floor. The analog filters are used to shape this noise floor and provide a continuous-time signal processed by the RF front-end. The RF front-end up-converts the IF signal to RF carrier frequency using the local oscillator, followed by amplification using a power amplifier.

\begin{figure}[!t]
	\vspace{-0.2cm}
\centering
	\captionsetup{justification=centering}
		\vspace{-0.2cm}
	\includegraphics[width=0.9\linewidth]{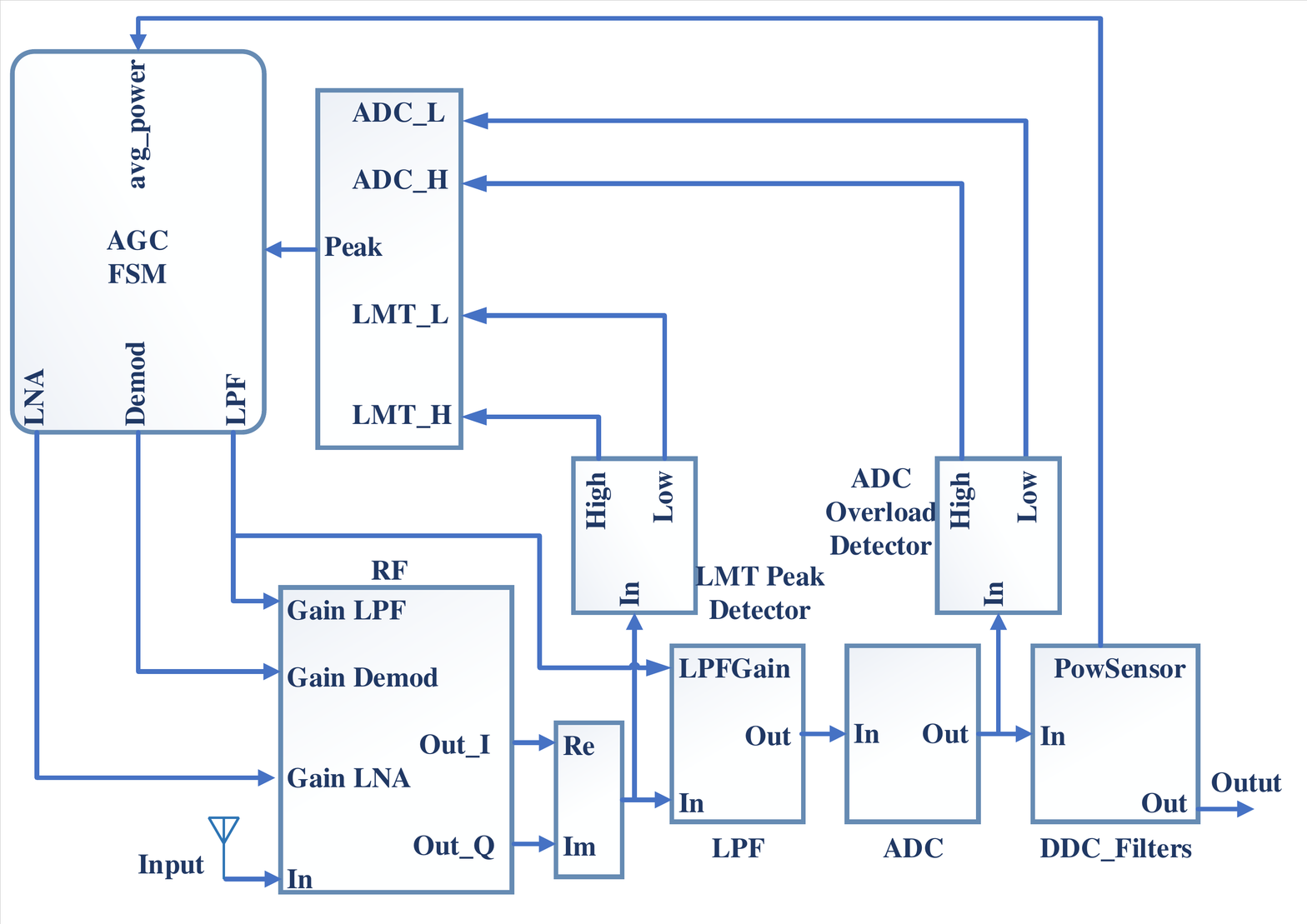}
	\centering
	\caption{Analog Front End: RF Receiver.}
		\vspace{-0.2cm}
	\label{tx3}
\end{figure}

At the receiver side, the RF front-end down-converts the signal centered on the same LO frequency to IF using a quadrature demodulator, as shown in the Fig.~ \ref{tx3}. The RF front-end has mainly three components: low noise amplifier (LNA), quadrature demodulator (Mixer), and trans-impedance amplifier (TIA), and the chain is indicated as LMT. The gains of each component are tunable and controlled by the AGC. The analog filters provide a continuous-time signal to the ADC. The ADC models a high-sampling rate third-order delta-sigma modulator. The low-pass digital down conversion filters convert the highly sampled signal at the output of the ADC to the baseband. The output of the AFE  is passed to the OFDM receiver in Zynq.
The integration of the AFE with the transceiver in Fig.~\ref{bl_diag} and its parameters as per the LDACS specification are discussed in section \ref{design}.

\begin{figure}[!b]
	\vspace{-0.2cm}
\centering
	\captionsetup{justification=centering}
	\includegraphics[scale=0.25]{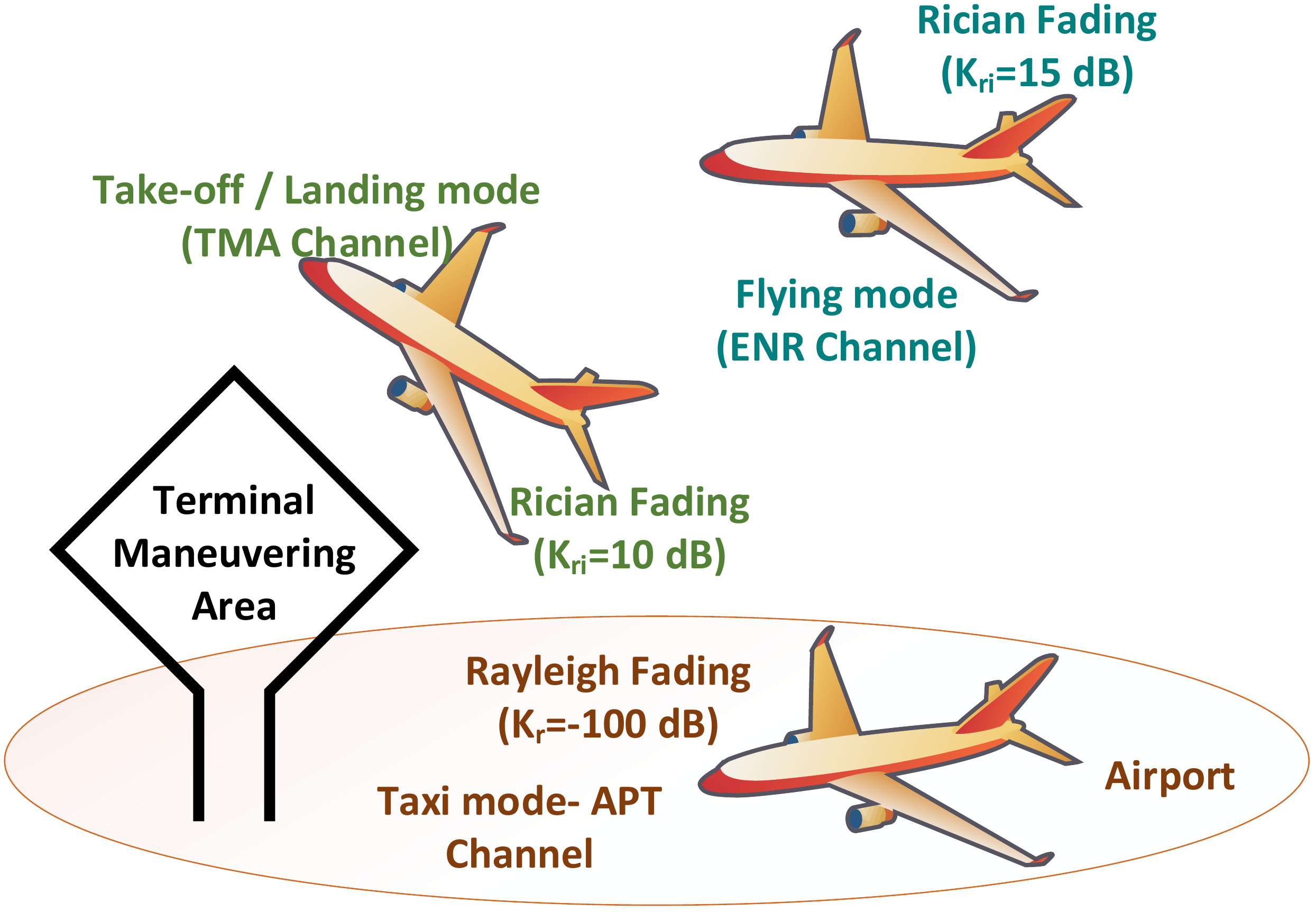}
	\centering
		\vspace{-0.2cm}
	\caption{Various LDACS channels and their parameters for different positions of the aircraft.}
	\label{channel}
		\vspace{-0.2cm}
\end{figure}

\subsection{LDACS Specific Wireless Channels and DME Interference}
As shown in Fig.~\ref{channel}, three channels that are specific to the LDACS environment are considered, and they are Airport (APT), Terminal Maneuvering Area (TMA),  En-routing (ENR). The channels are modeled as wide sense stationary with uncorrelated scattering and characterized using three properties: fading, delay paths, and Doppler frequency \cite{15}. The channel parameters are given in Table ~\ref{ch} \cite{15, 16, 17, 18}. Note that the Doppler frequency is obtained as $F_D=F_c \frac{v}{c}$ where $F_c$ is the carrier frequency and is at most 1215 MHz, $v$ is the velocity of the aircraft in $m/s$ (1 Knots True Airspeed (KTAS)= 0.5144 $m/s$) and $c=3*10^8 m/s$. 

\begin{table}[]
	\vspace{-0.2cm}
\captionsetup{justification=centering}
	\caption{Channel Parameters}
		\vspace{-0.2cm}
	\label{ch}

\resizebox{\columnwidth}{!}{
 \begin{tabular}{|c|c|c|c|c|c|}
\hline
\textbf{Scenario} & \textbf{\begin{tabular}[c]{@{}c@{}}Max \\ Delay ($\mu$s)\end{tabular}} & \textbf{\begin{tabular}[c]{@{}c@{}}Acceleration\\ $(m/s^2)$\end{tabular}} & \textbf{Harmonics} & \textbf{\begin{tabular}[c]{@{}c@{}}Velocity\\ (KTAS)\end{tabular}} & \textbf{\begin{tabular}[c]{@{}c@{}}Doppler\\ Frequency (Hz)\end{tabular}}        \\ \hline
APT               & 3                                                                      & 5                                                                         & 8                  & 200                                                                & \begin{tabular}[c]{@{}c@{}}$(1215e6)\frac{200*.5144}{3e8}$ \\ = 413\end{tabular} \\ \hline
TMA               & 20                                                                     & 50                                                                        & 8                  & 300                                                                & \begin{tabular}[c]{@{}c@{}}$(1215e6)\frac{300*.5144}{3e8}$\\  = 624\end{tabular} \\ \hline
ENR               & 15                                                                     & 50                                                                        & 25                 & 600                                                                & \begin{tabular}[c]{@{}c@{}}$(1215e6)\frac{600*.5144}{3e8}$\\ =1250\end{tabular}  \\ \hline
\end{tabular}}
	\vspace{-0.2cm}
\end{table}

Along with these specific LDACS real-time channels, DME interference is also taken into account. DME is measuring equipment used for navigation purposes and has major interference on LDACS as LDACS is deployed between two DME channels. The DME signal is composed of Gaussian pulse pairs given as:
\begin{equation}
    S=e^{\frac{-\alpha t^{2}}{2}}+e^{\frac{-\alpha (t-\Delta t)^{2}}{2}}
\end{equation}

where, \textbf{$\Delta t = 12 \mu s$} denotes the spacing between the pulses and $\alpha$ is the pulse width of $4.5 \times 10^{-11} s^{-2}$. All the experimental results presented in this paper considers the DME interference. {\color{red} The DME interference signal is scaled by a factor of 0.2 so that the power level of the DME signal closely matches the LDACS signal.}

\subsection{Receiver}
At the receiver, the preamble detection block detects the beginning of the data frames using auto-correlation and extract it for subsequent processing. For cyclic prefix removal, the starting 11 samples are discarded out of the 75 incoming samples. The remaining 64 samples are given as input to the 64 point FFT block. Out of 64 symbols at the output of the FFT block, 48 data symbols are extracted by a selector. The output data symbols are demodulated using the BPSK demodulator. The deinterleaver then deinterleaves the bits using the predefined sequence followed by decoding using a Viterbi decoder using the same generator polynomial as a convolutional encoder in the transmitter. The descrambler uses the corresponding descrambling sequence to retrieve the 24 bits of a frame. A similar process is repeated for each frame. The next section presents the HW-SW co-design approach used for transceiver design and implementation.

\section{Hardware-Software Co-design Approach}
The HW-SW co-design approach gives the flexibility to choose which part of the transceiver is best suited to be implemented on PL and PS of the ZSoC. In this section, we present design details of various transceiver configurations (V1-V10), shown in Fig.~\ref{bl_diag} realized using the HW-SW co-design approach. The data transfer between PS and PL  plays an important role in this approach, and corresponding details are summarized in Table~\ref{dt}.
\begin{figure}[!h]
	\vspace{-0.2cm}
\centering
\captionsetup{justification=centering}
        \includegraphics[scale=0.3]{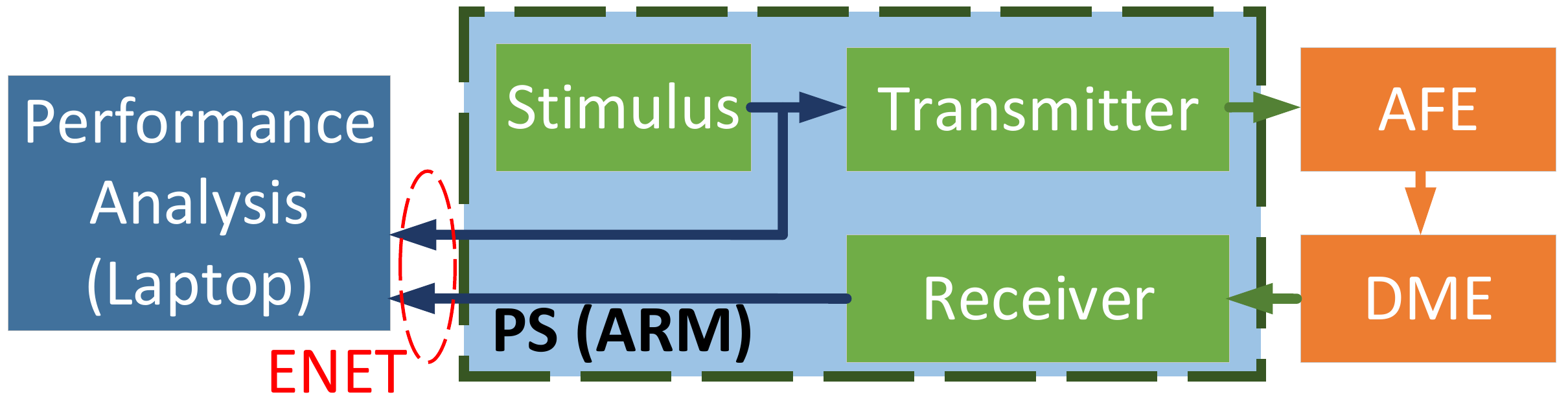}
        	\vspace{-0.2cm}
        \caption{Configuration V1 of the transceiver.}
        \label{tv}
        	\vspace{-0.2cm}
    \end{figure}
We begin with the configuration V1 in which the complete transceiver is implemented on the PS, as shown in Fig.~ \ref{tv}, and hence, there is no data transfer between PS to PL as shown in Table~\ref{dt}. The stimulus model generates 32-bit unsigned integers out of which 24 are data bits (single frame), 2 are valid and reset signals, and remaining are zero-padded bits. Each data bit is modulated and processed to obtain an OFDM symbol with 75 samples (64 subcarriers + 11 samples as CP). Each sample can be represented in the form of an 8/16/32-bit fixed-point data type. {\color{red} Each OFDM symbol in a frame comprises 75 samples (64 + 11 CP) and corresponding symbol period is $75 \mu s$ assuming 1 sample takes 1 $ \mu s$. We refer this as time per frame symbol (tpfs).} With 36 data frames, 2 pilot frames, and additional delays due to frame synchronizations, one simulation runs for $43*tpfs$ duration. The performance analysis model compares the transmitted and received bits for subsequent BER and throughput analysis. The realization of this architecture on ZSoC is done using MATLAB HDL coder and verifier, along with Embedded Coder toolboxes. Please refer to \cite{appendix,sasha} for detailed steps involved in the HW-SW co-design.

\begin{table}[!h]
	\vspace{-0.2cm}
\captionsetup{justification=centering}
    \caption{Data Transfer between PS and PL (transmitter Side)}
    \label{dt}
    	\vspace{-0.2cm}
   \resizebox{\columnwidth}{!}{
\begin{tabular}{|c|c|c|c|}
\hline
\textbf{Model Variants} & \textbf{Data Type} & \textbf{Size of 1 element} & \textbf{No. of elements} \\ \hline
V1                      & -                  & -                          & -                        \\ \hline
V2 (FOFDM)             & Signed Fixed Point & 8/16/32 bits               & 150                       \\ \hline
V3                      & Signed Fixed Point & 8/16/32 bits               & 75                       \\ \hline
V4 (WOLA-OFDM)          & Signed Fixed Point & 8/16/32 bits               & 91                      \\ \hline
V5                      & Signed Fixed Point & 8/16/32 bits               & 48                       \\ \hline
V6                      & Signed Fixed Point & 8/16/32 bits               & 48                       \\ \hline
V7                      & Boolean            & 1 bit                      & 48                       \\ \hline
V8                      & Boolean            & 1 bit                      & 24                       \\ \hline
V9                      & Boolean            & 1 bit                      & 24                       \\ \hline
V10                     & Boolean            & 1 bit                      & 24                       \\ \hline
\end{tabular}}
	\vspace{-0.2cm}
\end{table}

\begin{figure}[!h]
    \vspace{-0.2cm}
    \centering
        \captionsetup{justification=centering}
            \vspace{-0.2cm}
        \includegraphics[scale=0.275]{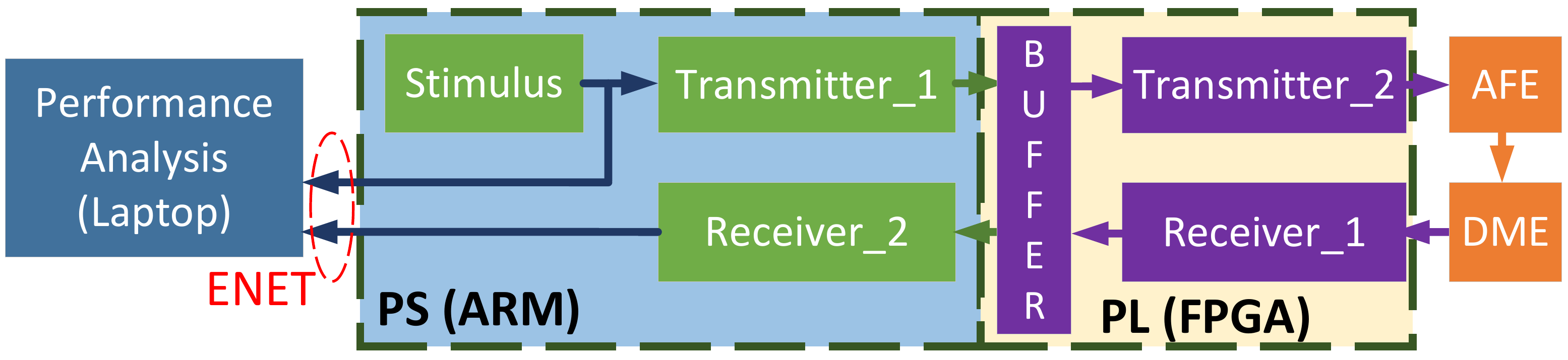}
        \caption{Configurations V2-V9 of the transceiver.}
        \label{tv1}
            \vspace{-0.2cm}
    \end{figure}

In configuration V2, the filtering operation is moved to PL, and hence, it is applicable only for FOFDM. As shown in Fig.~\ref{tv1}, the transmitter and receivers are divided into two sections, one for PS and other for PL. For V2, the output of transmitter\_1 is the frame consisting of 150 complex filtered OFDM samples, each of which can be represented in 8/16/32-bit fixed-point format. One such frame, along with valid and reset signals are interfaced with AXI-compatible buffer realized in PL. The buffering is necessary for subsequent sample-based processing in PL. Similarly, unbuffering is needed while passing the data from PL to PS after the filtering operation of the receiver in PL (Receiver\_1). Note that the sampling time of the blocks in PS is $75 \mu s$ while the sampling time of the blocks in PL is $1 \mu s$.

Configurations V3-V9 are similar to V2, where few more blocks are moved from PS to PL. For instance, in V3, preamble addition and detection blocks are realizing in PL along with filtering (in FOFDM). The configuration V4 realizes the windowing, overlap and add block along with the preamble addition and detection in PL, and the rest of the blocks are implemented on PS. This configuration is only applicable in WOLA-OFDM. In configuration V5-V6, IFFT and CP addition operations are also moved to PL, and hence, frame size is reduced from 75 to 48, as shown in Table~\ref{dt}. Similarly, in configuration V7, data modulation and demodulator blocks are moved to PL, which means Boolean data being transferred between PL and PS. For configurations V8-10, the number of data elements are reduced from 48 to 24 since channel encoder and decoders with a coding rate of $\frac{1}{2}$ are moved to PL. In final configuration V10, an entire transceiver is realized on PL except for stimulus block. It can be observed that each configuration needs to be designed carefully to synchronize the data transfer between PS and PL. Furthermore, the architecture of the block changes when it is moved between PS and PL due to frame and sample-based processing. For PL implementation of each block, we have added pipelining inside the block as well as between the blocks. This demands additional synchronization efforts between PS and PL due to the change in latency.

\section{Experimental Setup and Result Analysis} \label{res}
In this section, we present the details of the experimental setup and analyze different results to compare the performance and complexity of the proposed transceivers.  

\subsection{Testbed Setup and Configuration} \label{design}
\begin{figure}[h!]
\centering
	\vspace{-0.2cm}
    \captionsetup{justification=centering}
    \includegraphics[width=\linewidth]{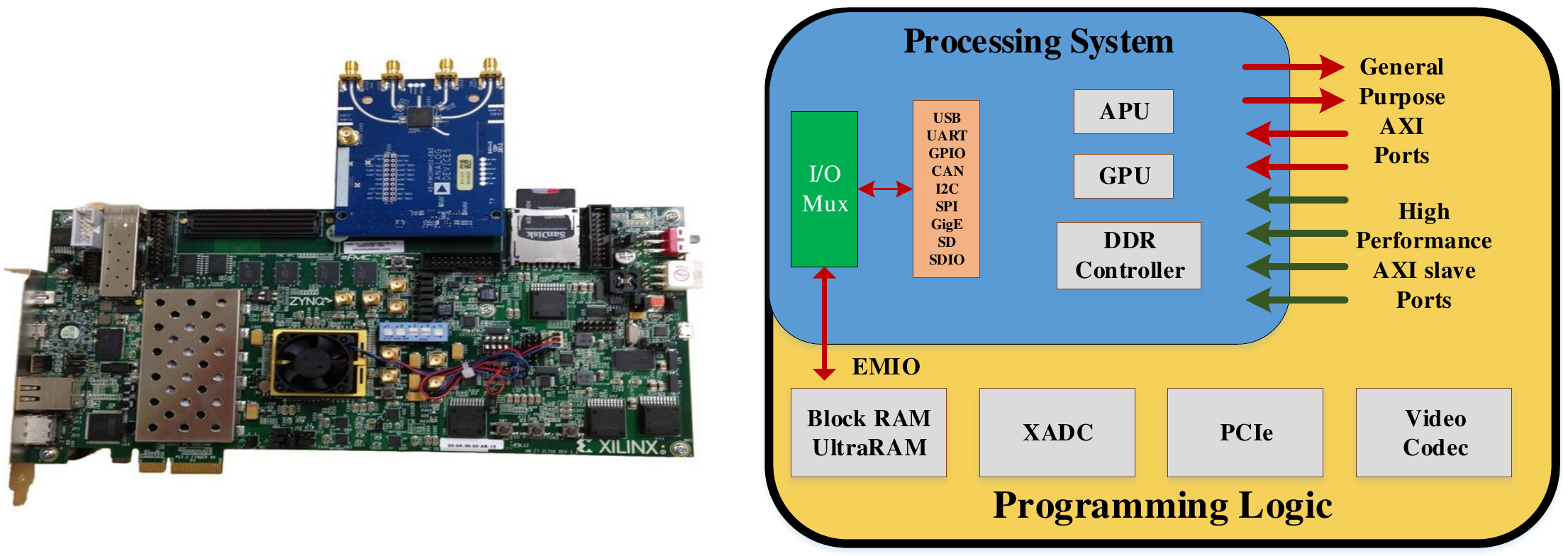}
    	\vspace{-0.2cm}
    \caption{Xilinx ZC706 evaluation board along with its important architectural features \cite{zynq}.}
    \label{Zyn}
    	\vspace{-0.2cm}
\end{figure}

In this paper, we have used the Xilinx ZSoC ZC706 evaluation board shown in Fig. ~\ref{Zyn} for implementation of the proposed transceivers and its specifications are briefly given in the Table.~ \ref{spec} \cite{zynq_book}. It consists of dual-core cortex A9 Advanced RISC Machines (ARM) as the software component (PS) and Xilinx 28nm Kintex 7-series as the hardware component (PL) \cite{zynq}. It is a processor centered device in which PS always boots first and is fully autonomous to PL. Both PS and PL communicate with each other using the Advanced eXtensible Interface (AXI) protocol. There are 9 AXI ports between PS and PL, and in this project, we use four ports for communication between PS and PL. Among various AXI protocols, we use AXI-stream for communication between PS and PL and AXI-Lite for communication between various signal processing blocks realized in the PL.

\begin{table}
\centering
	\vspace{-0.2cm}
    \captionsetup{justification=centering}
  \caption{Specifications of Zynq Board}
  	\vspace{-0.2cm}
    \label{spec}
  \begin{tabular}{|c|c|}
        \hline
     \textbf{Device}& ZC706 \\
     \hline
     \textbf{PL}& Kintex-7 \\
     \textbf{Registers}& 4,37,200 \\
     \textbf{LUTs}& 2,18,600 \\
     \textbf{DSP slices}& 900 \\
     \textbf{BRAM blocks}& 545\\
     \textbf{Processor}& ARM Cortex 9 \\
     \hline
\end{tabular}
	\vspace{-0.2cm}
\end{table}

For the design and implementation of the transceivers, we have used MATLAB 2017b and Vivado 2016.4. These are augmented with various MATLAB toolboxes such as Embedded coder and HDL coder/verifier to target the implementation on the PS and PL, respectively. To design and configure the AFE, we have used RF Toolbox along with communication and signal processing toolboxes, hardware support packages provided by Mathworks.

The AFE is programmed to meet the desired sampling and carrier frequency requirements of the LDACS. The custom digital and analog filters are designed and configured with the help of the RF Toolbox of the Matlab/Simulink. For the LDACS transceiver, the passband and stopband normalized frequencies are 0.33 and 0.41, respectively. The stopband attenuation is 80 dB, and the desired baseband sampling rate is 1.1 MHz. The filter at the receiver is identical to the transmitter. The local oscillator frequency is set to 985 MHz as the LDACS is deployed in the range of 960-1164 MHz, and for such up-conversion, various rate changer blocks are added in the design. The output of the AFE receiver is scaled by an appropriate factor (0.00019 to be exact) so that the power level of the signal at AFE receiver output closely matches the signal at AFE transmitter input. The AFE transceiver also introduces the phase noise due to transmission at RF frequency, and hence, it demands phase error estimation and correction at the receiver. For the proposed transceiver, we have used pilot signals in LDACS for phase estimation, and accordingly, correction is applied to all received samples. {\color{red}Next, we present the experimental results demonstrating the PSD and BER performance of the proposed transceivers using the discussed ZSoC based testbed.}

\subsection{Power Spectral Density (PSD) Comparison}
We begin with the PSD comparison for OFDM, WOLA-OFDM, and FOFDM based LDACS transceivers and analyzed their out-of-band (OOB) emission. Since higher OOB emission leads to higher interference to the legacy DME users, the transceivers should have lower OOB emission, and it should not exceed the desired interference constraints of the DME. Here, we assume that single LDACS transmitter is active in 1 MHz of the spectral gap between adjacent DME channels. 

\begin{figure}[h!]
    \centering
    \vspace{-0.2cm}
    \subfloat[]{\includegraphics[scale=0.45]{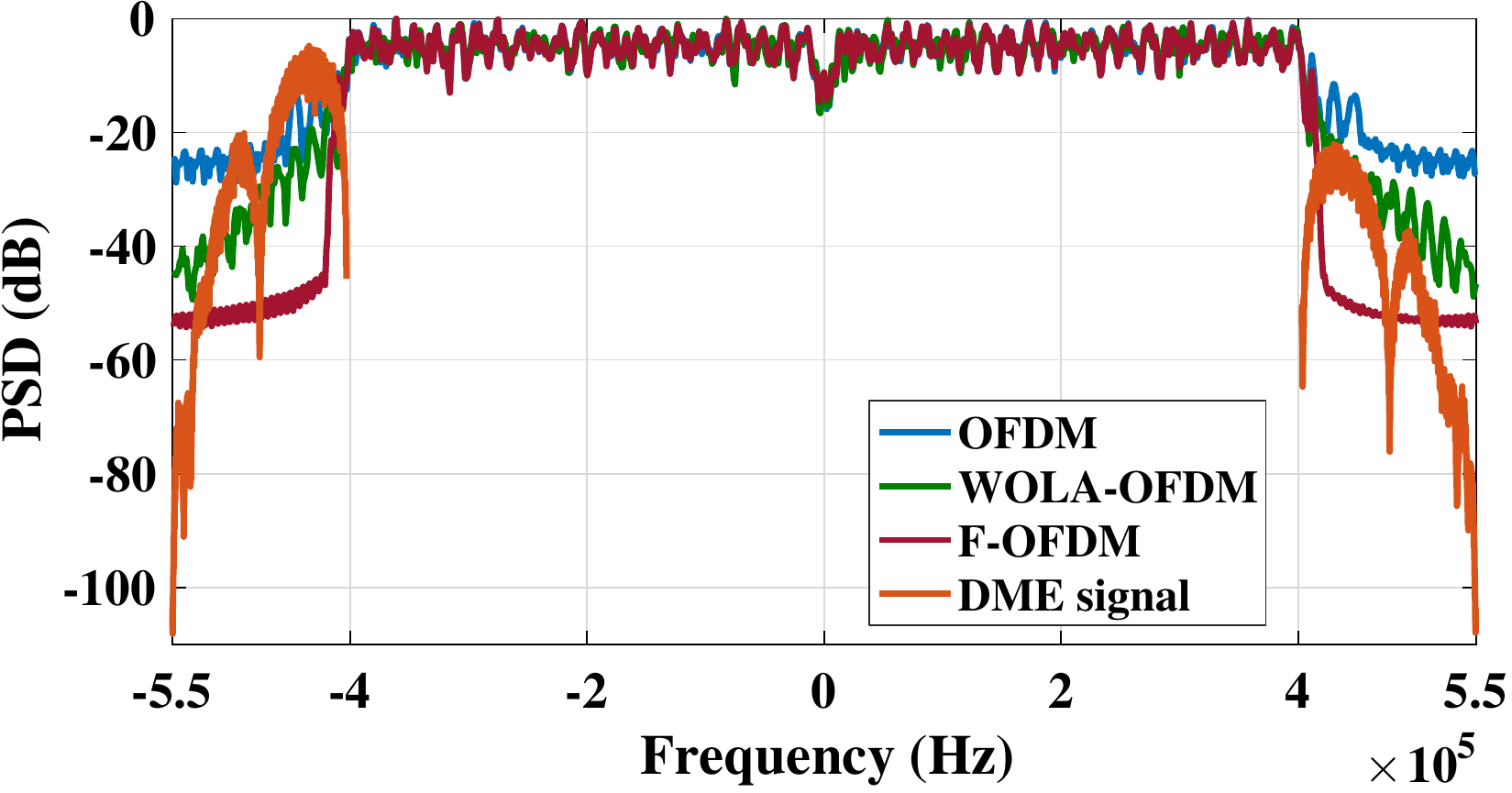}%
        \label{ps8}}
\vspace{-0.3cm}
    \subfloat[]{\includegraphics[scale=0.45]{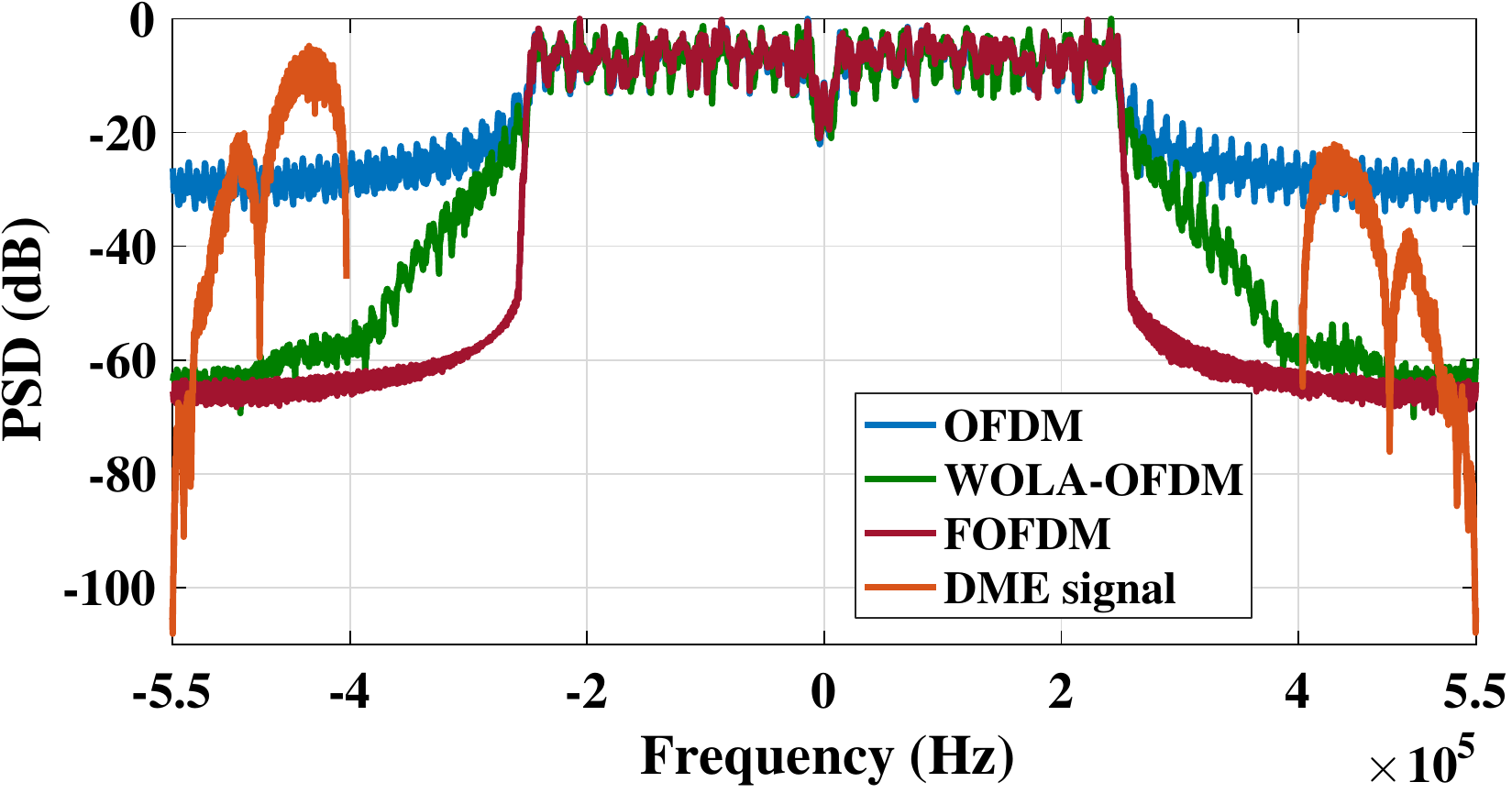}%
     \label{ps4}}
     \vspace{-0.2cm}
        \caption{The PSD comparison of various waveforms for two different transmission bandwidths, (a) 732kHz, and (b) 498kHz.}
        \vspace{-0.2cm}
            \label{psd}
\end{figure}

The PSD comparisons of OFDM, FOFDM, and WOLA-OFDM for 2 transmission bandwidths 1) 732 kHz and 2) 498 kHz are presented in Fig.~\ref{psd} (a) and (b) respectively. The legacy DME transmission is shown using orange color. Note that 498 kHz is the maximum possible bandwidth of existing OFDM based LDACS beyond which it fails to meet the interference constraints of DME. Though FOFDM can achieve 800 kHz bandwidth, we have chosen 732 kHz because it can be achieved using the frame structure same as that of 498 kHz, making it compatible with legacy LDACS \cite{NA}. For all the transceivers, word-length (WL) is fixed and equal to 32 bits. It can be observed that the FOFDM has approximately 40 dB lower OOB emission and hence, much lower interference to the legacy DME signals. This allows FOFDM to increase the transmission bandwidth from the standard 498 kHz (maximum possible in OFDM) to 732 kHz leading to significant improvement of approximately $50 \%$ in the spectral utilization over existing OFDM based LDACS.

\begin{figure}[!h]
\vspace{-0.2cm}
    \centering
    \subfloat[]{\includegraphics[scale=0.45]{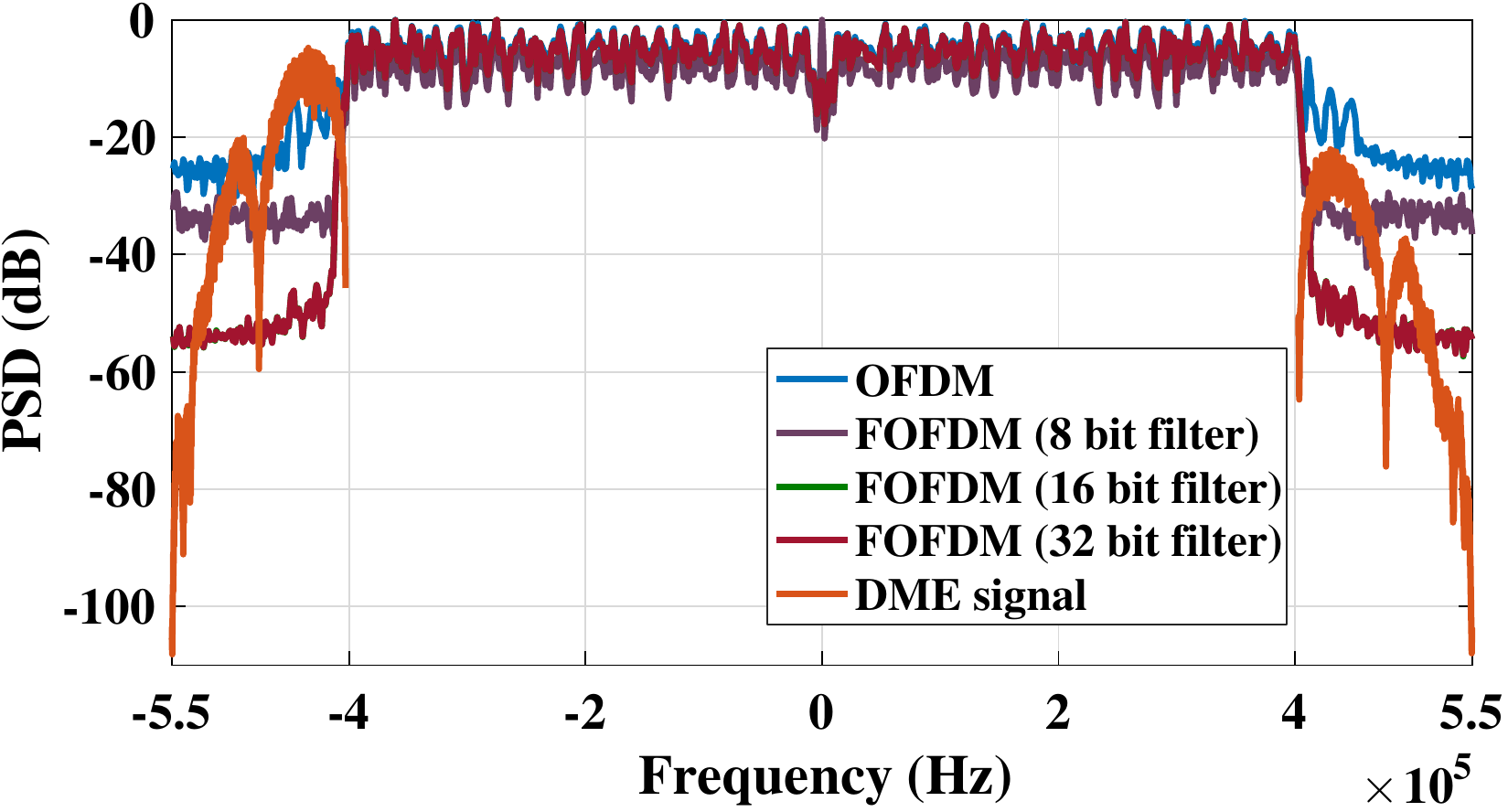}%
        \label{psfi}}
\vspace{-0.3cm}
    \subfloat[]{\includegraphics[scale=0.45]{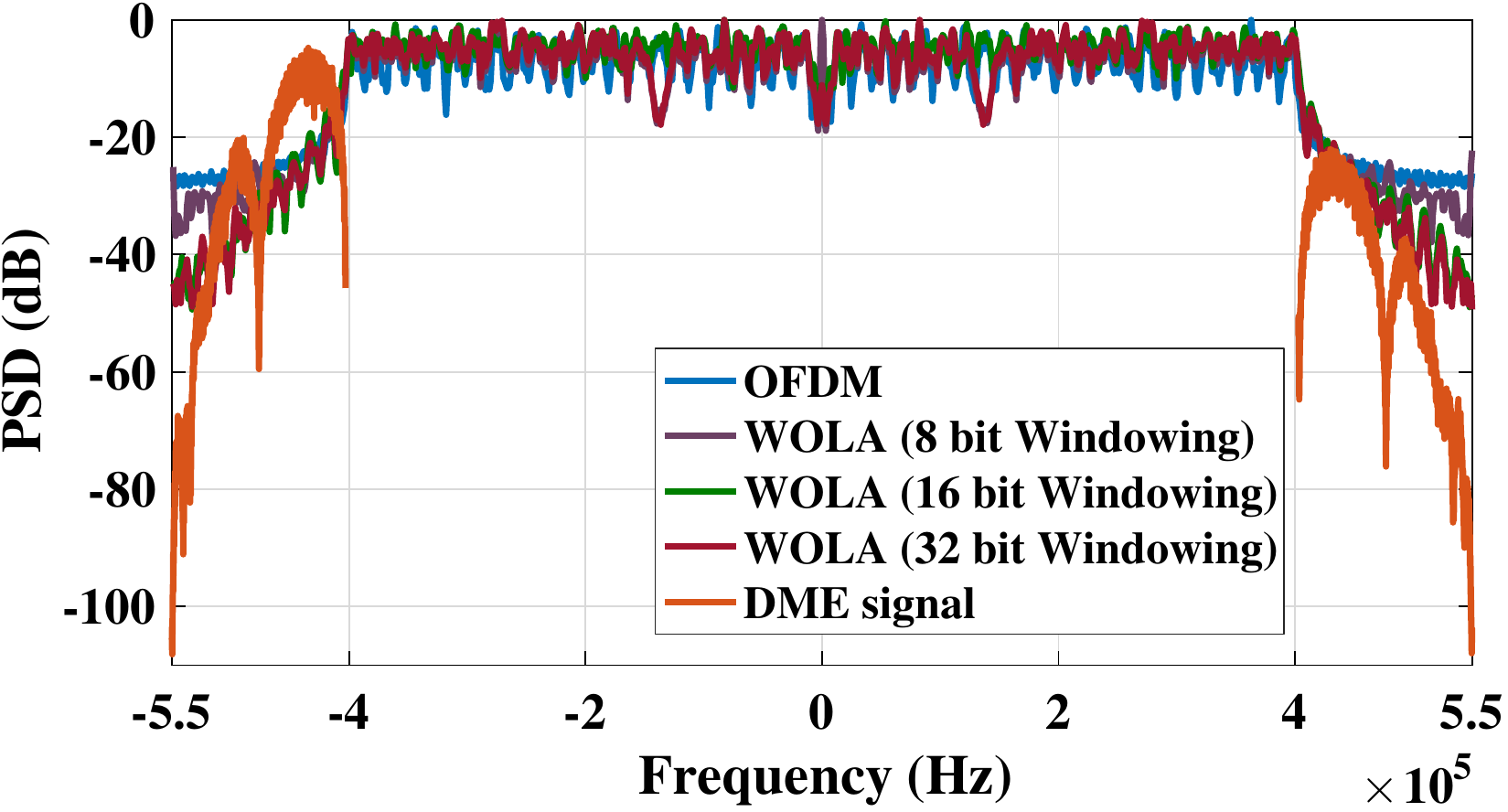}%
     \label{psw}}
     \vspace{-0.2cm}
        \caption{The PSD comparison of different fixed length implementation of (a) Filter and (b) Windowing.}
            \label{psd12}
            \vspace{-0.2cm}
\end{figure}

Next, we compare the performance of all transceivers by varying the WL. First, we change the WL of windowing and filtering blocks of the transceiver to 8/16 while keeping the WL of the rest of the transceiver to 32. As expected, there will be no change in the performance of OFDM as it does not involve windowing and filtering. The PSD of FOFDM and WOLA-OFDM for different WLs are shown in Fig \ref{psd12} (a) and (b). It can be observed that the PSD for WLs of 16 and 32 are almost identical, while there is significant degradation when WL is 8. Thus, it is possible to reduce the WL to 16 without compromising on the PSD performance.

\begin{figure}[!h]
    \centering
    \captionsetup{justification=centering}
    \includegraphics[scale=0.45]{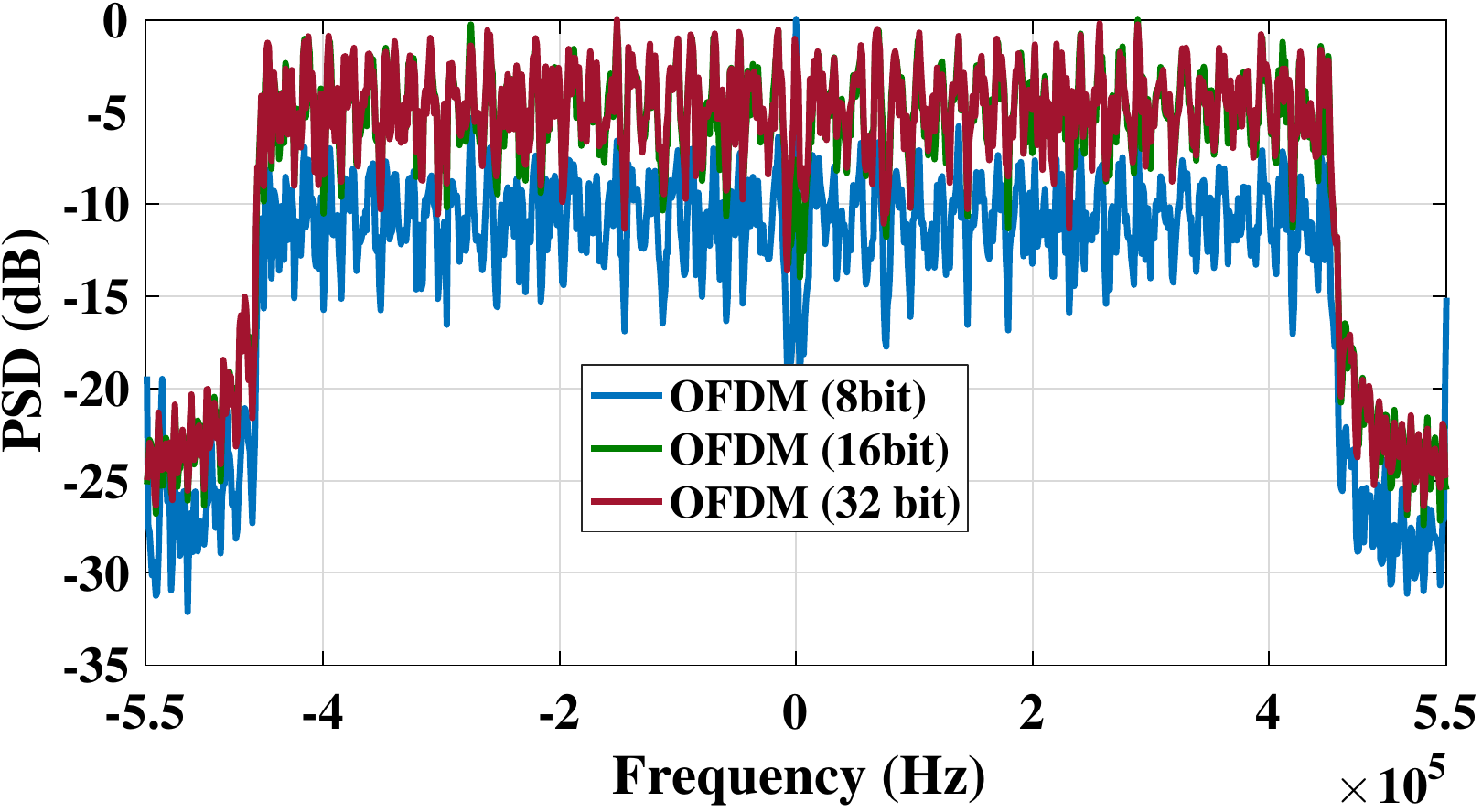}
    \vspace{-0.3cm}
    \caption{The PSD comparison of various waveforms for different fixed lengths.}
    \label{psf}
    	\vspace{-0.2cm}
\end{figure}

Next, we also analyzed the PSD performance when WL of the complete transceiver is reduced to 8 and 16 from 32. For illustration, we have shown the PSD of the OFDM in Fig.~\ref{psf}. Due to limited space constraints and to avoid repetitive results, we omitted the FOFDM and WOLA-OFDM transceivers. For all the transceivers, we observed that the PSD is almost identical for WL of 16 and 32, but there is significant degradation when WL is reduced to 8. 
 
To summarize, we observed that the FOFDM offers superior PSD and hence, lower interference to legacy DME when compared to other transceivers. This allows FOFDM to have wider transmission bandwidth, which is desired for the future air to ground communication. However, better PSD at the cost of poor BER performance is not acceptable for wireless transceivers. Hence, we study the BER performance of various transceivers in the next sub-section.

\subsection{Bit Error Rate Comparison}
For BER analysis, we consider end-to-end transceiver with LDACS channels (ENR, APT and TMA), DME interference, and RF impairments due to the AFE. We consider two transmission bandwidths: 1) 732 kHz and 2) 498 kHz. All BER results are obtained from hardware with at least 1000 frames of data.

\begin{figure}[!h]
    \centering
    \subfloat[]{\includegraphics[scale=0.5]{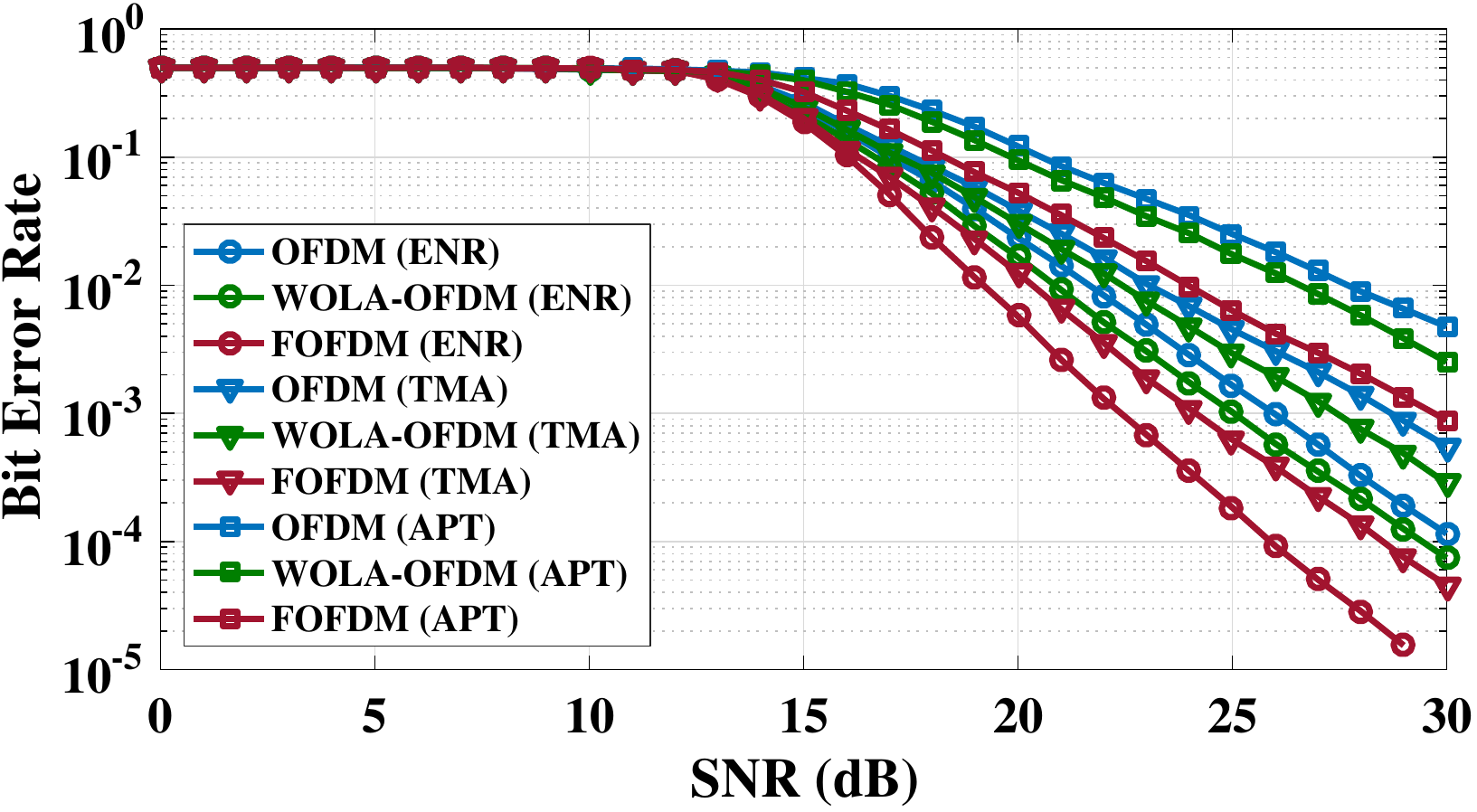}%
        \label{bea}}
\\\vspace{-0.3cm}
    \subfloat[]{\includegraphics[scale=0.5]{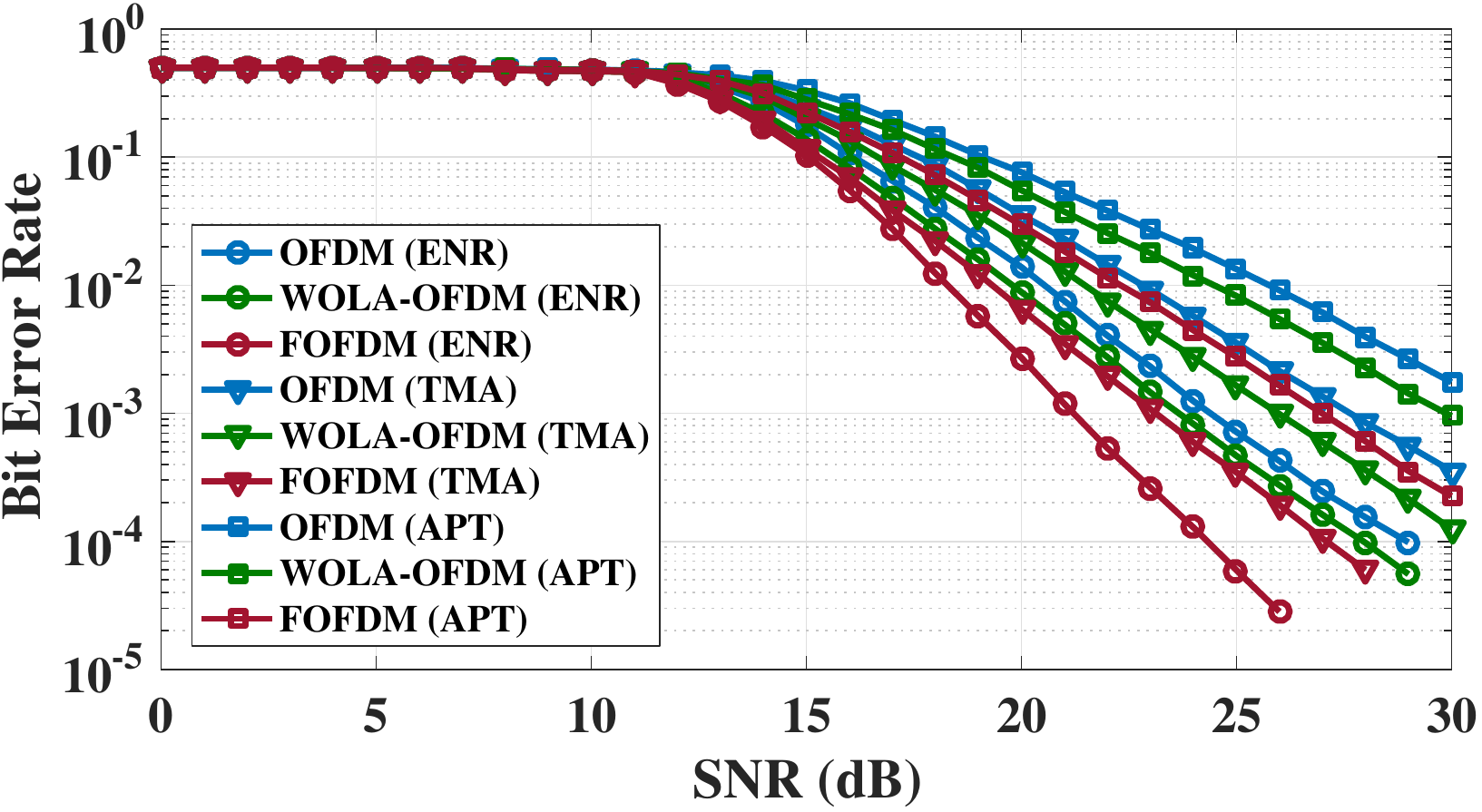}%
        \label{beb}}
        \caption{The BER comparison of various waveforms for two different transmission bandwidths, (a) 732kHz, and (b) 498kHz and three different channels.}
            \label{ber}
\end{figure}

{\color{red} As shown in Fig.~\ref{ber}, FOFDM offers better BER performance than OFDM and WOLA-OFDM for a wide range of SNRs. This is mainly due to the ability of FOFDM to reduce the effect of DME interference due to inherent filtering operation at the transmitter and receiver.} Note that though BER performance of WOLA-OFDM and OFDM is acceptable for 732 kHz, they cannot be deployed due to severe interference to DME.

\begin{figure}[!h]
    \centering
    \captionsetup{justification=centering}
    \includegraphics[scale=0.5]{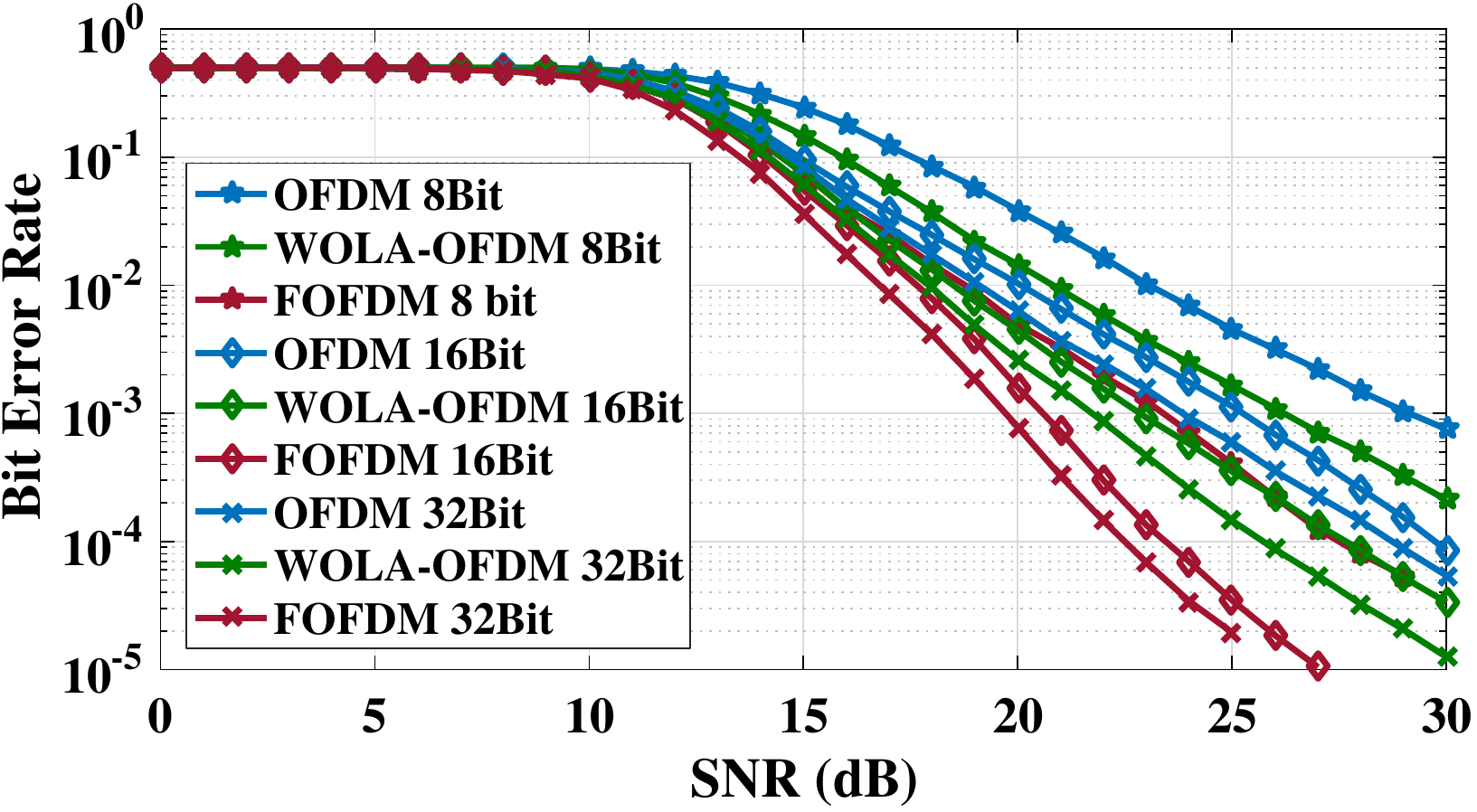}
    \vspace{-0.3cm}
    \caption{The BER comparison of various waveforms for different fixed lengths}
    \label{fix}
\end{figure}

Similar to PSD analysis, we compare the BER performance for three different WLs, 32, 16, and 8. As shown in Fig.~\ref{fix}, BER performance degrades with the decrease in WL for all the transceivers. However, FOFDM offers significantly better performance than others. In fact, the BER of FOFDM with WL of 16 is significantly better than that of WOLA-OFDM with WL of 32. Similarly, the BER of FOFDM with WL of 8 is significantly better than that of OFDM and WOLA-OFDM with WL of 32 and 16, respectively.

\begin{figure}[!h]
    \centering
    \captionsetup{justification=centering}
    \includegraphics[scale=0.5]{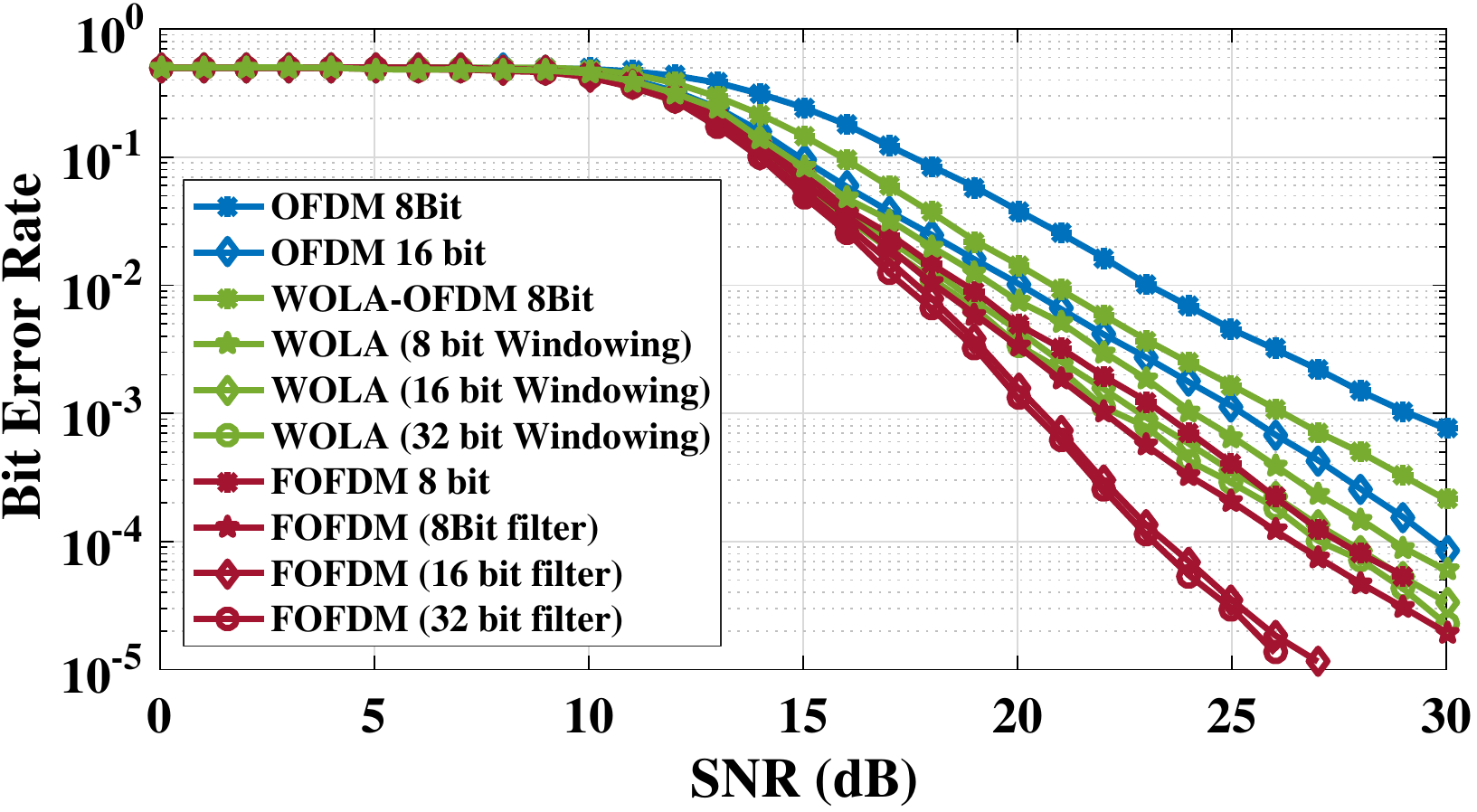}
    \vspace{-0.3cm}
    \caption{The BER comparison of various waveforms for different fixed lengths of filter and windowing operation}
    \label{berf}
\end{figure}

Next, we study the effect of WL of windowing and filtering blocks on the BER. Since the PSD and BER performance of transceivers with the WL of 16 and 32 are comparable, we have used the transceiver with WL of 16 for the results shown in Fig~\ref{berf}. It can be observed that the FOFDM with filtering operation using WL of 16 and 32 offers similar performance while its performance degrades when the WL is reduced to 8. The same trend is also observed for WOLA-OFDM. Thus, the selection of WL is an important criterion for transceiver, and higher WL may not guarantee a higher gain in performance. In terms of BER and PSD, FOFDM not only offers better performance but also leads to higher transmission bandwidth. However, this gain in performance should not come at a high cost in terms of complexity. To analyze this, we present the area and power complexity of these transceivers in the next sub-section.

\begin{table*}[!t]
	\centering
		\captionsetup{justification=centering}
		\caption{Resource Utilization and Power Consumption of Transceiver on ZSoC}
		\vspace{-0.2cm}
	\renewcommand{\arraystretch}{1}
	\begin{tabular}{|C{1.3cm}|C{1.3cm}|C{1cm}|C{1cm}|C{1cm}|C{1cm}|C{1cm}|C{1cm}|C{1cm}|C{1cm}|C{1cm}|} 	
		\hline
		\textbf{Parameter} & \textbf{Waveform} & \textbf{V2} & \textbf{V3} & \textbf{V4} & \textbf{V5}& \textbf{V6} & \textbf{V7} & \textbf{V8} & \textbf{V9} & \textbf{V10}\\
		\hline
		\textbf{No. of Flip-Flops} & OFDM & N/A & 14200 (3.25\%) & N/A & 31617 (7.21\%) & 31945 (7.29\%) & 32628 (7.45\%) & 33982 (7.75\%) & 37738 (8.61\%) & 38193 (8.72\%)\\
		& WOLA-OFDM & N/A & 14200 (3.25\%) & 23018 (5.26\%) & 33015 (7.54\%) & 34785 (7.93\%) & 38254 (8.75\%) & 41945 (9.58\%) & 43015 (9.83\%) & 44971 (10.02\%)\\
    	& FOFDM & 10100 (2.31\%) & 29954 (6.84\%) & N/A & 37285 (8.51\%) & 39120 (8.92\%) & 40015 (9.13\%) & 41184 (9.40\%) & 44015 (10.04\%) & 46253 (10.56\%)\\
		\hline
		\textcolor{blue}{\textbf{No. of DSP48}} & OFDM & N/A & 534 (59.33\%) & N/A & 570 (63.33\%) & 570 (63.33\%) & 570 (63.33\%) & 570 (63.33\%) & 570 (63.33\%) & 570 (63.33)\\
		& WOLA-OFDM & N/A & 534 (59.33\%) & 554 (61.56\%) & 570 (63.33\%) & 570 (63.33\%) & 570 (63.33\%) & 570 (63.33\%) & 570 (63.33\%) & 570 (63.33\%)\\
		& FOFDM & 296 (32.89\%) & 785 (87.22\%) & N/A & 812 (90.22\%) & 812 (90.22\%) & 812 (90.22\%) & 812 (90.22\%) & 812 (90.22\%) & 812 (90.22\%) \\
		\hline
		\textbf{No. of LUT as Memory}& OFDM & N/A & 396 (0.56\%) & N/A & 865 (1.23\%) & 881 (1.25\%) & 918 (1.30\%) & 922 (1.31\%) & 941 (1.34\%) & 994 (1.35\%)\\
		& WOLA-OFDM & N/A & 396 (0.56\%) & 685 (0.972\%) & 940 (1.34\%) & 945 (1.35\%) & 972 (1.38\%) & 982 (1.39\%) & 1050 (1.48\%) & 1102 (1.56\%)\\
        & FOFDM & 64 (0.09\%) & 411 (0.583\%) & N/A & 894 (1.27\%) & 913 (1.29\%) & 936 (1.32\%) & 943 (1.34\%) & 964 (1.37\%) & 995 (1.41\%)\\
		\hline
		\textcolor{blue}{\textbf{No. of LUT as Logic}}& OFDM & N/A & 22083 (10.10\%) & N/A & 31687 (14.50\%) & 31985 (14.63\%) & 32555 (14.89\%) & 33509 (15.328\%) & 35509 (16.24)& 36657 (16.77\%)\\
		& WOLA-OFDM & N/A & 22083 (10.10\%) & 30513 (13.96\%) & 33218 (15.195\%) & 34824 (15.96\%) & 36156 (16.54\%) & 37599 (17.20\%)  & 41621 (19.04\%) & 44376 (20.30\%)\\
		& FOFDM & 5350 (2.45\%) & 25361 (11.61\%) & N/A & 32811 (15.01\%) & 34495 (15.78\%) & 35391 (16.19\%) & 37052 (16.95\%) & 40660 (18.60\%) & 42539 (19.46\%)\\
		\hline
		\textbf{No. of}& OFDM & N/A & 35 & N/A & 683 & 745 & 1144 & 1217 & 1882 & 1930\\
		\textbf{MUXes}& WOLA-OFDM & N/A & 35 & 872 & 1254 & 1501 & 1784 & 1835 & 2575 & 2725 \\
		          & FOFDM & 25 & 57 & N/A & 835 & 1152 & 1401 & 1523 & 1985 & 2102\\
		\hline
		
		\textcolor{blue}{\textbf{Dynamic}} & OFDM & N/A & 0.045 & N/A & 0.285 & 0.295 & 0.297 & 0.299 & 0.301 & 0.304\\
		\textcolor{blue}{\textbf{Power}} & WOLA-OFDM & N/A & 0.073 & 0.161 & 0.294 & 0.296 & 0.299 & 0.301 & 0.302 & 0.306\\
	\textcolor{blue}{\textbf{in Watt}} & FOFDM & 0.112 & 0.205 & N/A & 0.434 & 0.493 & 0.494 & 0.496 & 0.500 & 0.509\\
				
		\hline
	\end{tabular}
		\vspace{-0.2cm}
	\label{result}
\end{table*}

\subsection{Resource Utilization and Power Consumption}
In this subsection, we compare the resource utilization and power consumption of the proposed OFDM, WOLA-OFDM, and FOFDM architectures for ten different configurations. Since the bandwidth of the transceiver is tunable, the results are shown in Table~\ref{result} corresponds to 732 kHz bandwidth, which has higher complexity than 498 kHz bandwidth. To begin with, we consider the WL of 16 in Table~\ref{result}. All results are obtained after realizing the transceiver on ZC706 from Xilinx.

As shown in Table~\ref{result}, the comparison is made in terms of the number of flip-flops, DSP48 (embedded multipliers), look-up-table (LUT) for memory, LUT for logical and arithmetic operations, multiplexers and dynamic power consumption of the PL. The static power consumption of PS (1.566 W) and PL (0.247W) is, as expected, identical for all configurations and hence, not shown in the table.

Since V1 configuration is realized completely in PS, PL resource utilization results are omitted. In V2, FOFDM resource utilization is due to the filtering block realized in PL. As expected, multiply-accumulate operations in the filter are mapped to DSP48 to get the best possible performance. In V3, preamble addition and detection block is moved to PL, and due to in-built auto-correlation operations, it is one of the most complicated block as evident from the increase in the resource utilization compared to V2. Similarly, a significant increase in resource utilization and power consumption is observed in V5, where FFT/IFFT is moved from PS to PL. 

To summarize, FOFDM incurs 27\% higher DSP48 than others due to MAC-based filtering, which can be shifted to LUT if needed. For example, windowing operation in WOLA-OFDM is realized using a combination of DSP48 and LUTs. The utilization of the rest of the resources is almost identical in all three waveforms. The IFFT/FFT block consumes the highest power, followed by filtering in the FOFDM. Due to limited space constraints, we have skipped some results. For completeness of the discussion, we briefly mention the observations: 1) The power consumption of the FOFDM increases slightly if we reduce the number of DSP48 at the cost of LUT as logic, 2) Resource utilization and power consumption increases with the rise in the WL, 3) The process of pipelining the transceiver architecture involves addition of registers at the appropriate locations so as to reduce the critical path delay. The reduction in critical path delay allows the transceiver to be clocked at higher frequency. Thus, pipelining offers trade-off between resource (number of FFs) utilization and clock period. For instance, the critical path delay with and without pipelining for OFDM, WOLA-OFDM and FOFDM transceivers are \{9.75 ns, 10.25 ns, 12.5 ns\}and \{259 ns, 265.83 ns, 271.23 ns\}, respectively. Furthermore, pipelining incurs additional latency due to newly added registers \cite{Vuk}.

\begin{figure}[!h]
     \centering
     \vspace{-0.2cm}
     \subfloat[]{\includegraphics[scale=0.25]{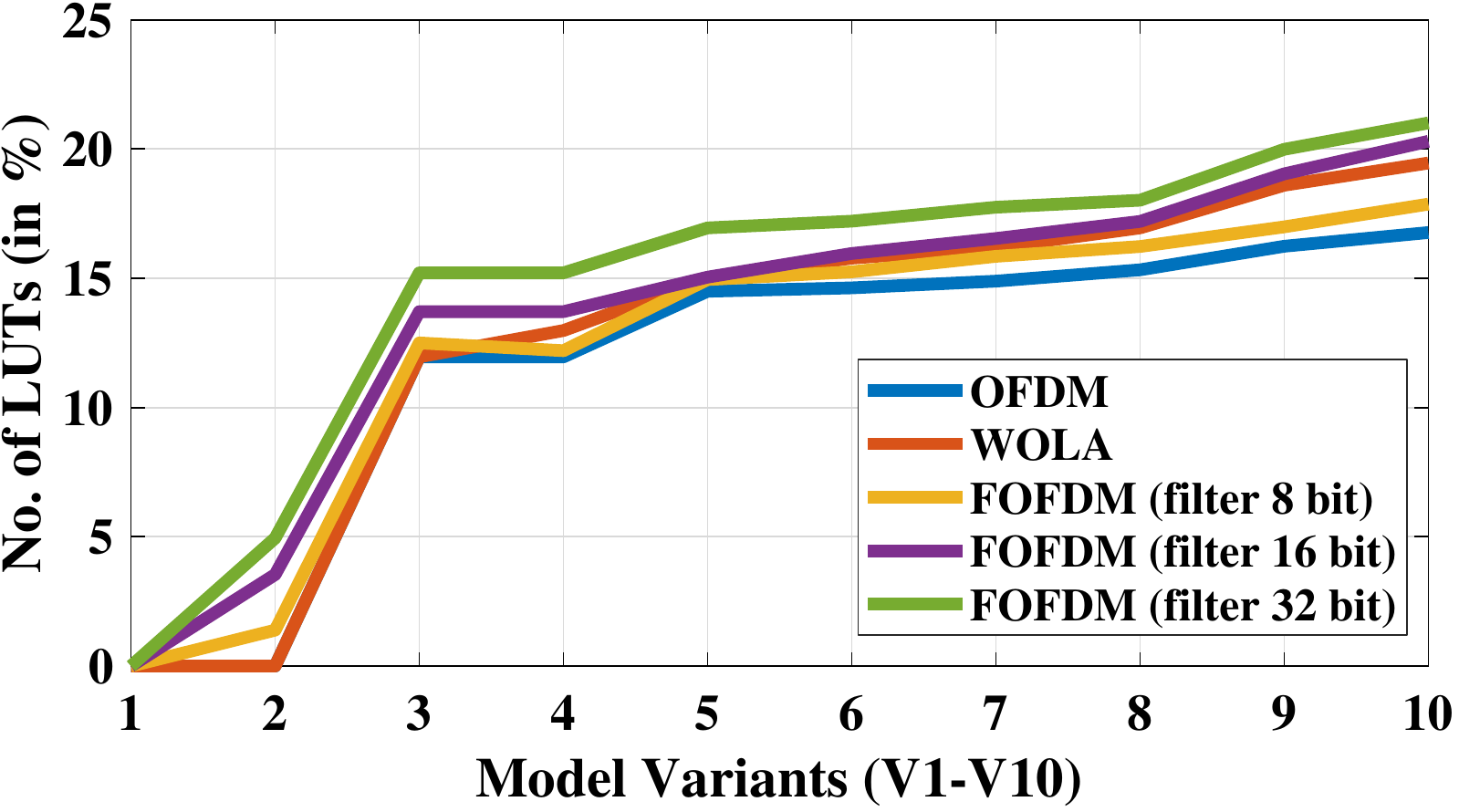}%
         \label{v}}
     \subfloat[]{\includegraphics[scale=0.25]{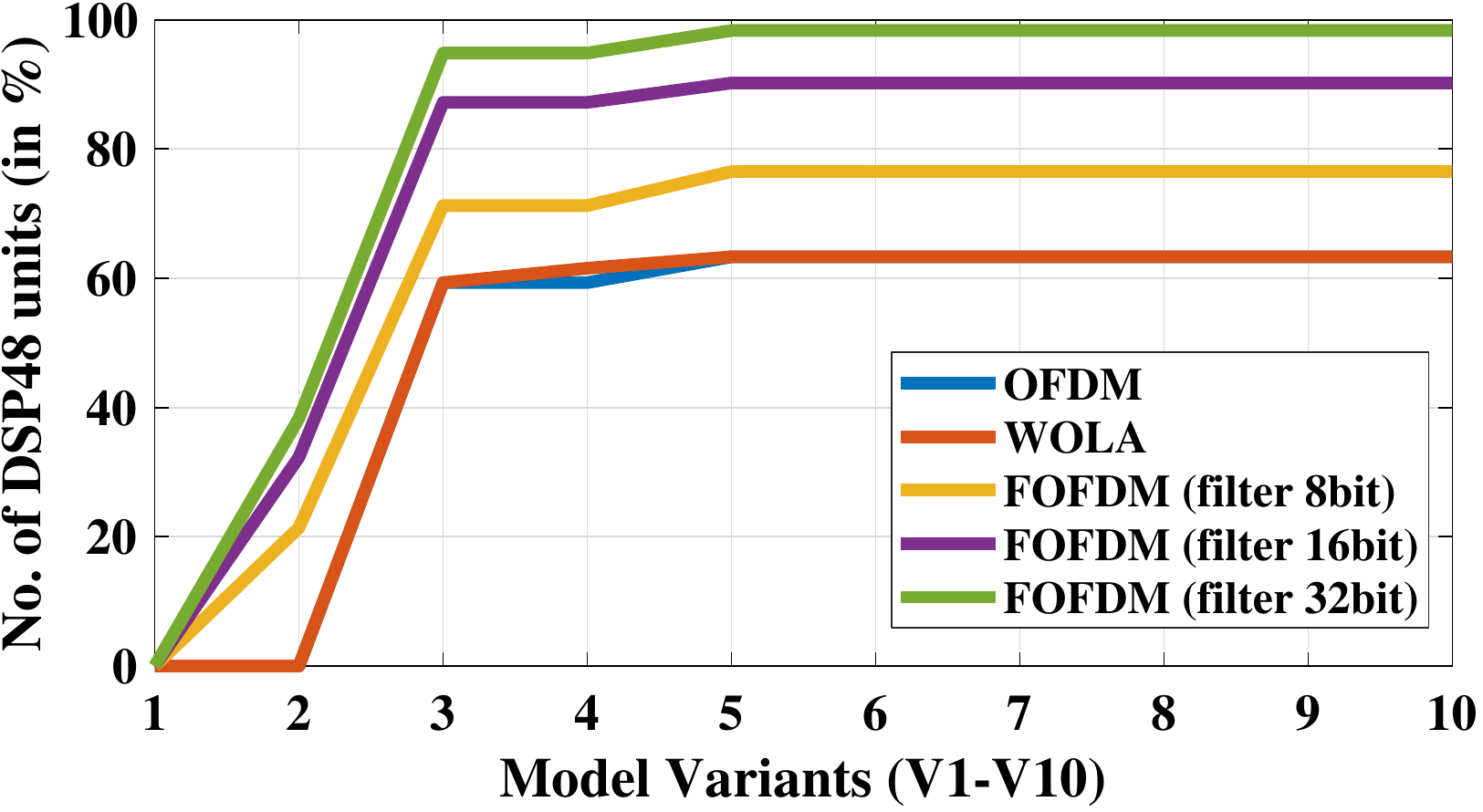}%
         \label{vr4}}
         \vspace{-0.2cm}
     \caption{Analysis of resource utilization on ZC706 for different model variants and fixed lengths, (a) Number of LUTs and (b) Number of DSP'48 units.}
     \vspace{-0.2cm}
     \label{de11}
 \end{figure}

In Fig.~\ref{berf}, we discussed the effect of WL of filter coefficients in the filtering block of FOFDM on BER. In the case of resource utilization, we observed the increase in the utilization with WL, as shown in Fig~\ref{de11}. For WOLA-OFDM, different WL of windowing coefficients is not feasible for air to ground communications due to poor PSD and BER performance.

The above-discussed results show that the FOFDM offers better sidelobe attenuation and better BER performance in trade-off to resource utilization. Filter designed by considering 8 bit fixed WL performs worse than 16/32 bit filter in terms of PSD and BER but better in terms of resource utilization. The FOFDM has a higher usage of resources compared to OFDM and WOLA-OFDM but still uses less than 50\% of the Zynq ZC702 resources except for DSP48. This makes the FOFDM based LDACS as an appealing substitute to the future air to ground communication.

\section{Conclusion}
In this paper, we proposed the design and implementation of end-to-end LDACS transceiver on Xilinx ZC706 FPGA via a hardware-software co-design approach. We considered OFDM based LDACS and improved the performance using windowing and/or filtering. The proposed approach offers flexibility to choose the configuration along with the word-length for a given area, power, and delay constraints. The transceiver architectures are then integrated with analog front-end to endorse its performance in the presence of various RF impairments, DME interference, and LDACS specific wireless channels.  Detailed experimental results are presented to analyze the area, power, PSD, and BER performance for OFDM, WOLA-OFDM, and FOFDM, having three word-lengths of 8/16/32 bit. The results show that the transceivers with the WL of 16 and 32 bit offer similar performance while the performance degrades for 8 bit WL. The Filtered OFDM based LDACS performs much better in terms of out of band emission (approximately 40 dB) and has significantly better BER performance, which allows adapting a wider transmission bandwidth up to 732 kHz with a slight penalty in terms of resource utilization and power consumption. This paper provides profound performance analysis and results to present the flexibility of end-to-end LDACS transceiver and proposed wide bandwidth FOFDM based LDACS offers an attractive solution for the next generation air to ground communication.

\footnotesize{
\section*{Acknowledgment}
This work is supported by the Visvesvaraya Ph.D. fellowship and DST Inspire Faculty Fellowship granted by Govt. of India. We would like to thank the authors of the paper \cite{Be1,Be2} for their help in the implementation of OFDM on ZSoC.
}

\begin{thebibliography}{1}
\bibitem{STAT}
STATFOR, the EUROCONTROL Statistics and Forecast Service,  \lq\lq Challenges of Growth, Task 4: European Air Traffic in 2035,\rq\rq \emph{EUROCONTROL,} June.~2013. \\
http://www.eurocontrol.int/statfor.

\bibitem{Schnell}
M. Schnell, U. Epple, D. Shutin and N. Schneckenburger, \lq\lq LDACS: future aeronautical communications for air-traffic management,\rq\rq in ~\emph{IEEE Communications Magazine,} vol.~52, no.~5, pp.~104--110, May.~2014.

\bibitem{SESAR}
SESAR Joint Undertaking, \lq\lq NextGen – SESAR State of Harmonisation,\rq\rq~\emph{Federal Aviation Administration} June.~2016.



\bibitem{Zavani}
R. Zayani, Y. Medjahdi, H. Shaiek and D. Roviras, \lq\lq WOLA-OFDM: A Potential Candidate for Asynchronous 5G\rq\rq , in \emph{ 2016 IEEE Globecom Workshops (GC Wkshps),} Washington, DC, 2016, pp. 1-5.

\bibitem{Abdoli}
J. Abdoli, M. Jia and J. Ma, \lq\lq Filtered OFDM: A new waveform for future wireless systems\rq\rq , ~\emph{2015 IEEE 16th International Workshop on Signal Processing Advances in Wireless Communications (SPAWC),} Stockholm, 2015, pp. 66-70.

\bibitem{sasha}
S. Garg, N. Agrawal, S. J. Darak and P. Sikka, \lq\lq Spectral coexistence of candidate waveforms and DME in air-to-ground communications: Analysis via hardware software co-design on Zynq SoC,\rq\rq ~\emph{2017 IEEE/AIAA 36th Digital Avionics Systems Conference (DASC),} St. Petersburg, FL, 2017, pp. 1-6.

\bibitem{Ta1}
M.~Sajatovic, B.~Haindl, M.~Ehammer, Th.~Gräupl, M.~Schnell, U.~Epple, and S.~Brandes, \lq\lq L-DACS1 System Definition Proposal: Deliverable D2,\rq\rq~ in~\emph{Technical Report, Eurocontrol,} no.~1.0, Feb~2009.

\bibitem{Ta2}
S.~Brandes, et al. \lq\lq Physical layer specification of the L-band Digital Aeronautical Communications System (L-DACS1), \rq\rq ~\emph{IEEE Integrated Communications, Navigation and Surveillance Conference,} pp.~1-12, Arlington, USA, May.~2009.

\bibitem{Ta3}
G.~Snjezana, M.~Schnell, and U.~Epple. \lq\lq The LDACS1 physical layer design,\rq\rq~ \emph{INTECH Open Access Publisher,} 2011.

\bibitem{Ta5}
T.~H.~Pham, V.~A.~Prasad and A.~S.~Madhukumar, \lq\lq A Hardware-Efficient Synchronization in L-DACS1 for Aeronautical Communications,\rq\rq~in \emph {IEEE Transactions on Very Large Scale Integration (VLSI) Systems,} vol.~26, no.~5, pp.~924-932, May~2018.

\bibitem{Ta6}
T.~H.~Pham, A.~P.~Vinod, and A.~S.~Madhukumar, \lq\lq An efficient data aided synchronization in L-DACS1 for aeronautical communications,\rq\rq ~ \emph{Proc. ACM Int. Conf. Data Mining, Commun. Inf. Technol. (DMCIT),} pp.~1–5, Phuket, Thailand, May~2017.

\bibitem{Ta7}
S.~Shreejith, A.~Ambede, A.~P.~Vinod and S.~A.~Fahmy, \lq\lq A power and time efficient radio architecture for LDACS1 air-to-ground communication,\rq\rq ~\emph{IEEE/AIAA 35th Digital Avionics Systems Conference (DASC),}  pp.~1-6, Sacramento, USA, Sept.~2016.

\bibitem{Ta8}
S.~Shreejith, M.~Libin, A.~P.~Vinod and F.~Suhaib, \lq\lq Efficient spectrum sensing for aeronautical LDACS using low-power correlators\rq\rq,~in \emph{ IEEE Transactions on Very Large Scale Integrated Systems,} vol.~26, no.~6, pp.~1183-1191, June~2018.

\bibitem{Ta9}
A.~Ambede, A.~P.~Vinod and A.~S.~Madhukumar, \lq\lq Design of a low complexity channel filter satisfying LDACS1 spectral mask specifications for air-to-ground communication,\rq\rq ~\emph{Integrated Communications Navigation and Surveillance (ICNS),} pp.~1-7, Herndon, USA, Apr.~2016.

\bibitem{Ta10}
S.~Dhabu, A.~P.~Vinod and A.~S.~Madhukumar, \lq\lq Low complexity fast filter bank-based channelization in L-DACS1 for aeronautical communications,\rq\rq ~\emph{IEEE 13th International New Circuits and Systems Conference (NEWCAS),} pp.~1-4 ,Grenoble, France, Jun.~2015.




\bibitem{Jonathan}
J. van de Belt, P. D. Sutton and L. E. Doyle, \lq\lq Accelerating software radio: Iris on the Zynq SoC, \rq\rq ~\emph{2013 IFIP/IEEE 21st International Conference on Very Large Scale Integration (VLSI-SoC)}, Istanbul, 2013, pp. 294-295.

\bibitem{ryan}
R. Marlow, C. Dobson and P. Athanas, \lq\lq An enhanced and embedded GNU radio flow, \rq\rq ~\emph{2014 24th International Conference on Field Programmable Logic and Applications (FPL)}, Munich, 2014, pp. 1-4.

\bibitem{Baris}
B. Özgül, J. Langer, J. Noguera and K. Visses,  \lq\lq Software-programmable digital pre-distortion on the Zynq SoC, \rq\rq ~\emph{2013 IFIP/IEEE 21st International Conference on Very Large Scale Integration (VLSI-SoC)}, Istanbul, 2013, pp. 288-289.

\bibitem{john}
J. Pendlum, M. Leeser and K. Chowdhury, \lq\lq Reducing Processing Latency with a Heterogeneous FPGA-Processor Framework, \rq\rq ~\emph{2014 IEEE 22nd Annual International Symposium on Field-Programmable Custom Computing Machines}, Boston, MA, 2014, pp. 17-20.

\bibitem{Shree}
S. Shreejith, B. Banarjee, K. Vipin, and S. A. Fahmy, \lq\lq Dynamic Cognitive Radios on the Xilinx Zynq Hybrid FPGA,\rq\rq ~\emph{Cognitive Radio Oriented Wireless Networks - 10th International Conference, CROWNCOM,} Doha, Qatar, 2015, pp. 427–437.


\bibitem{Be1}
B. Drozdenko, M. Zimmermann, Tuan Dao, M. Leeser and K. Chowdhury, \lq\lq High-level hardware-software co-design of an 802.11a transceiver system using Zynq SoC,\rq\rq ~\emph{2016 IEEE Conference on Computer Communications Workshops (INFOCOM WKSHPS),} San Francisco, CA, 2016, pp. 682-683.

\bibitem{Be2}
B. Drozdenko, M. Zimmermann, T. Dao, K. Chowdhury and M. Leeser, \lq\lq Hardware-Software Codesign of Wireless Transceivers on Zynq Heterogeneous Systems,\rq\rq in ~\emph{IEEE Transactions on Emerging Topics in Computing.}

\bibitem{Ta4}
H.~Jamal; D.~W.~Matolak, \lq\lq FBMC and LDACS Performance for Future Air to Ground Communication Systems,\rq\rq~in \emph{IEEE Transactions on Vehicular Technology ,} vol.~66, no.~6, pp.~5043-5055, June~2017.

\bibitem{gfdm}
N.~Michailow et al., \lq\lq Generalized Frequency Division Multiplexing for 5th Generation Cellular Networks,\rq\rq~in \emph{IEEE Transactions on Communications,} vol.~62, no.~9, pp.~3045-3061, Sept.~2014.

\bibitem{NA}
N. Agrawal, S. J. Darak and F. Bader, \lq\lq New Spectrum Efficient Reconfigurable Filtered-OFDM Based L-Band Digital Aeronautical Communication System,\rq\rq in \emph{IEEE Transactions on Aerospace and Electronic Systems,} vol. 55, no. 3, pp. 1.

\bibitem{appendix}
N. Agrawal and S. Garg, S. J. Darak and F. Bader \lq\lq Spectral Coexistence of LDACS and DME: Analysis via Hardware Software Co-Design in Presence of Real Channels and RF Impairments \rq\rq available at http://arxiv.org/abs/1910.04649.

\bibitem{wol}
R. Zayani, Y. Medjahdi, H. Shaiek and D. Roviras, \lq\lq WOLA-OFDM: A Potential Candidate for Asynchronous 5G,\rq \rq \emph{IEEE Globecom Workshops (GC Wkshps),} Washington, DC, 2016, pp. 1-5.

\bibitem{SJ}
S. J. Darak, A. P. Vinod, E. M. K. Lai, J. Palicot and H. Zhang, \lq\lq Linear-Phase VDF Design With Unabridged Bandwidth Control Over the Nyquist Band,\rq\rq in ~\emph{IEEE Transactions on Circuits and Systems II: Express Briefs,} vol. 61, no. 6, pp. 428-432, June 2014.

\bibitem{SJ2}
S. J. Darak, A. P. Vinod, R. Mahesh and E. M. K. Lai, \lq\lq A reconfigurable filter bank for uniform and non-uniform channelization in multi-standard wireless communication receivers,\rq\rq ~\emph{17th International Conference on Telecommunications (ICT'17),} Doha, 2010, pp. 951-956.

\bibitem{15}
Eurocontrol, FAA, \lq\lq Communications Operating Concept and Requirements for the Future Radio System,\rq\rq COCR V2.0. 

\bibitem{17}
B-VHF project, \lq\lq Software Implementation of Broadband VHF channel models,\rq\rq  D-17, May~2006.

\bibitem{16}
B-VHF project, \lq\lq Expected B-AMC system performance,\rq\rq D-5,  May~2006.

\bibitem{18}
A. Haider, \lq\lq Comparison of proposals for the Future Aeronautical Communication System LDACS, \rq\rq ~\emph{Master Thesis}, ILMENAU University of Technology and Institute of Communication and Navigation German Aerospace Center, 2012.





\bibitem{zynq_book}
L.H. Crockett, R.A. Elliot, M.A. Enderwitz and R.W. Stewart, \lq\lq The Zynq Book: Embedded Processing with the Arm Cortex-A9 on the Xilinx Zynq-7000 All Programmable Soc\rq\rq , in \emph{Strathclyde Academic Media,} 2014.

\bibitem{zynq}
\lq\lq Zynq-7000 All Programmable SoC Overview\rq\rq , \emph{Application Processor Unit}, 2012.

\bibitem{Vuk}
V. Marojevic, X. Revés Ballesté and A. Gelonch, \lq\lq A Computing Resource Management Framework for Software-Defined Radios,\rq\rq in \emph{IEEE Transactions on Computers,} vol. 57, no. 10, pp. 1399-1412, Oct. 2008.
\end{thebibliography}

\end{document}


\title{Appendix for Spectral coexistence of LDACS DME via Hardware Software Co-design approach}
\maketitle
\chapter{HDL Implementation details for OFDM based transceiver architecture}
\section{Architecture}
The appendix gives the details on the PS and PL implementation of the basic building blocks of OFDM based transceiver system. In our work, the OFDM transceiver system is based on the IEEE 802.11a standard. The transceiver consists of blocks such as scrambler, convolutional encoder, interleaver, binary phase shift keying (BPSK) modulator, fast Fourier transform (FFT) and cyclic prefix adder in the transmitter with additional blocks for windowing and filtering for the the corresponding WOLA-OFDM and FOFDM implementations. The implementation details for the stimulus(i.e the information source), transmitter and the receiver are explained below.

\subsection*{Stimulus Subsystem} 
The stimulus subsystem, reads the bitstream to be transmitted from the MATLAB workspace. The input is a binary stream containing 864 bits. Out of the 864 bit, 24 bits are transmitted per OFDM frame. Thus, overall 36 OFDM frames are transmitted. The further operations have been model for a single frame. These operations are then repeated as a new set of 24 bits are read. This is done with the help of a free-running counter keeping a track of the number of frames. With the help of the selector block, we select 24 bits from the incoming stream as the input to the transmitter.

\subsection*{Transmitter}
\subsubsection*{Scrambler}
\begin{figure}[H]	
	\centering
	\includegraphics[scale=0.5]{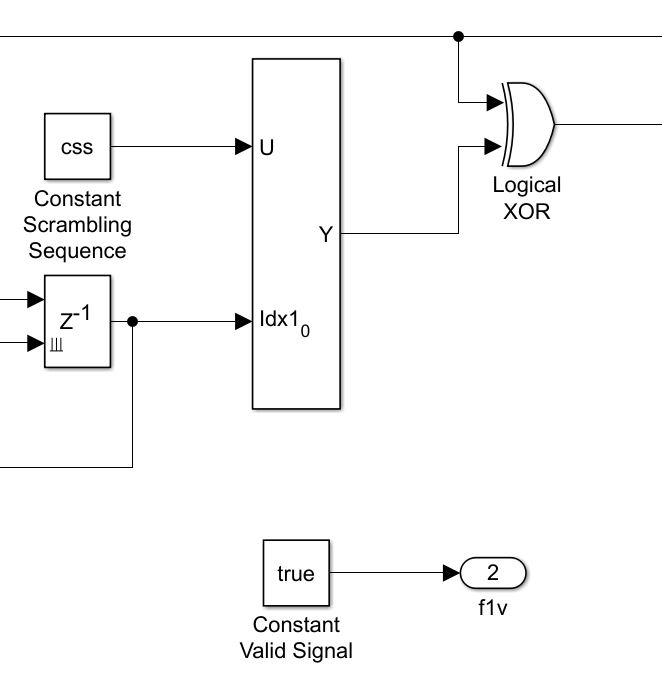}
	\caption{PS implementation of Scrambler}
	\label{scram_1}
\end{figure}

\begin{figure}[H]	
	\centering
	\includegraphics[scale=0.5]{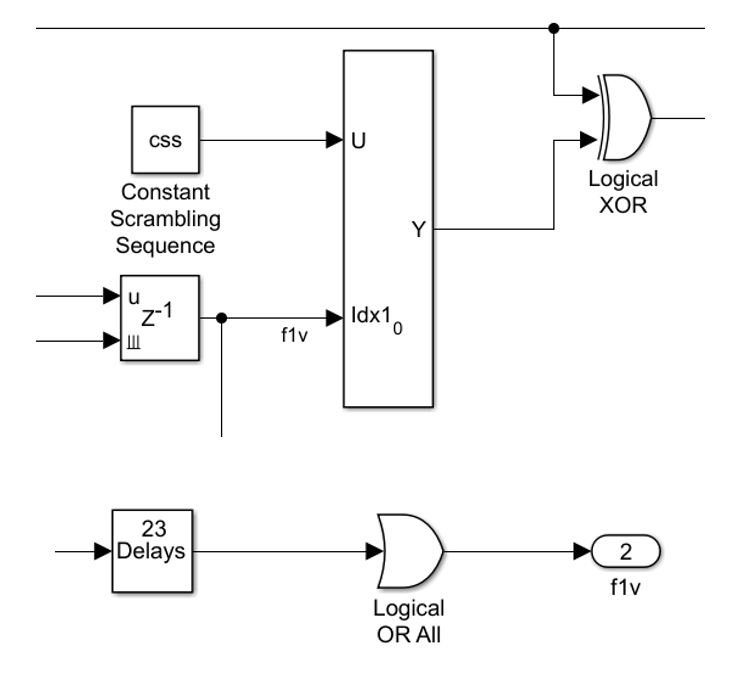}
	\caption{PL implementation of Scrambler}
	\label{scram_2}
\end{figure}

The 24 bit input stream is scrambled according to a predefined constant scrambling sequence by performing a bit-wise XOR operation. The selector block is used to select the corresponding bit of the scrambling sequence for each incoming data bit. The difference in the PS(Fig.~\ref{scram_1}) and PL(Fig.~\ref{scram_2}) implementation of the scrambler block is in the generation of valid signal due to addition of PS-PL boundary in the PL model. For the PS implementation, the valid signal is constant as true. While for the PL implementation, due to presence of PS-PL boundary prior to the scrambling block, appropriate valid signal is generated once all the 24 valid bits have been received. 

\subsubsection{Convolutional Encoder}
\begin{figure}[H]	
	\centering
	\includegraphics[scale=0.4]{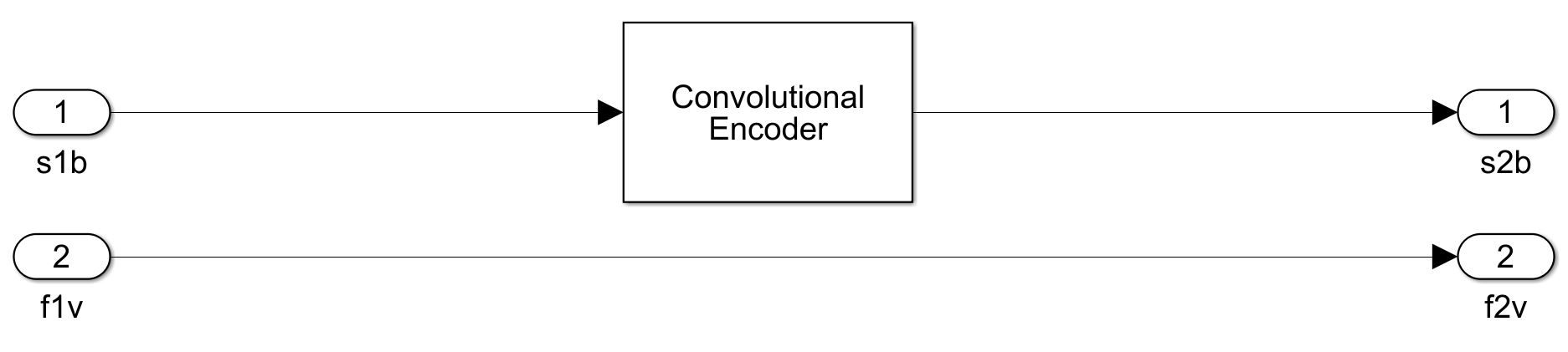}
	\caption{PS implementation of Convolutional Encoder}
	\label{ce_1}
\end{figure}

\begin{figure}[H]	
	\centering
	\includegraphics[scale=0.35]{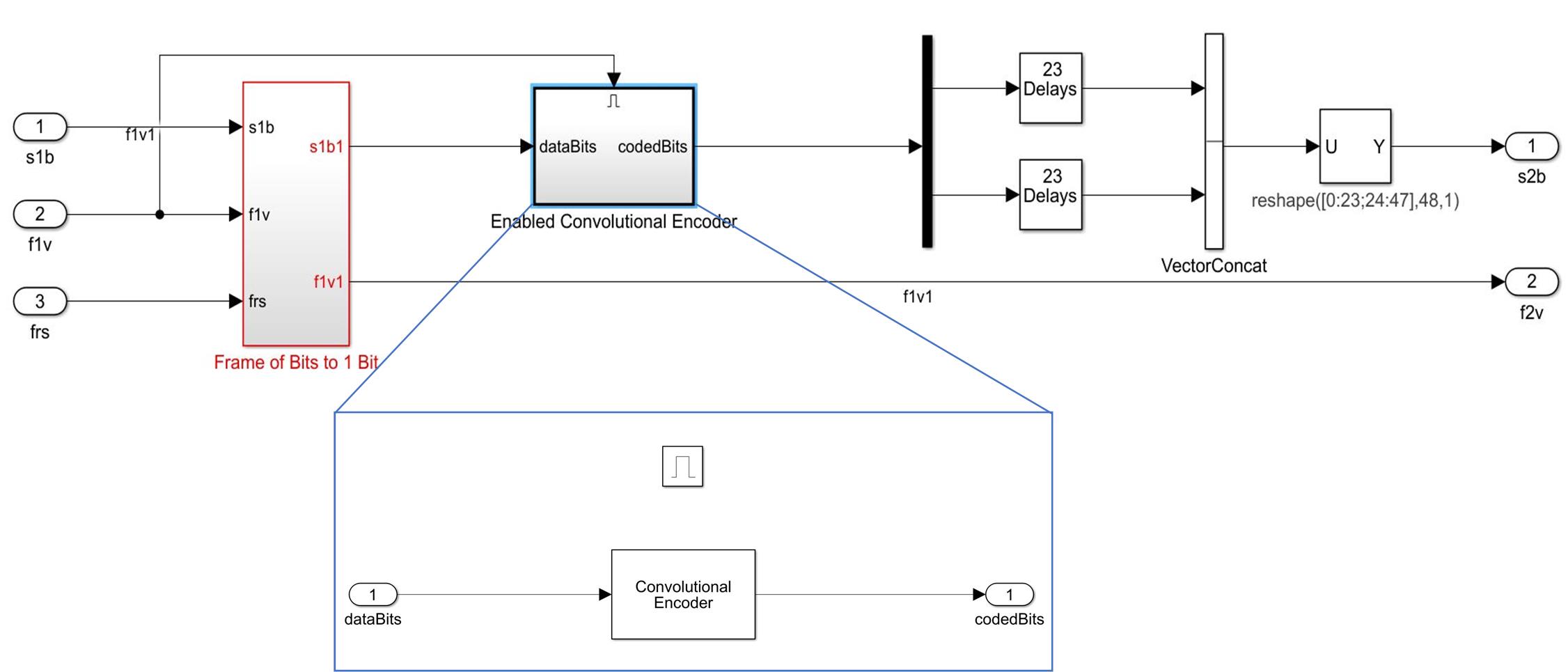}
	\caption{PL implementation of Convolutional Encoder}
	\label{ce_2}
\end{figure}

The scrambled sequence and the valid sequence are then forwarded to the convolutional encoder block. A $1/2$ rate convolutional encoder with $p_1=133$ and $p_2=171$ as the generator polynomials has been used to add error detection and correction capability at the receiver. The entire sequence is encoded using simulink convolutional encoder block for the PS implementation (Fig.~\ref{ce_1}). Since, the PL implementation is sample based, a frame to sample conversion is require prior to the convolutional encoder and a vector concatenation after encoding to retrieve the entire frame as shown in Fig.~\ref{ce_2}.

\subsubsection{Interleaver}
\begin{figure}[H]	
	\centering
	\includegraphics[scale=0.4]{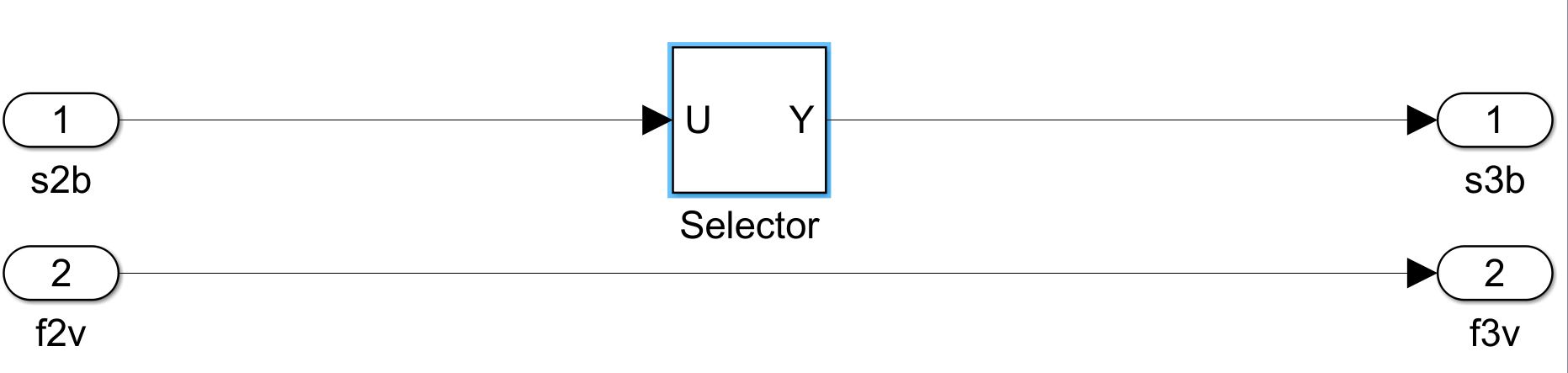}
	\caption{PS/PL implementation of Interleaver}
	\label{int}
\end{figure}

For the interleaver, selector block is used along with the interleaving index. The implementation is similar in both PS and PL as shown in Fig.~\ref{int}.

\subsubsection{BPSK Modulation}
\begin{figure}[!ht]	
	\centering
	\includegraphics[scale=0.3]{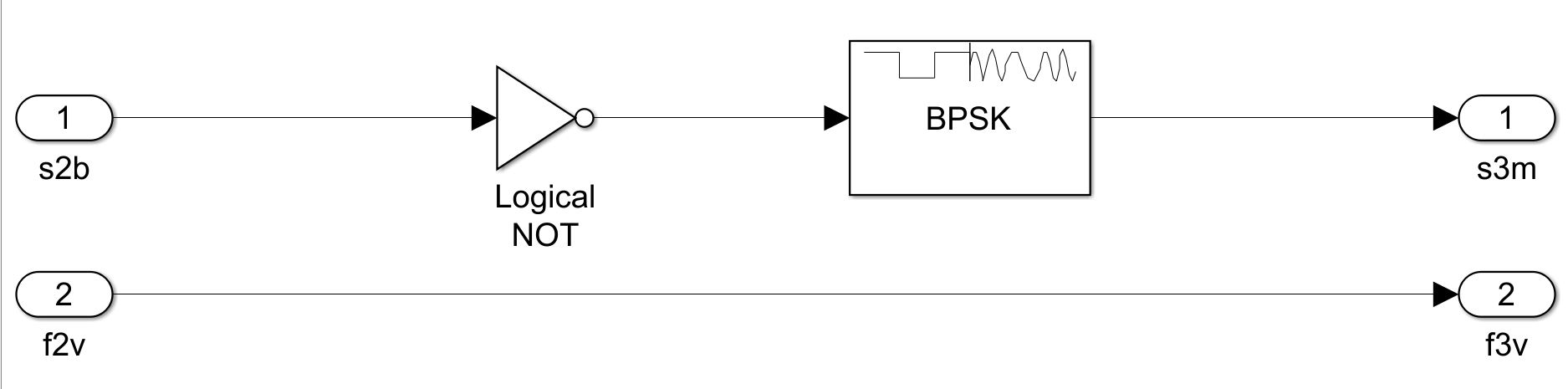}
	\caption{PS implementation of BPSK Modulation}
	\label{bpsk_1}
\end{figure}

\begin{figure}[!ht]	
	\centering
	\includegraphics[scale=0.3]{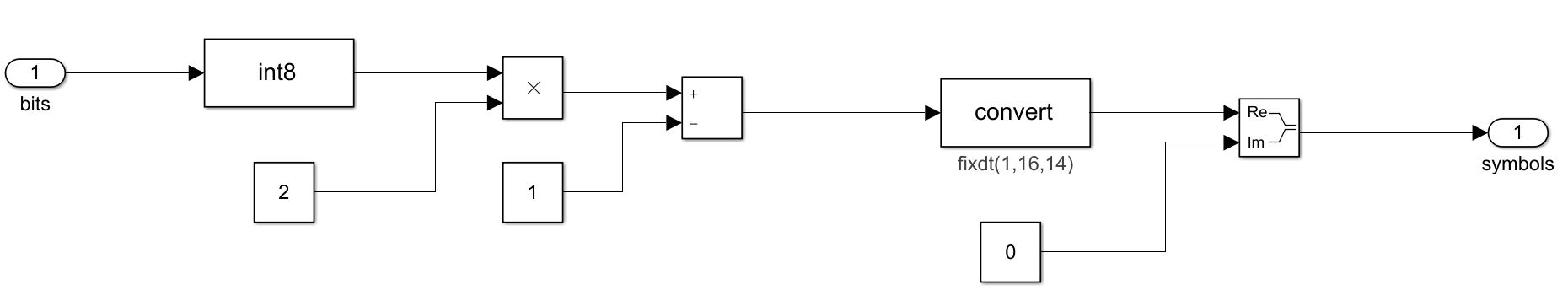}
	\caption{PL implementation of BPSK Modulation}
	\label{bpsk_2}
\end{figure}

BPSK modulation used in the architecture has $-1$ and $+1$ as the constellation points. The pre-defined BPSK baseband modulator block of the simulink communication toolbox is used in the PS implementation (Fig.~\ref{bpsk_1}). The phase offset is set to zero. Similarly, other modulation schemes can also be used. We have designed our own model for PL implementation of the BPSK modulation wherein we shift the amplitude of the incoming bit to the respective constellation points using multiplication and subtraction. The complex BPSK symbols are generated by assigning the imaginary part to be zero (Fig.~\ref{bpsk_2}). However an HDL coder block for BPSK modulation is also available.

\subsubsection{Frame Generation}
As per the LDACS specifications, 64 subcarriers are used out of which 50 subcarriers are active on which data and pilots are transmitted. The number of pilots and data subcarriers depends on the symbol index. For example: The number of data subcarriers at symbol index 0 and 1 are 36 and 48 respectively, also the pilot subcarriers are 14 and 2 respectively. There are total of 54 symbols in 1 frame and symbols at indexes 0, 51, 52 and 53 have totally different pilot positions and other symbols have 5 different pilot patterns which repeats after every 5 symbols.
\begin{figure}[!ht]	
	\centering
	\includegraphics[width=\linewidth]{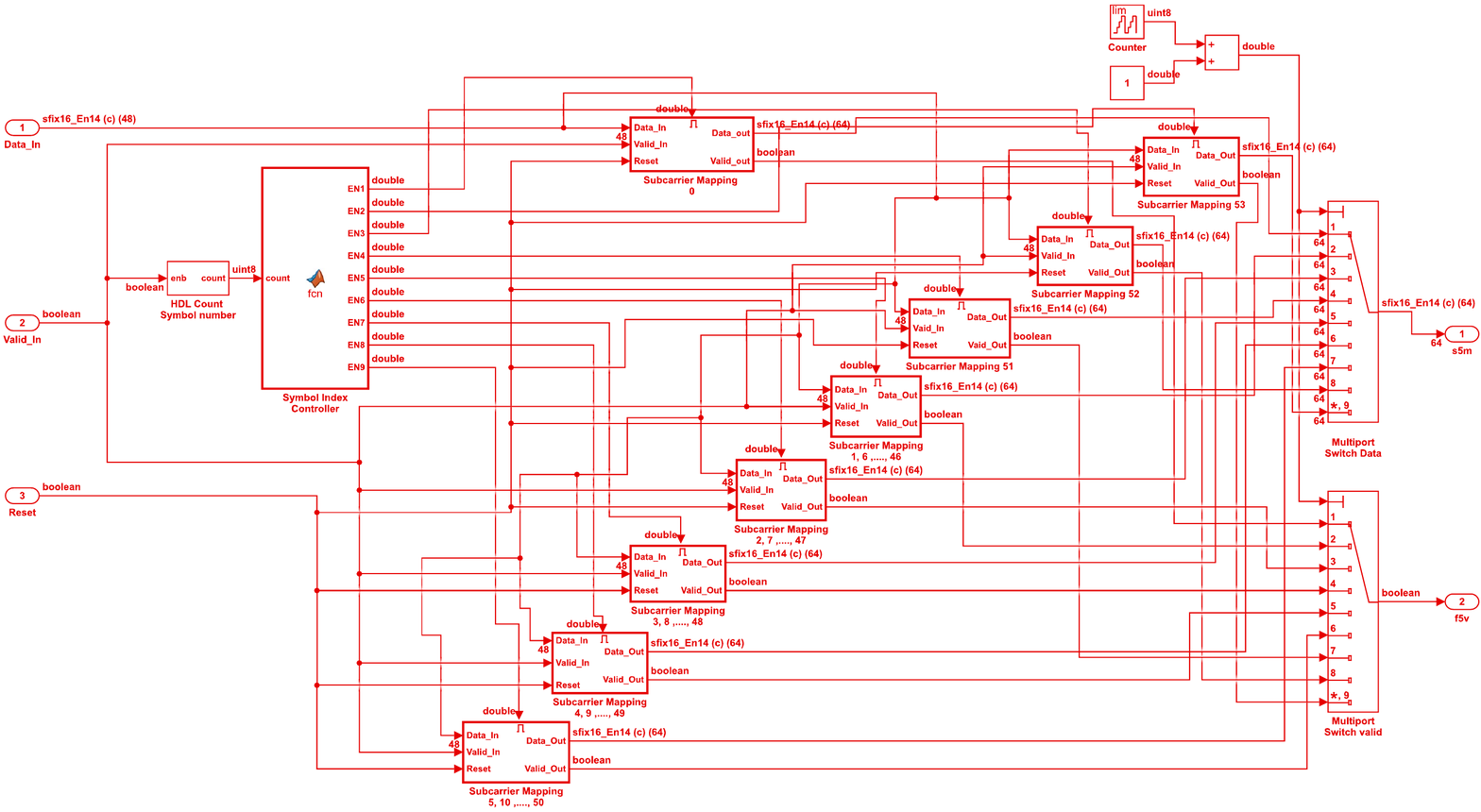}
	\caption{PS/PL implementation of Symbol to subcarrier mapping}
	\label{fr}
\end{figure}
The architecture in Fig.~\ref{fr} consists of an HDL counter to count symbol number, a Symbol Index Controller (HDL Compatible User Defined Function in Matlab), and the subcarrier mapping for different symbol indexes as major blocks. Here, the input valid signal increments the HDL counter value, and the counter counts till 53, as there are total 54 symbols in one frame. Using the counter input, which corresponds to OFDM symbol index, the symbol index controller generates nine enable signals EN1-EN9. These EN1-EN9 signals are used to enable the corresponding symbol to subcarrier mapping. We can understand this with an example, Let HDL counter gives the counter value as 1, the symbol index controller will enable the output EN5 and the EN5 will enable the subcarrier mapping designed for symbol index 1. The EN5 will also be enabled when the counter values are 6, 11, \ldots, 46 as these symbols follows the same pilot pattern as symbol index 1. There are 7 different pilot patterns generated similar to the reference MATLAB codes shared by the German Aerospace Center (DLR) provided on https://www.ldacs.com/software/. These pilot patterns are used to map the pilots in the corresponding subcarrier mapping according to the symbol index. The subcarrier mapping is similarly designed for symbols 53 and 54, which have zeros and synchronization symbols instead of pilot symbols.

\begin{figure}[!h]
    \centering
    \subfloat[]{\includegraphics[scale=0.23]{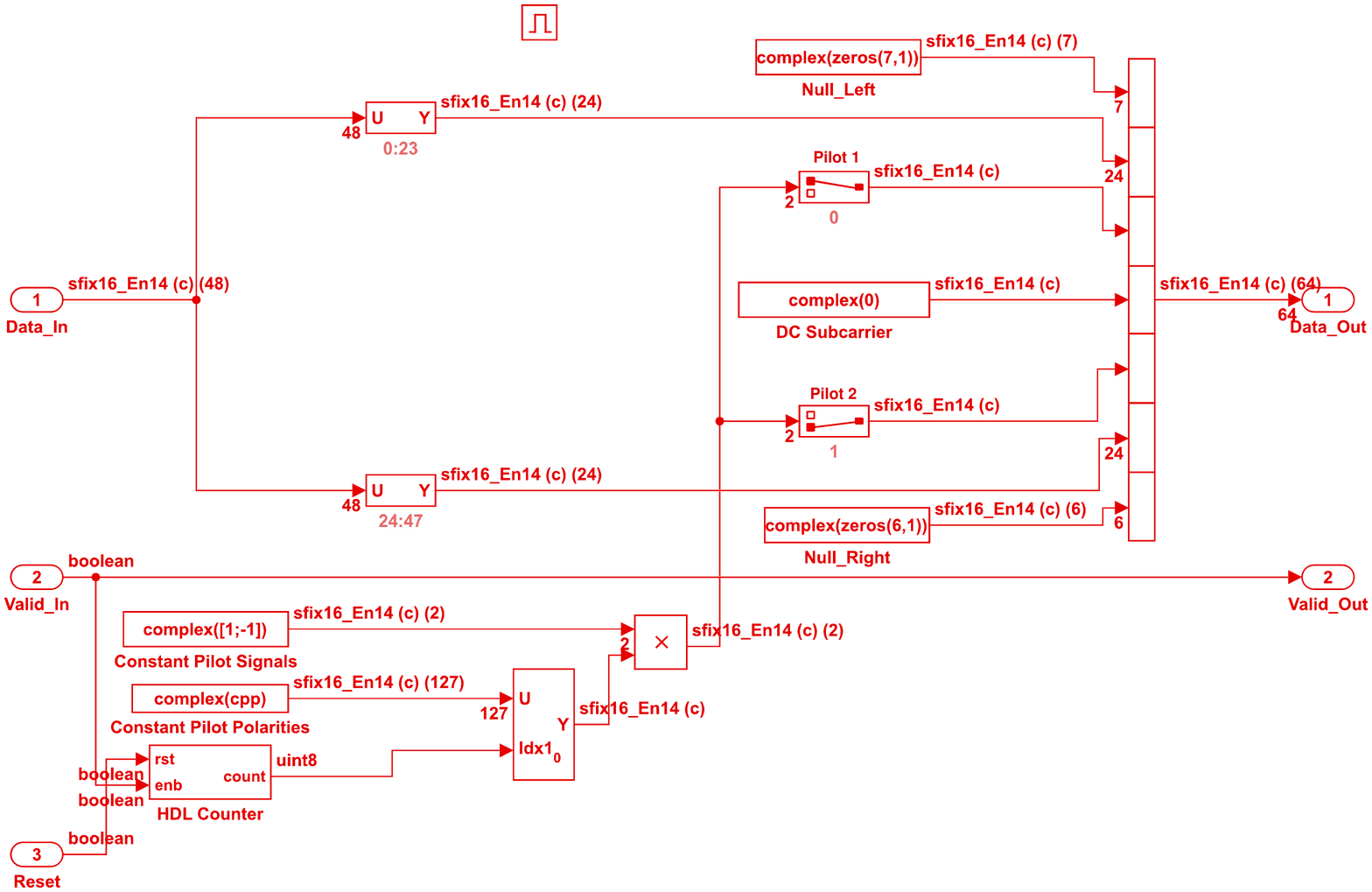}}
\vspace{-0.3cm}
\hspace{0.5cm}
    \subfloat[]{\includegraphics[scale=0.28]{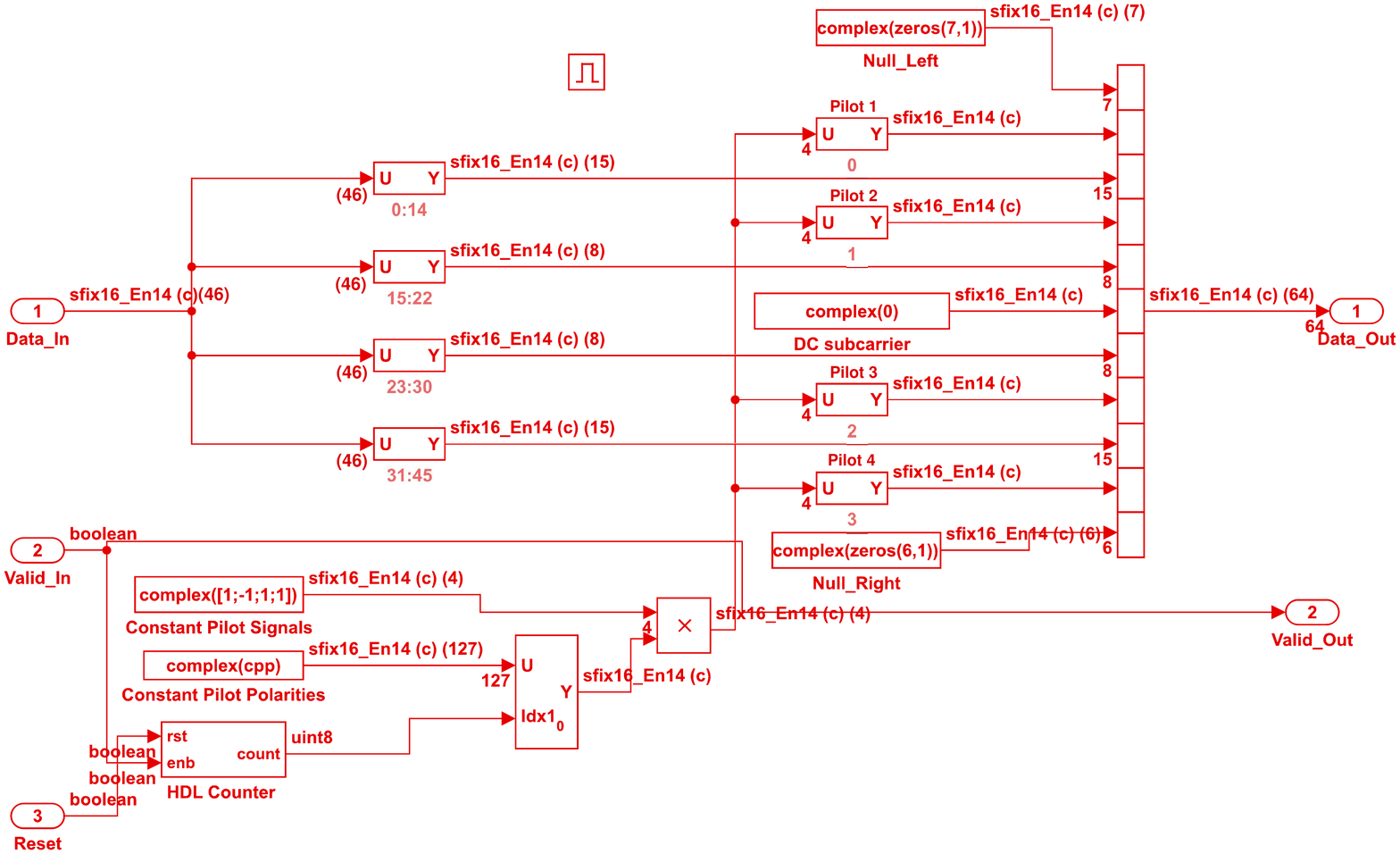}}
     \vspace{-0.2cm}
        \caption{PS/PL implementation of Symbol to Subcarrier mapping for (a) symbol index 1, 6, 11 \ldots 46, and (b) symbol indexes 3, 8, 13 \ldots 48}
        \vspace{-0.2cm}
            \label{sym}
\end{figure}

Here, we are showing the subcarrier mapping of symbol indexes 1, 6, 11, \ldots, 46 having 48 data , two pilots positioned at $32^{nd}$ and $34^{th}$ subcarrier,
, 1 DC (in middle), and 13 Null subcarriers (7 left subcarrier + 6 right subcarrier position) (as shown in the appendix), and symbol indexes 3, 8, 13, \ldots, 48 having 46 data, four pilot positioned at $8^{th}$, $24^{th}$, $42^{nd}$ and $58^{th}$ subcarriers, 1 DC (in middle), and 13 Null subcarriers (7 left +6 right) in Fig.~\ref{sym} (a) and (b) respectively.

\subsubsection{Preamble Addition}
\begin{figure}[!ht]	
	\centering
	\includegraphics[scale=0.3]{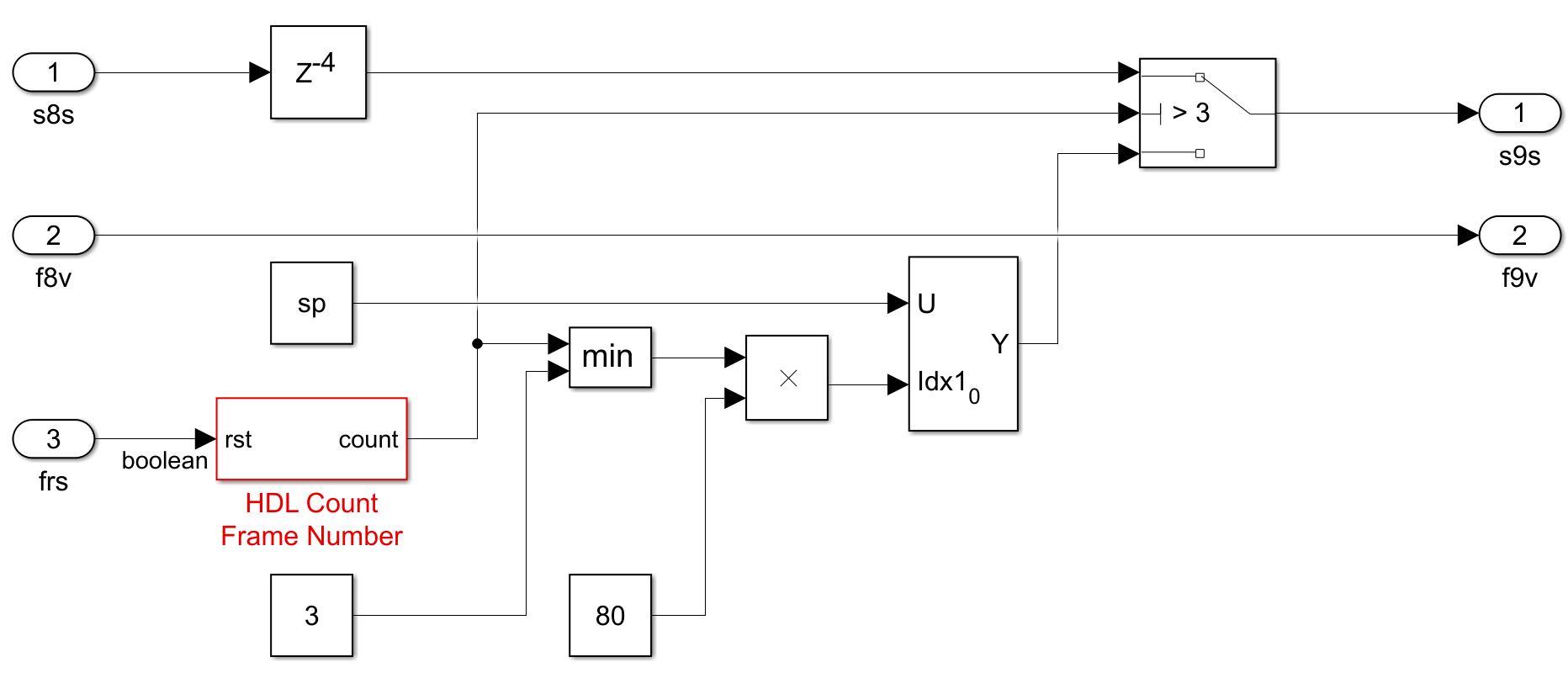}
	\caption{PS implementation of Preamble Addition}
	\label{pre_1}
\end{figure}

\begin{figure}[!ht]	
	\centering
	\includegraphics[scale=0.3]{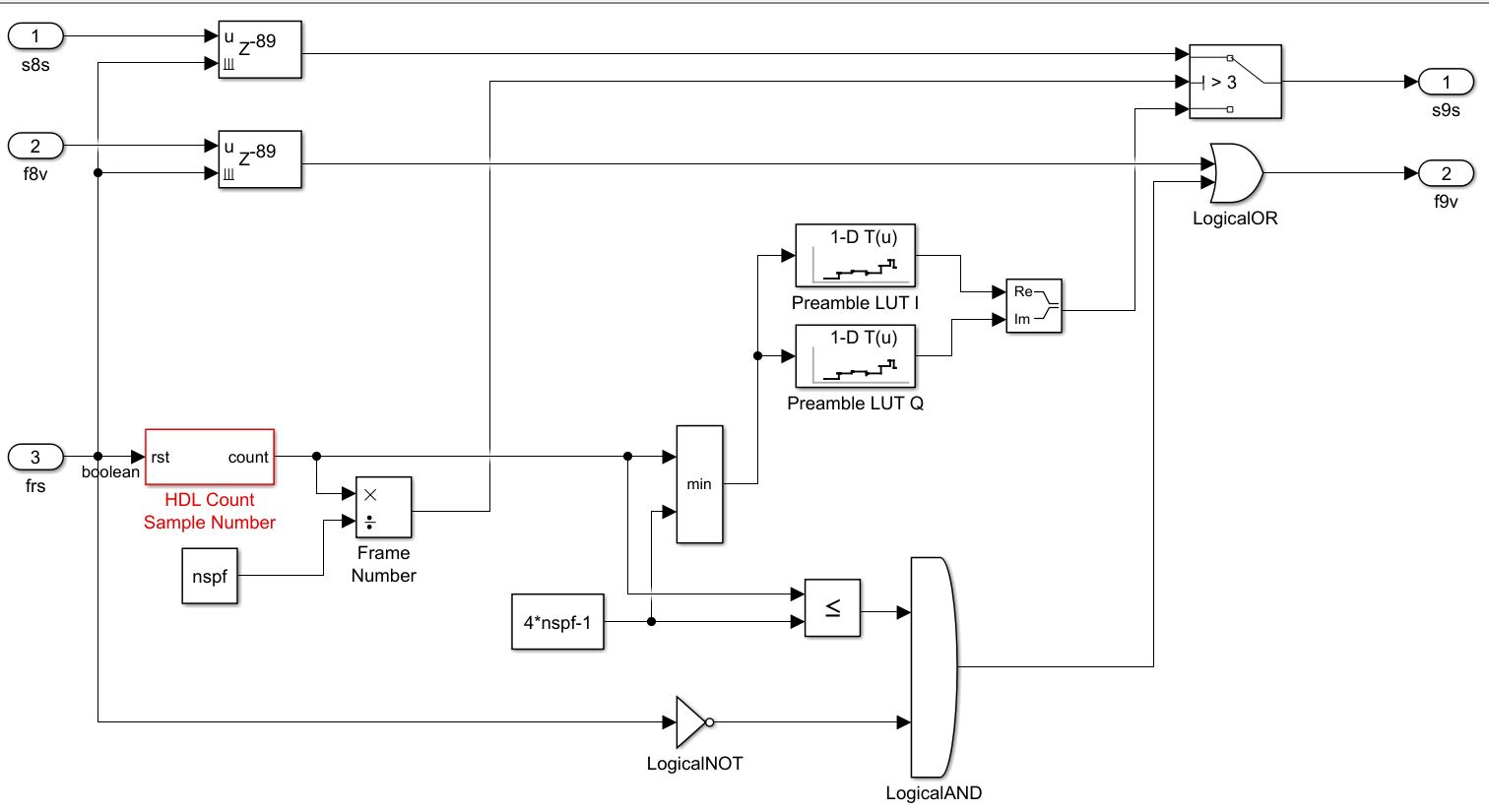}
	\caption{PL implementation of Preamble Addition}
	\label{pre_2}
\end{figure}

The preamble sequence is predefined and has both long and short preamble sequence. For the PS implementation Fig.~\ref{pre_1}, a counter is used to detect the frame number. For the first 4 frames, the preamble (both short and long) is transmitted. The preamble sequence is read from the workspace. Of the 320 samples, 80 samples to be transmitted are selected depending on the frame number. For the PL implementation, the preamble sequence is stored in LUTs each for short and long preamble. The data and the valid signal to be transmitted are then decided depending on the sample number detected using a counter Fig.~\ref{pre_2}.

\pagebreak
\vspace{10cm}
\subsection*{Receiver}
\subsubsection{Preamble Detection}
\begin{figure}[!ht]	
	\centering
	\includegraphics[scale=0.5]{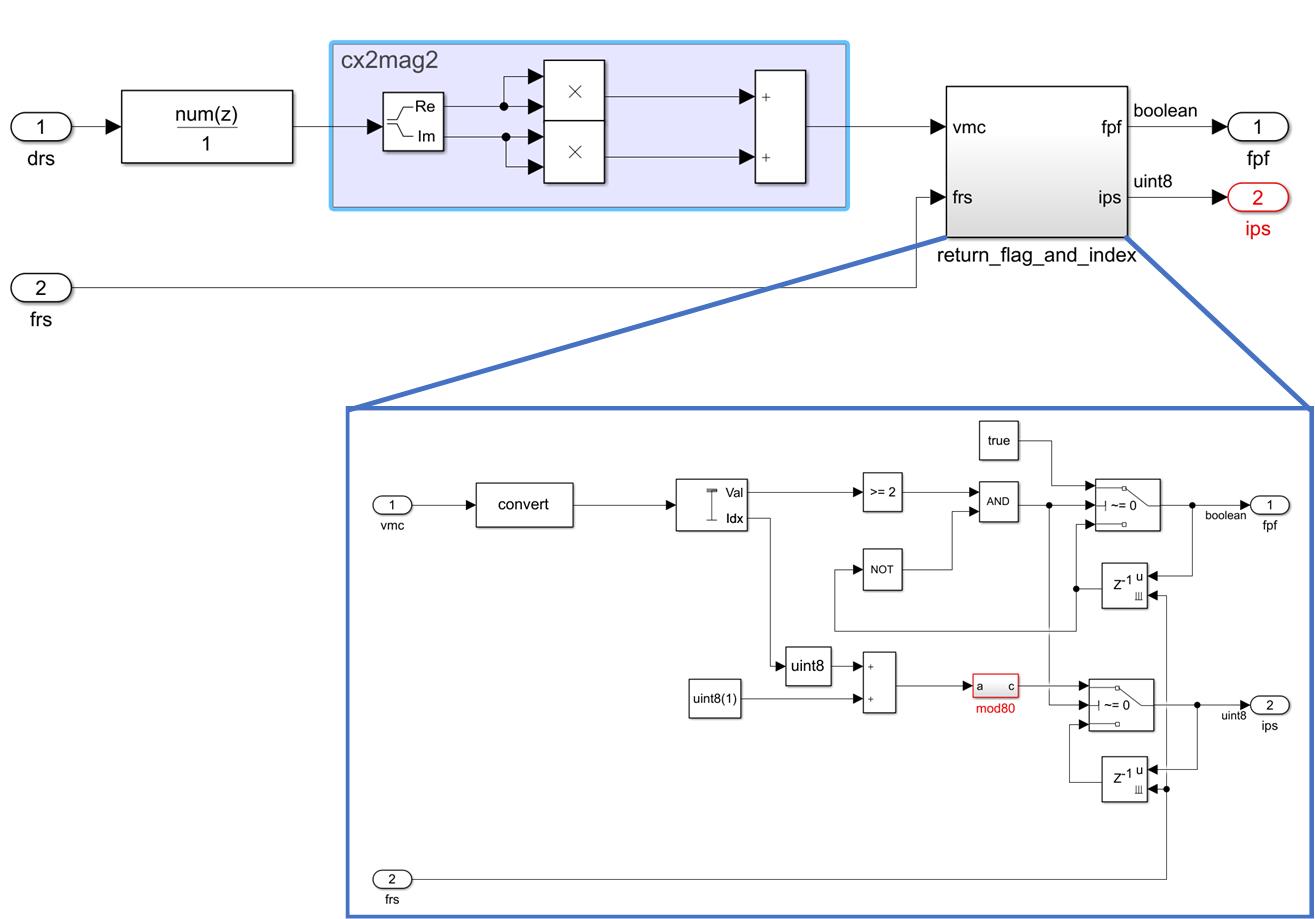}
	\caption{PS/PL implementation of Preamble Detection}
	\label{preD_2}
\end{figure}

The valid signal from the transmitter now enables the receiver functionality. The first step in the receiver is preamble detection which uses auto-correlation to detect the data frames. The auto-correlation is performed using a filter and a magnitude detector to detect the peak Fig.~\ref{preD_2}. The implementation of preamble detection is same in both PS and PL.


\subsubsection{BPSK Demodulator}
\begin{figure}[!ht]	
	\centering
	\includegraphics[scale=0.4]{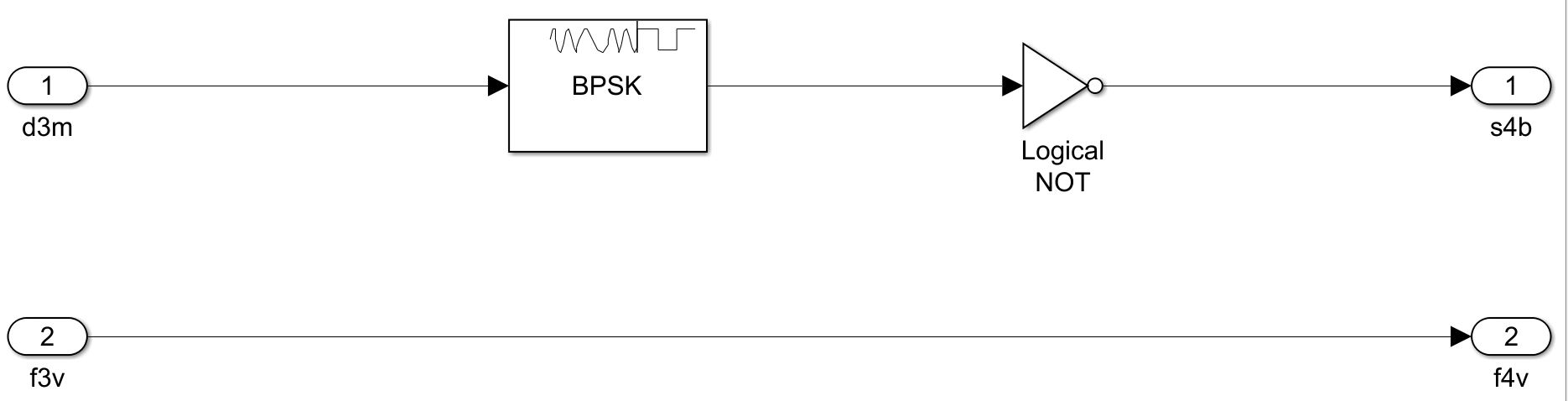}
	\caption{PS implementation of BPSK Demodulation}
	\label{bpskD_1}
\end{figure}

\begin{figure}[!ht]	
	\centering
	\includegraphics[scale=0.5]{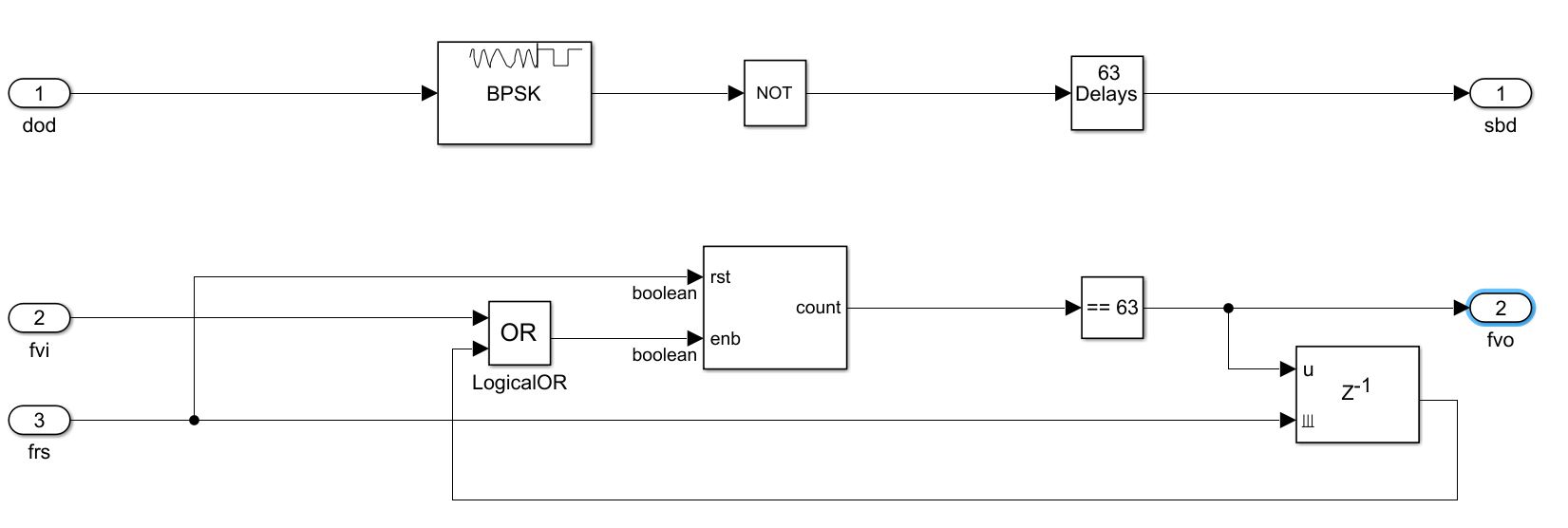}
	\caption{PL implementation of BPSK Demodulation}
	\label{bpskD_2}
\end{figure}

Using the simulink BPSK baseband demodulator block we retrieve the bi stream from BPSK symbols in both the PS and PL implementations as shown in (Fig.~\ref{bpskD_1}) and (Fig.~\ref{bpskD_1}). However, delays are used in the PL implementation to generate an appropriate valid signal to keep the data integrity and get the complete frame of 64 bits required for the deinterleaving process ahead (Fig.~\ref{bpskD_2}). 

\subsubsection{Deinterleaver}
The bitstream is deinterleaved using the pre-defined deinteleaver sequence and a selector block similar to the interleaving process discussed above. 

\subsubsection{Viterbi Decoder}
The viterbi decoder block from communications system toolbox is used to decode the data. For the PS implementation, the complete frame is fed to the decoder to generate the output frame consisting of the decoded bits (Fig.~\ref{vd_1}). While in the PL implementation, sample counter is used to monitor the incoming bits corresponding to each output sample bit as shown in Fig.~\ref{vd_1}. 
\begin{figure}[H]	
	\centering
	\includegraphics[scale=0.4]{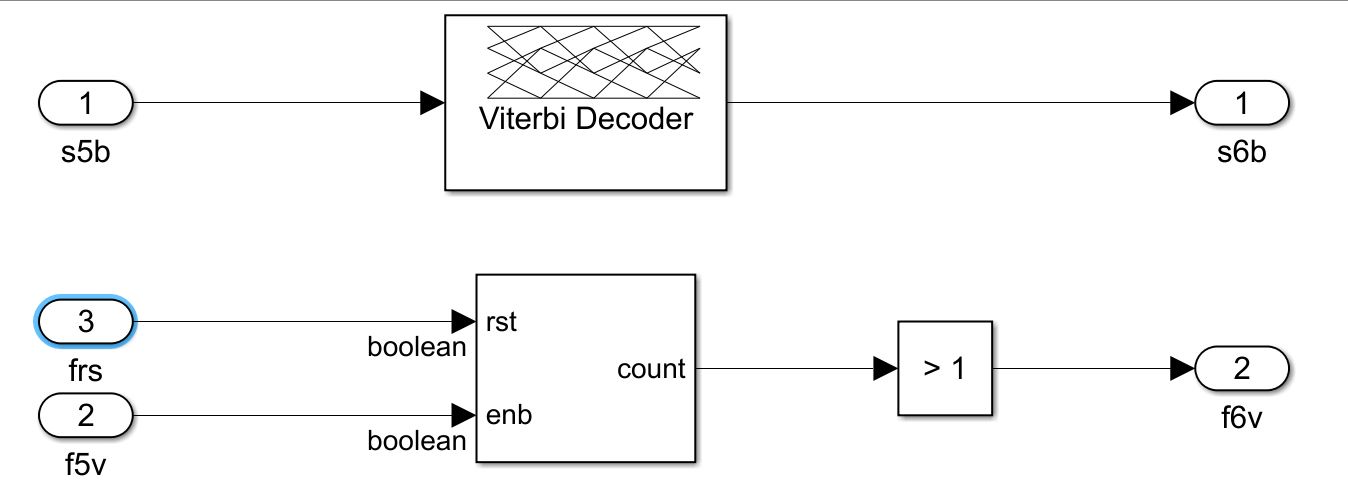}
	\caption{PS implementation of Viterbi Decoder}
	\label{vd_1}
\end{figure}

\begin{figure}[H]	
	\centering
	\includegraphics[scale=0.3]{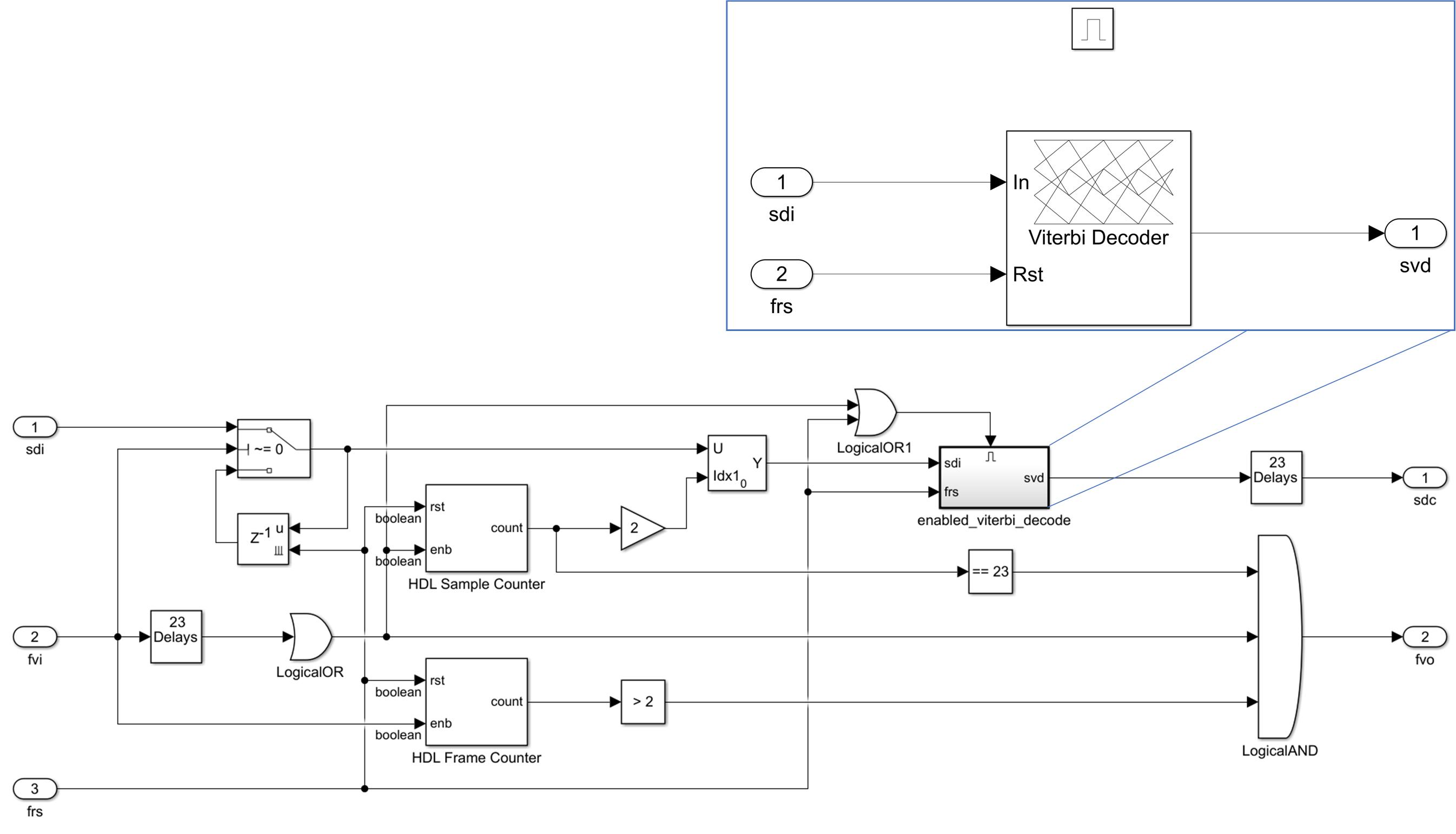}
	\caption{PL implementation of Viterbi Decoder}
	\label{vd_2}
\end{figure}

\subsubsection{Descrambler}
A predefined descrambling sequence is then used to retrieve the original 24 bits using an XOR operation in both PS and PL. While no additional functionality is required in the PS, in PL the decrambler in enabled upon receiving the appropriate valid signal and input bit. 
Fig.~\ref{vd_1}. 
\begin{figure}[H]	
	\centering
	\includegraphics[scale=0.5]{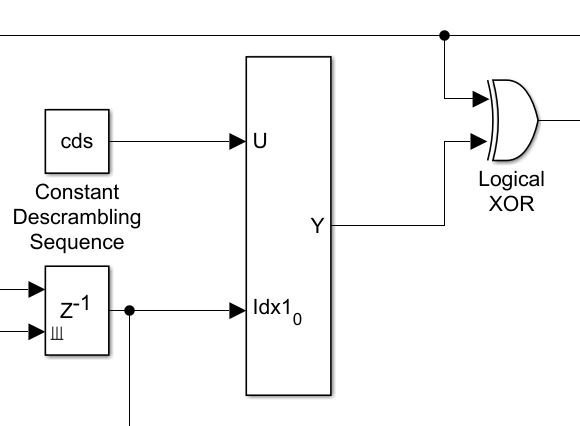}
	\caption{PS implementation of Descrambler}
	\label{vd_1}
\end{figure}

\begin{figure}[H]	
	\centering
	\includegraphics[scale=0.5]{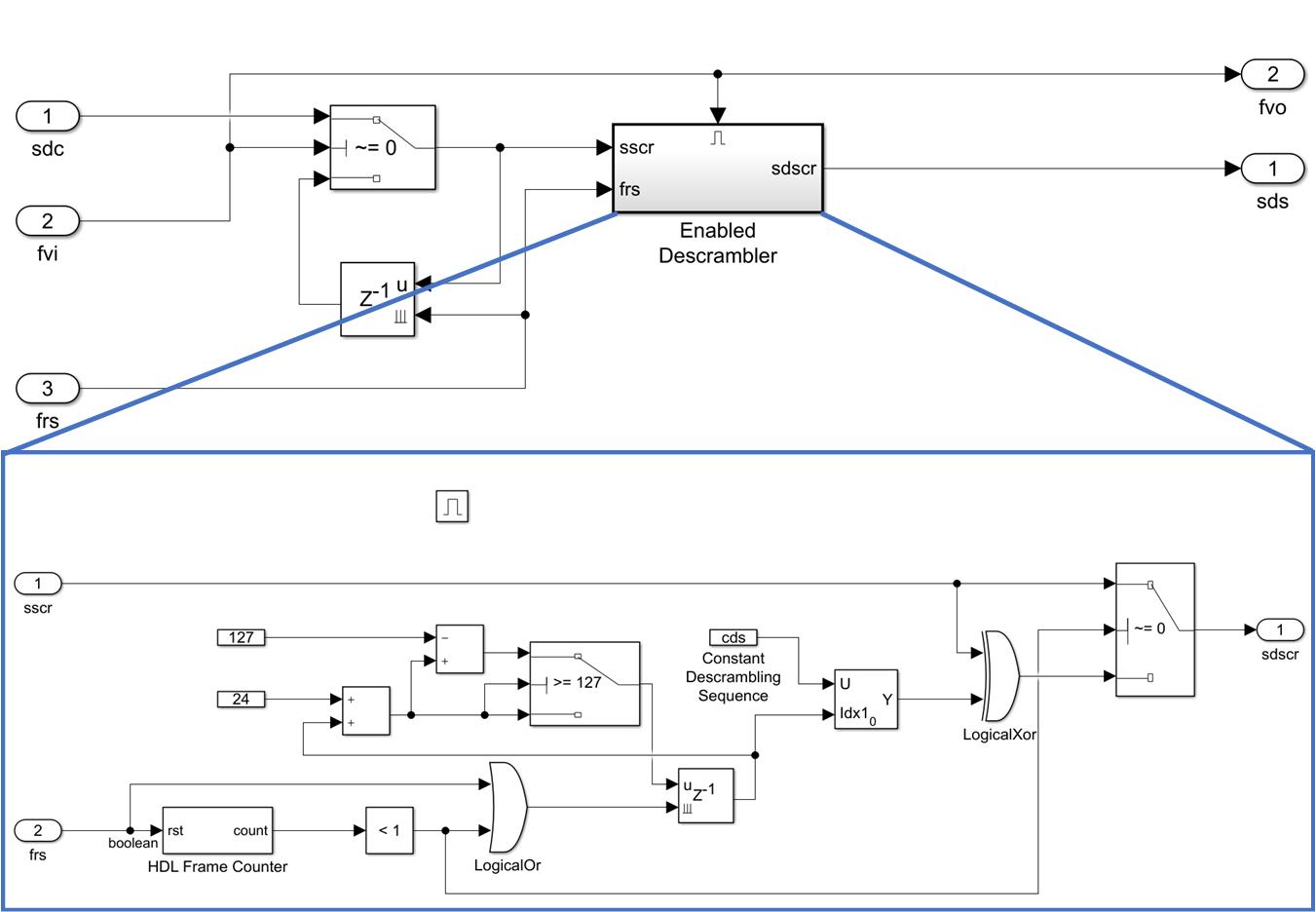}
	\caption{PL implementation of Descrambler}
	\label{vd_2}
\end{figure}

next, we discuss the workflow for Hardware - software co-design approach used for the transceiver implementation.

\subsection*{\textbf{Hardware - Software co-design workflow}}
To design and simulate the transceiver models Hardware - Software Co-Design approach is being used. It is an important approach to implement any algorithm on ZSoC as it utilizes the heterogeneity of PS and PL. This approach also gives the flexibility to choose which part of the system is best suited to be implemented on PL and which on PS. PS makes easy and faster decision-making operations on the other hand PL reduces power consumption and increases speed. The steps for hardware-software Co-Design approach are as follows: 

\begin{enumerate}
    \item 
    Designing a Simulink model for transceivers and set the parameters like number of samples per frame, sampling frequency, total FFT size, Active subcarriers, and subcarrier spacing. All the blocks present in the Simulink library are not hardware synthesizable. So, while designing the Simulink model these blocks need to be avoided.
    \item Differentiate the subsystem of the model which is going to implement on the PL believing that all the other subsystems will target to implement on PS. PL works in sample mode, and PS works in frame mode, which requires an appropriate sample to frame and frame to sample conversion at the boundary of PS-PL interface. Fig.~ \ref{flow} shows the design have $N$ functional blocks. Transmitter subsystem consisting of blocks $1_{T}, 2_{T} ,3_{T}...i_{T}$ are implemented on PS, and remaining blocks are implemented on PL. Similar process is used for the receiver operations. Note that, the output to the host computer will come back through the PS.
    
    \begin{figure*}[!h]
    \centering
        \captionsetup{justification=centering}
        \includegraphics[width=\linewidth]{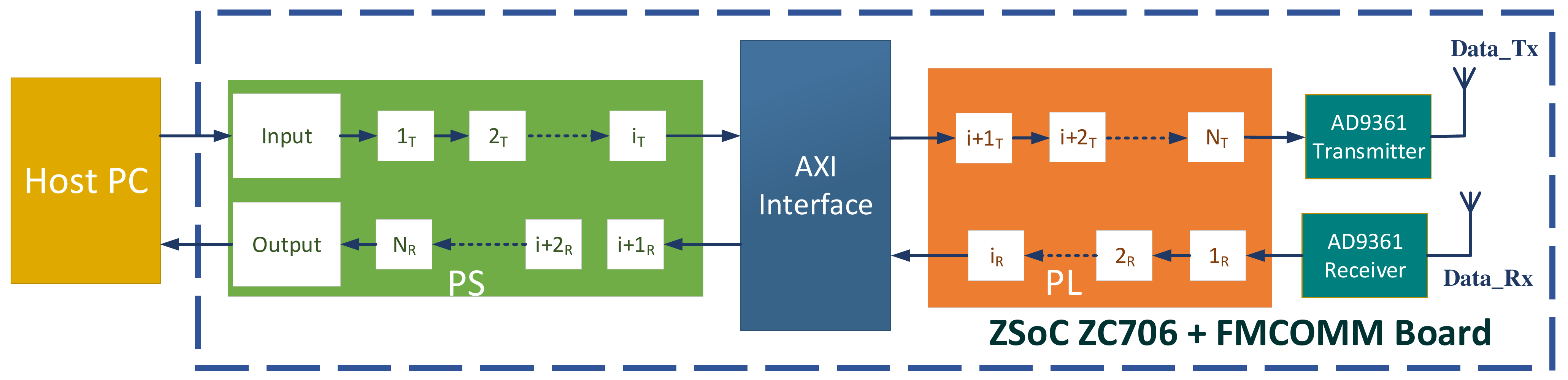}
        \caption{Hardware - Software Co-Design approach for algorithm implementation}
        \label{flow}
    \end{figure*}
    
    \item
    Then, run the HDL Workflow Advisor to auto-generate an IP Core block for the transceiver design as shown in Fig.~\ref{Workflow}. It automatically generates a Vivado block diagram to combine the DUT with all the AXI interface components and creates an interface model to interact with the PL. It then uses the HDL coder and Xilinx Vivado for synthesis, implementation and bitstream generation. This bitstream is then used to program the PL.
    
    \begin{figure}[!h]
    \centering
	\includegraphics[scale=0.5]{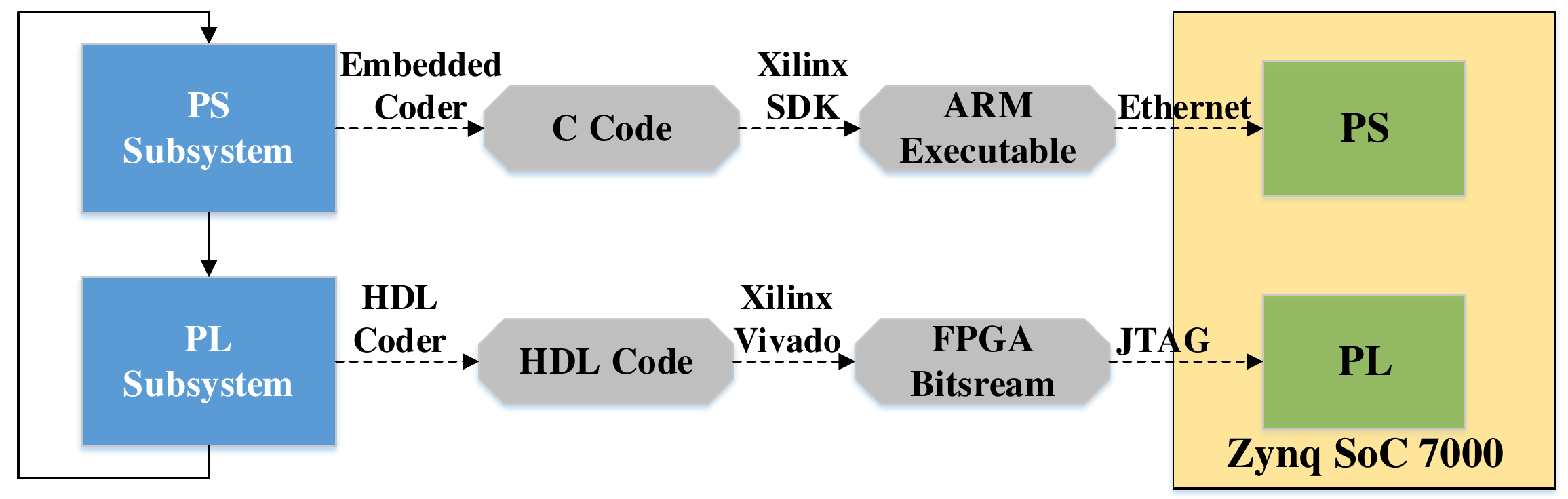}
	\caption{Hardware-Software work flow for ZSoC using HDL and Embedded coders of Matlab/Simulink and Xilinx Vivado.}
	\label{Workflow}
\end{figure}

    \item
    Finally, by setting the generated interface model to run in external mode, Simulink uses Embedded Coder to generate C code for all the processing blocks. Xilinx Vivado SDK then converts this C code to ARM executable code. When we run the simulation, it launches the executable on PS via Ethernet. 
\end{enumerate}